\documentclass[paperpaper,twocolumn,english,superscriptaddress,aps,prx,floatfix,amsfonts,amssymb]{revtex4-1}
\usepackage{float}
\usepackage{epsfig}
\usepackage{graphicx}% Include figure files
\usepackage{dcolumn}% Align table columns on decimal point
\usepackage{bm}% bold math
\usepackage{appendix}
\usepackage{subfigure}
\usepackage{color}
\usepackage{natbib}

\begin{document} 
\title{Ising spin-$1/2$ $XXZ$ chain's quantum problems beyond the spinon paradigm}

%\date{}

\author{J. M. P. Carmelo}
\affiliation{Center of Physics of University of Minho and University of Porto, LaPMET, P-4169-007 Oporto, Portugal}
\affiliation{CeFEMA, Instituto Superior T\'ecnico, University of Lisbon, LaPMET, P-1049-001 Lisbon, Portugal}
\author{P. D. Sacramento}
\affiliation{CeFEMA, Instituto Superior T\'ecnico, University of Lisbon, LaPMET, P-1049-001 Lisbon, Portugal}
\begin{abstract}
{\bf Author to whom correspondence should be addressed: carmelo@fisica.uminho.pt}\\
This paper is dedicated to David K. Campbell's 80th Birthday.
Spin chains are correlated quantum models of great interest in quantum systems and materials exhibiting quasi-one-dimensional 
magnetic properties. Here we review results on quantum problems associated with spin chains that are 
beyond the usual spinon paradigm. Alternatively, we use a representation valid in the thermodynamic limit, $N\rightarrow\infty$,
in terms of the $N$ spin-$1/2$ physical spins of the spin-$1/2$ $XXZ$ chain in its whole Hilbert space.
It was originally introduced for the isotropic point in Ref. \onlinecite{Carmelo_15A}, co-authored by
David, and more recently extended to spin anisotropies $\Delta >1$ in Refs. \onlinecite{Carmelo_23} and \onlinecite{Carmelo_22}. 
The physical-spins representation accounts for the spin-$1/2$ $XXZ$ chain's 
continuous $SU_q(2)$ symmetry parametrized by $q = \Delta + \sqrt{\Delta^2-1}\in ]1,\infty]$ and
associated with $q$-spin $S_q$. Specifically, in this review we consider 
two quantum problems that are beyond the spinon representation: (a) Spin Bethe strings of length $n$ that have no spinon representation,
contribute to the dynamical properties of the spin-$1/2$ $XXZ$ chain with anisotropy $\Delta >1$ and
for $n=1,2,3$ were experimentally identified and realized in the zigzag materials SrCo$_2$V$_2$O$_8$ 
and BaCo$_2$V$_2$O$_8$; (b) The spin stiffness associated with ballistic spin transport at arbitrary finite temperature,
which involves a huge number of energy eigenstates, many of which are 
generated in the thermodynamic limit from ground states by an infinite number of elementary processes. 
As found in Refs. \onlinecite{Carmelo_23} and \onlinecite{Carmelo_22}, the use of the continuous $SU_q(2)$ symmetry
reveals that for anisotropy $\Delta > 1$ the Bethe strings of length $n=1,2,3,...$ describe a number $n$ of physical-spins
$S_q=0$ singlet pairs that for $n>1$ are bound within a $S_q=0$ singlet configuration. Their contribution
to the spin dynamical structure factor of both the spin-$1/2$ $XXZ$ chain in a longitudinal magnetic field
and the spin chains in SrCo$_2$V$_2$O$_8$  is one of the issues addressed in this paper.
In addition, the $SU_q(2)$ symmetry imposes that only $2S_q$ out of the $N$ physical spins are the spin carriers. 
We also review recent results of Ref. \onlinecite{Carmelo_24} concerning the vanishing
of the contributions to finite-temperature ballistic spin transport at zero magnetic field.
Within the physical-spins representation, this merely follows from the absolute value of the elementary spin currents carried by
the $M=2S_q$ spin carriers of all finite-$S_q$ states that contribute to the spin stiffness being finite. 
Finally, we discuss deviations of the zigzag materials BaCo$_2$V$_2$O$_8$ and SrCo$_2$V$_2$O$_8$  
from the one-dimensional physics described the spin-$1/2$ $XXZ$ chain, due to selective interchain couplings. 
\end{abstract}
%\pacs{}
\maketitle
{\bf One-dimensional systems are renowned for their ability to host ground states and phases 
markedly different from their higher-dimensional counterparts. Heisenberg spin-$1/2$ chains are the 
archetype of quantum integrable one-dimensional models describing magnetic properties of a wide 
range of systems. Here we consider the so called Heisenberg spin-$1/2$ $XXZ$ chain 
in the gapped Ising regime of spin anisotropy $\Delta >1$ whose
complex Bethe strings of length two and three were experimentally identified and realized in the 
zigzag materials SrCo$_2$V$_2$O$_8$ and BaCo$_2$V$_2$O$_8$.
We review results on quantum problems associated with that spin chain
beyond the spinon paradigm, namelly results of Refs. \onlinecite{Carmelo_23} and \onlinecite{Carmelo_22}
on such Bethe strings and very recent results of Ref. \onlinecite{Carmelo_24} 
on spin transport. Our study reviews and
relies on a representation in terms of the spin-$1/2$ $XXZ$ chain's $N$ physical spins that 
captures hidden underlying symmetries. Its use reveals that Bethe $n$-strings
involve a number $n$ of singlet pairs of physical spins and that
at zero magnetic field spin transport is non-ballistic for spin anisotropy $\Delta >1$ 
and all temperatures. We also discuss deviations of the 
materials BaCo$_2$V$_2$O$_8$ and SrCo$_2$V$_2$O$_8$ from the one-dimensional 
physics described the spin-$1/2$ $XXZ$ chain in a longitudinal magnetic field, 
due to selective interchain couplings.}

\section{Introduction}
\label{SECI}

This paper is dedicated to David K. Campbell's 80th Birthday, collaborator of both authors 
in the study of several one-dimensional (1D) many-particle quantum models. In the case
of J. M. P. C, for more than thirty years \cite{Carmelo_93,Carmelo_94,Carmelo_94A,Carmelo_94B,Carmelo_95,Baeriswyl_95,Carmelo_95A,Peres_97,Carmelo_98,Peres_99,
Carmelo_01,Carmelo_15,Carmelo_15A,Carmelo_19,Carmelo_19A,Carmelo_23}.
Such quantum models are non-perturbative systems whose
low-energy one-particle properties cannot be described by Fermi-liquid theory
\cite{Carmelo_93,Carmelo_94,Carmelo_94A,Carmelo_94B,Carmelo_95,Baeriswyl_95,Carmelo_95A,Peres_97,Carmelo_98,Peres_99,
Carmelo_01,Carmelo_15,Carmelo_15A,Carmelo_19,Carmelo_19A,Carmelo_23,Baeriswyl_87}.

The spin-$1/2$ $XXZ$ chain \cite{Heisenberg,Bethe} is a 1D model of spins $1/2$ with a coupling constant $J$
and spin anisotropy $\Delta$ \cite{Gaudin_71}. At the isotropic point, $\Delta =1$, it was the first quantum system ever to be 
solved by the Bethe ansatz in 1931 \cite{Bethe}. This spin-chain model remains of great interest due 
to its underlying richness. Quantum spin chains are actually some of the most intensively studied 1D models
\cite{Carmelo_15,Carmelo_15A,Carmelo_23,Gaudin_71,Carmelo_22,Carmelo_24,Gaudin_14,Takahashi_71,Takahashi_99,
Imambekov_12,Carmelo_18,Carmelo_17,Carmelo_20}. They represent strongly fluctuating quantum many-body systems because of their amenability to exact analysis and because of the sustained interest in materials exhibiting quasi-1D magnetic properties
\cite{Hikihara_04,Kimura_07,Okunishi_07,Kimura_08,Canevet_13,Okutani_15,Klanjsek_15,Grenier_15,Dupont_16,Shen_19,Han_21,Scheie_21,Cui_22,Shen_22}. For instance, the complex strings in the spin-$1/2$ $XXZ$ chain's Bethe-ansatz solution 
were for anisotropy $\Delta \approx 2$ experimentally identified and realized in quasi-1D materials
\cite{Carmelo_23,Wang_18,Wang_19,Bera_20}.

In the last three to four decades, representations in terms of spinons and 
similar quasi-particles such as psinons and antipsinons \cite{Karbach_02} have been widely used to successfully describe the 
static and dynamical properties of both spin-chain models and the physics of the materials they describe. 
Hence such representations became the paradigm of the spin-chains physics.
Those spinon and alike representations apply to physical quantities controlled by excited states of both low 
and high energy that in the thermodynamic limit are generated from the ground state by a finite number of 
elementary processes. 

However, there are quantum problems associated with spin chains that are beyond the spinon paradigm. 
This review focuses on two such quantum problems. To achieve that goal, we review and use a physical-spins representation of the spin-$1/2$ $XXZ$ 
chain in its whole Hilbert space for anisotropy $\Delta > 1$ that accounts for the model symmetries irreducible representations.
It was originally introduced in Ref. \onlinecite{Carmelo_15A} 
for the isotropic point. Recently, it was extended to spin anisotropies $\Delta >1$ \cite{Carmelo_23,Carmelo_22}.

The first such quantum problems beyond the spinon paradigm refers to spin Bethe strings and their contribution 
to dynamical properties. Complex $n$-strings of length $n>1$ \cite{Gaudin_71,Gaudin_14,Takahashi_71,Takahashi_99}
are associated with Bethe-ansatz quantum numbers that describe spin configurations that have 
no spinon representation and exist in some energy eigenstates.
In spite of their theoretical origin, as mentioned above they were experimentally identified and 
realized in the zigzag materials SrCo$_2$V$_2$O$_8$ and BaCo$_2$V$_2$O$_8$
\cite{Carmelo_23,Wang_18,Wang_19,Bera_20}. 

As found in Refs. \onlinecite{Carmelo_23} and \onlinecite{Carmelo_22}, 
the use of the spin-$1/2$ $XXZ$ chain's continuous $SU_q(2)$ symmetry accounted by the physical-spins representation
reveals that for anisotropy $\Delta > 1$ the Bethe strings of length $n=1,2,3,...$ describe a number $n$ of physical-spins
$S_q=0$ singlet pairs that for $n>1$ are bound within a $S_q=0$ singlet configuration.
Here $S_q$ is for $\Delta >1$ the $q$-spin associated with the continuous $SU_q(2)$ symmetry 
parametrized by $q = \Delta + \sqrt{\Delta^2-1}\in ]1,\infty]$ \cite{Carmelo_22,Pasquier_90,Prosen_13}.

The second quantum problem beyond the spinon paradigm discussed and reviewed in this paper refers to the 
finite-temperature spin stiffness for zero magnetic field and spin anisotropy $\Delta >1$. 
It is associated with ballistic spin transport. For a recent review on the 
current understanding of transport in 1D lattice models, in particular, in the 
present paradigmatic example of the spin-$1/2$ $XXZ$ chain, see Ref. \onlinecite{Bertini_21}.

Spin transport in that model was studied in the limit of infinite temperature by many authors \cite{Znidaric_11,Ljubotina_17,Medenjak_17,Ilievski_18,Gopalakrishnan_19,Weiner_20,Gopalakrishnan_23}.
At zero magnetic field it was found in that limit to be anomalous 
super-diffusive at the isotropic point, $\Delta =1$, and normal diffusive for $\Delta >1$.

Concerning experiments, an evidence for super-diffusion in the spin-$1/2$
$XXX$ chain came from a neutron scattering experiment on the quasi-1D magnet KCuF$_3$ \cite{Scheie_21}.
These experimental results are indeed inconsistent with either ballistic or diffusive scaling. While KCuF$_3$ realizes the isotropic 
spin-$1/2$ $XXX$ chain, a variety of other quasi-1D magnets exist that realize anisotropic spin-$1/2$ $XXZ$ spin 
chains \cite{Shen_22}. Spin transport in a tunable spin-$1/2$ $XXZ$ chain has also been realized in
systems of ultra-cold atoms \cite{Jepsen_20}. 

Concerning theoretical results on the spin-$1/2$ $XXZ$ chain for anisotropy $\Delta >1$,
in this paper we are interested in spin transport at zero magnetic field for {\it finite} temperatures, which until recently remained
an unsolved problem. The complexity of that quantum problem stems from many of the energy eigenstates 
that contribute to spin transport having at zero magnetic field
expectation values of the spin-current operator that diverge in
the thermodynamic limit. This is what renders it very difficult to solve.

Nonetheless, as it happens often in Physics, very complex problems
drastically simplify if one uses a suitable physical description that
captures hidden underlying symmetries. In the present case, the continuous $SU_q(2)$ symmetry 
accounted for the physical-spins representation imposes that only $2S_q$ out of the 
$N$ physical spins are the spin carriers. This result was used very recently in Ref. \onlinecite{Carmelo_24}
to show that, in contrast to the corresponding spin-current operator expectation values, 
the absolute values of the elementary spin currents carried by a single spin
carrier are for $\Delta >1$ always finite. And this finiteness was used in
that reference to show that for $\Delta >1$ the contributions to ballistic spin transport
exactly vanish at zero magnetic field and {\it all} finite temperatures. 
The behavior of the zero-temperature spin stiffness is also briefly
discussed \cite{Carmelo_24,Shastry_90,Zotos_99}.

The studies of Ref. \onlinecite{Carmelo_24} indicate that finite-temperature spin transport is 
normal diffusive at zero field magnetic for anisotropy $\Delta >1$, in agreement with previous expectations \cite{Ljubotina_17}.

It is interesting that the spin-$1/2$ $XXZ$ chain in a longitudinal magnetic field
describes for anisotropy $\Delta\approx 2$ many of the static properties
\cite{Hikihara_04,Kimura_07,Okunishi_07,Kimura_08,Canevet_13,Okutani_15,Klanjsek_15,Grenier_15,Dupont_16,Shen_19,Han_21,Scheie_21,Cui_22,Shen_22} and dynamical properties \cite{Carmelo_23,Wang_18,Wang_19,Bera_20} of
quasi-1D materials such as BaCo$_2$V$_2$O$_8$ and SrCo$_2$V$_2$O$_8$.
In this paper we also review recent results on {\it deviations} of these zigzag materials from the 1D 
physics described by that spin chain, due to selective interchain couplings \cite{Carmelo_23}.

The paper is organized as follows. In Sec. \ref{SECII} the continuous $SU_q (2)$ symmetry's irreducible representations
of the spin-$1/2$ $XXZ$ chain for anisotropy $\Delta >1$ \cite{Carmelo_22,Pasquier_90,Prosen_13} and the physical-spins 
representation that applies to its whole Hilbert space \cite{Carmelo_15A,Carmelo_23,Carmelo_22} are reviewed. 
Physical quantities needed for our studies that are associated with subspaces spanned by states generated from the 
ground state by a finite number of elementary processes is the topic of Sec. \ref{SECIII}. In Sec. \ref{SECIV}
the low-temperature dynamical properties of both the spin-$1/2$ $XXZ$ chain in a longitudinal
magnetic field and the zigzag material SrCo$_2$V$_2$O$_8$ are briefy reviewed \cite{Carmelo_23,Carmelo_22}.
The spin carriers and their elementary spin currents, the role of such elementary currents in the 
finite-temperature spin transport being non-ballistic at zero magnetic field for $\Delta >1$
\cite{Carmelo_24}, and the zero-temperature spin stiffness for anisotropy $\Delta \geq 1$ and spin 
density $m=2S^z/N\in [0,1]$ are the subjects of Sec. \ref{SECV}. 
The effects of selective interchain couplings in the deviations from
the 1D physics of the quasi-1D materials BaCo$_2$V$_2$O$_8$ and SrCo$_2$V$_2$O$_8$
is the topic studied in Sec. \ref{SECVI}. Finally, the concluding remarks are presented in Sec. \ref{SECVII}.

Secs. \ref{SECIIA}, \ref{SECIIB}, and \ref{SECIII} of this paper, which review results presented in Refs. \cite{Carmelo_23} and \cite{Carmelo_22},
are rather technical. While the results reviewed in these sections provide useful and interesting physical information on
the quantum problem symmetries and the physical-spins representation, both for the whole Hilbert space
and specific subspaces of interest for the problems under review in the ensuing sections, readers may proceed 
directly to Sec. \ref{SECIV} and use Secs. \ref{SECII} and \ref{SECIII} for consultation, when equations or concepts given in
these two sections are mentioned elsewhere in the paper.

\section{The quantum problem's $\Delta >1$ continuous $SU_q (2)$ symmetry and the physical-spins
representation}
\label{SECII}

For spin anisotropy $\Delta=\cosh\eta \geq 1$ and thus $\eta \geq 0$, spin densities $m= 2S^z/N \in [0,1]$, exchange integral $J$, 
and length $L\rightarrow\infty$ for $N/L$ finite, the Hamiltonian of the spin-$1/2$ $XXZ$ chain in a longitudinal magnetic 
field $h$ reads,
\begin{equation}
\hat{H} = J\sum_{j=1}^{N}\left({\hat{S}}_j^x{\hat{S}}_{j+1}^x + {\hat{S}}_j^y{\hat{S}}_{j+1}^y + 
\Delta\,{\hat{S}}_j^z{\hat{S}}_{j+1}^z\right) + g\mu_B\,h\,\hat{S}^z \, .
\label{HD1}
\end{equation}
Here $\hat{S}^z$ is the $z$ component of the spin-$1/2$ operator $\hat{\vec{S}}_{j}$ at site $j=1,...,N$ with 
components $\hat{S}_j^{x,y,z}$ and $\mu_B$ is the Bohr magneton. For simplicity, our results refer to the
numbers $N$ and $2S_q$ being even integers, yet in the thermodynamic limit they are
valid as well for these numbers being odd integers. In this paper we use natural units in which the Planck constant and the lattice 
spacing are equal to one, so that $N=L$. We consider spin densities $m= 2S^z/N \in [0,1]$.

At zero temperature and anisotropy $\Delta >1$ the quantum problem described by the Hamiltonian,
Eq. (\ref{HD1}), has a spin-insulating quantum phase for magnetic fields $h\in [0,h_{c1}]$,
a spin-conducting quantum phase for fields $h\in [h_{c1},h_{c2}]$, and a fully polarized ferromagnetic
quantum phase for fields $h>h_{c2}$. The critical magnetic fields $h_{c1}$ and $h_{c2}$ 
are given by \cite{Carmelo_23,Carmelo_22,Takahashi_99},
\begin{eqnarray}
h_{c1} & = & {2J\over\pi\,g\mu_B}\sinh \eta\,K (u_{\eta})\,\sqrt{1 - u_{\eta}^2}
\nonumber \\
h_{c2} & = & {J (\cosh\eta +1 )\over g\mu_B} = {J (\Delta +1 )\over g\mu_B} \, .
\label{criticalfields}
\end{eqnarray}
Here $K (u_{\eta})$ is the complete elliptic integral,
\begin{equation}
K (u_{\eta}) = \int_0^{\pi\over 2}d\theta {1\over \sqrt{1 - u_{\eta}^2\sin^2\theta}} \, .
\label{elliptic}
\end{equation}  
The dependence of $u_{\eta}$ on $\eta$ is defined by its inverse function,
$\eta = \pi K' (u_{\eta})/K (u_{\eta})$, where $K' (u_{\eta}) =K \left(\sqrt{1 - u_{\eta}^2}\right)$.
This also defines its dependence on the anisotropy $\Delta = \cosh\eta$.

Since $K (u_{\eta})$, $K' (u_{\eta})$, and $u_{\eta}$ appear in the expressions
of several physical quantities, including in that of the critical fields, Eq. (\ref{criticalfields}), we provide
here their limiting behaviors, which are given by,
\begin{eqnarray}
K (u_{\eta}) & = & {\pi^2\over 2}{1\over\eta} = {\pi^2\over 2}{1\over\sqrt{2(\Delta -1)}} \, ,
\hspace{0.20cm} K' (u_{\eta})  = {\pi\over 2} \hspace{0.20cm}{\rm and}
\nonumber \\
u_{\eta} & = & 1 - 2\,e^{-{\pi^2\over\eta}} = 1 - 2\,e^{-{\pi^2\over\sqrt{2(\Delta -1)}}}
\label{limitingKuS}
\end{eqnarray} 
for $\eta\ll 1$ and $(\Delta -1)\ll 1$ and,
\begin{eqnarray}
K (u_{\eta}) & = & {\pi\over 2}\left(1 + 4\,e^{-\eta}\right) = {\pi\over 2}\left(1 + {2\over\Delta}\right) \, ,
\nonumber \\
K' (u_{\eta}) &= & {\eta\over 2}\left(1 + 4\,e^{-\eta}\right) = {\ln (2\Delta)\over 2}\left(1 + {2\over\Delta}\right)
\hspace{0.20cm}{\rm and}
\nonumber \\
u_{\eta} & = & 4\,e^{-\eta/2} = 2\,\sqrt{2\over\Delta} \, ,
\label{limitingKuL}
\end{eqnarray} 
for $\eta\gg 1$ and $\Delta\gg 1$.

This gives $h_{c1}={2\pi J\over g\mu_B}e^{-{\pi^2\over 2\eta}}$ and 
$h_{c2} = {2J\over g\mu_B}\left(1 + {\eta^2\over 4}\right)$ where $\eta = \sqrt{2(\Delta -1)}$
for $\eta\ll 1$ and $(\Delta -1)\ll 1$ and $h_{c1}={J\over g\mu_B}\Delta$ and $h_{c2} = h_{c1} + {J\over g\mu_B}$
where $\Delta = {e^{\eta}\over 2}$ for $\eta\gg 1$ and thus $\Delta\gg 1$.
Hence $h_{c1}\rightarrow 0$ and $h_{c2} = {2J\over g\mu_B}$ for $\eta\rightarrow 0$ and $\Delta\rightarrow 1$.

The magnetic-field width $(h_{c2}-h_{c1})$ of the spin-conducting quantum phase continuously 
decreases upon increasing the anisotropy from $(h_{c2}-h_{c1})= {2J\over g\mu_B}$
for $\Delta\rightarrow 1$ to $(h_{c2}-h_{c1})= {J\over g\mu_B}$ for $\Delta\rightarrow\infty$.

\subsection{The $\Delta >1$ continuous $SU_q (2)$ symmetry and related physical-spins representation}
\label{SECIIA}

We start by briefly reviewing the $\Delta >1$ continuous $SU_q (2)$ symmetry \cite{Carmelo_22,Pasquier_90,Prosen_13}
of the Hamiltonian, Eq. (\ref{HD1}), at $h=0$. Its irreducible representations are associated with 
both its energy eigenstates for the whole magnetic-field interval $h\in [0,h_{c2}]$ and the physical-spins 
representation \cite{Carmelo_23,Carmelo_22} used in our studies and also briefly reviewed in the following. That continuous
symmetry is parametrized by $q = \Delta + \sqrt{\Delta^2-1}\in ]1,\infty]$, which can be expressed as
$q = e^{\eta}\in ]1,\infty]$.

For spin anisotropy $\Delta >1$, the spin projection 
$S^z$ remains a good quantum number whereas spin $S$ is not. 
It is replaced by the $q$-spin $S_q$ in the eigenvalue of the Casimir generator of the continuous $SU_q(2)$ 
symmetry of the spin-$1/2$ $XXZ$ chain \cite{Carmelo_22,Pasquier_90,Prosen_13}.
As justified below, the values of $q$-spin $S_q$ are exactly the same for anisotropy 
$\Delta >1$ and $\eta >0$ as spin $S$ for $\Delta =1$ and $\eta =0$.
This includes their relation to the values of $S^z$. Hence {\it singlet} and {\it multiplet} refer in this paper to 
physical-spins configurations with zero and finite $S_q$, respectively. 

For simplicity, in this section we use mostly the anisotropy parameter $\eta$ in $\Delta = \cosh\eta$.
We denote the energy eigenstates for $\eta =0$ and $\eta >0$
by $\vert l_{\rm r}^0,S,S^z\rangle$ and $\vert l_{\rm r}^{\eta},S_q,S^z\rangle$, respectively.
Here $l_{\rm r}^{\eta}$ stands for $\eta$ and all quantum numbers other than spin $S$ for $\eta =0$, $q$-spin $S_q$ for $\eta >0$, 
and $S^z$ for $\eta \geq 0$ needed to specify an energy eigenstate.
Except for $\eta$ itself, the set of quantum numbers described by $l_{\rm r}^{\eta}$ are independent of $\eta$.

A useful property is that the Hilbert space of the present quantum problem 
is the same for $\eta =0$ and $\eta >0$, respectively.
In then follows that there is a uniquely defined unitary transformation that refers to a one-to-one relation between
the two sets of $2^N$ $\eta =0$ and $\eta >0$ energy eigenstates, $\{\left\vert l_{\rm r}^0,S,S^z\right\rangle\}$ and
$\{\left\vert l_{\rm r}^{\eta},S_q,S^z\right\rangle\}$, respectively \cite{Carmelo_22}. 
Indeed, both of them refer to sets of complete and orthonormal energy eigenstates for the same Hilbert space
of dimension $2^N$.

That unitary transformation is associated with unitary operators 
$\hat{U}_{\eta}^{\pm}$ such that \cite{Carmelo_22},
\begin{eqnarray}
\left\vert l_{\rm r}^{\eta},S_q,S^z\right\rangle & = & \hat{U}_{\eta}^+\left\vert l_{\rm r}^0,S,S^z\right\rangle
\hspace{0.20cm}{\rm and}
\nonumber \\
\left\vert l_{\rm r}^0,S,S^z\right\rangle & = & \hat{U}_{\eta}^-\left\vert l_{\rm r}^{\eta},S_q,S^z\right\rangle \, ,
\label{Ophi}
\end{eqnarray}
where $S_q =S$. Fortunately, the specific involved form of the unitary operators $\hat{U}_{\eta}^{\pm}$ 
is not needed for our studies. Only that they are uniquely defined and
the transformations they generate, Eq. (\ref{Ophi}), are needed.

The values of $S_q$, $S^z$, and of the set of quantum numbers represented
by $l_{\rm r}^{\eta}$ in $\left\vert l_{\rm r}^{\eta},S_q,S^z\right\rangle$ other than $\eta$ remain invariant 
under $\hat{U}_{\eta}^{\pm}$, yet only at $\eta =0$ the number $S_q$ is the spin $S$.
That invariance implies that for any two $\eta > 0$ and $\eta = 0$ energy eigenstates,
respectively, related as in Eq. (\ref{Ophi}),
the $q$-spin $S_q$ has for $\eta >0$ exactly the same values as the spin $S$ at $\eta =0$.
$N$ is an even and odd integer number when the states $q$-spin $S_q$ is an integer and half-odd integer number, respectively.
As reported in Sec. \ref{SECI}, we consider for simplicity that $N$ and $2S_q$ are even integer numbers.
 
Standard counting of $q$-spin continuous $SU_q (2)$ symmetry irreducible representations
is under the unitary transformation, Eq. (\ref{Ophi}), similar to that of corresponding spin 
$SU(2)$ symmetry irreducible representations.
One then finds from the use of the two isomorphic algebras that 
when $N$ and $2S_q$ are even integer numbers (similar results apply when
$N$ and $2S_q$ are odd integer numbers \cite{Carmelo_22}) the following summation
gives , 
\begin{equation}
\sum_{l_{\rm r}^{\eta},S_q,S^z} =  \sum_{l_{\rm r}^{\eta}}\sum_{S_q=0}^{N/2}
\sum_{S^z=-S_q}^{S_q} = {\cal{N}}_{\rm singlet} (S_q) \, ,
\label{sums}
\end{equation}
in each fixed-$S_q$ subspace where,
\begin{eqnarray}
{\cal{N}}_{\rm singlet} (S_q) = {N\choose N/2-S_q}-{N\choose N/2-S_q-1} \, ,
\label{Nsinglet}
\end{eqnarray}
is that subspace number of independent singlet configurations of physical spins.
Including physical-spins multiplet configurations when $S_q>0$, the dimension is given by 
${\cal{N}}(S_q) = (2S_q+1)\,{\cal{N}}_{\rm singlet}  (S_q)$. Consistently, 
\begin{equation}
\sum_{S_q=0}^{N/2}\,{\cal{N}}(S_q) = 2^{N} \hspace{0.20cm}{\rm where}\hspace{0.20cm}
{\cal{N}}(S_q) = (2S_q+1)\,{\cal{N}}_{\rm singlet} (S_q) \, ,
\label{SumRuleSphi}
\end{equation}
gives the Hilbert-space dimension. It thus equals the total number of both irreducible representations 
and energy and momentum eigenstates. 

That the irreducible representations of the $\eta >0$ continuous $SU_q(2)$ symmetry \cite{Pasquier_90,Prosen_13}
are isomorphic to those of the $\eta=0$ $SU(2)$ symmetry, then requires that,
\begin{eqnarray}
\hat{S}^{\pm}_{\eta}\left\vert l_{\rm r}^{\eta},S_q,S^z\right\rangle & \propto &
\hat{U}_{\eta}^+\hat{S}^{\pm}\hat{U}_{\eta}^-\left\vert l_{\rm r}^{\eta},S_q,S^z\right\rangle
\nonumber \\
(\hat{\vec{S}}_{\eta})^2\left\vert l_{\rm r}^{\eta},S_q,S^z\right\rangle & \propto &
\hat{U}_{\eta}^+(\hat{\vec{S}})^2\hat{U}_{\eta}^-\left\vert l_{\rm r}^{\eta},S_q,S^z\right\rangle \, .
\label{SSSS}
\end{eqnarray}
To confirm this requirement, we express the generators of the $q$-spin continuous $SU_q(2)$ symmetry 
in terms of those of the spin $SU(2)$ symmetry. 
The exact relations under the unitary transformation, Eq. (\ref{Ophi}),
between the ladder operators $\hat{S}^{\pm}_{\eta}$ and square
operators $(\hat{\vec{S}}_{\eta})^2$ for $\eta >0$ and the
corresponding operators $\hat{S}^{\pm}$ and $(\hat{\vec{S}})^2$ at $\eta =0$ 
appearing in Eq. (\ref{SSSS}) can indeed be expressed as,
\begin{eqnarray}
&& \hat{S}^{\pm}_{\eta} = \sum_{l_{\rm r},S_q}
\sum_{S^z=-S_q+{(1\mp 1)\over 2}}^{S_q-{(1\pm 1)\over 2}}
\nonumber \\
&& \times \sqrt{{\sinh^2 (\eta\,(S_q+1/2)) - \sinh^2 (\eta\,(S^z \pm 1/2))\over
((S_q+1/2)^2 - (S^z \pm 1/2)^2)\sinh^2 \eta}}
\nonumber \\
&& \times \left\langle l_{\rm r}^{\eta},S_q,S^z\pm 1\right\vert
\hat{U}_{\eta}^+\hat{S}^{\pm}\hat{U}_{\eta}^-
\left\vert l_{\rm r}^{\eta},S_q,S^z\right\rangle
\nonumber \\
&& \times
\left\vert l_{\rm r}^{\eta},S_q,S^z\pm 1\right\rangle\left\langle l_{\rm r}^{\eta},S_q,S^z\right\vert \, ,
\label{OneSOphipm}
\end{eqnarray}
and
\begin{eqnarray}
&& (\hat{\vec{S}}_{\eta})^2 = \sum_{l_{\rm r}^{\eta},S_q,S^z}
{\sinh^2 (\eta\,(S_q+1/2)) - \sinh^2 (\eta/2)\over
((S_q+1/2)^2 - 1/4)\sinh^2 \eta}
\nonumber \\
& \times & \left\langle l_{\rm r}^{\eta},S_q,S^z\right\vert
\hat{U}_{\eta}^+(\hat{\vec{S}})^2\hat{U}_{\eta}^-
\left\vert l_{\rm r}^{\eta},S_q,S^z\right\rangle
\nonumber \\
&& \times \left\vert l_{\rm r}^{\eta},S_q,S^z\right\rangle\left\langle l_{\rm r}^{\eta},S_q,S^z\right\vert \, ,
\label{CasimirO}
\end{eqnarray}
respectively. Combining this latter summation over the operators 
$\left\vert l_{\rm r}^{\eta},S_q,S^z\right\rangle\left\langle l_{\rm r}^{\eta},S_q,S^z\right\vert$ 
with the expression of $\hat{S}^+_{\eta}\hat{S}^-_{\eta}$ obtained from the use of Eq. (\ref{OneSOphipm}),
straightforwardly leads to the following known result \cite{Carmelo_22,Pasquier_90,Prosen_13},
\begin{equation}
(\hat{\vec{S}}_{\eta})^2 = \hat{S}^+_{\eta}\hat{S}^-_{\eta} - {\sinh^2 (\eta/2)\over\sinh^2 \eta} 
+ {\sinh^2 (\eta\,(\hat{S}^z+1/2))\over\sinh^2 \eta}  \, .
\label{CasimirOknown}
\end{equation}

That of the commutator $[\hat{S}^{+}_{\eta},\hat{S}^{-}_{\eta}]$
in terms of the operator $\hat{U}_{\eta}^+[\hat{S}^{+},\hat{S}^{-}]\hat{U}_{\eta}^-$ is thus given by,
\begin{eqnarray}
[\hat{S}^{+}_{\eta},\hat{S}^{-}_{\eta}] & = & \sum_{l_{\rm r}^{\eta},S_q,S^z}
{\sinh (\eta\, 2S^z)\over 2S^z \sinh \eta}
\nonumber \\
& \times & \left\langle l_{\rm r}^{\eta},S_q,S^z\right\vert
\hat{U}_{\eta}^+[\hat{S}^{+},\hat{S}^{-}]\hat{U}_{\eta}^-
\left\vert l_{\rm r}^{\eta},S_q,S^z\right\rangle
\nonumber \\
& \times & \left\vert l_{\rm r}^{\eta},S_q,S^z\right\rangle\left\langle l_{\rm r}^{\eta},S_q,S^z\right\vert 
\nonumber \\
& = & {\sinh (\eta\,2\hat{S}^z)\over\sinh \eta} \, .
\label{commPPhiDelta1}
\end{eqnarray}

Important symmetries are associated with the following commutations involving
the momentum operator $\hat{P}$ and $\hat{S}^z$,
\begin{equation}
[\hat{P},\hat{U}_{\eta}^{\pm}]=0 \hspace{0.20cm}{\rm and}\hspace{0.20cm}
[\hat{S}^z,\hat{U}_{\eta}^{\pm}]=0 \, .
\label{commPSzU}
\end{equation}
It follows that both the momentum eigenvalues $P$ (given below in Eq. (\ref{P0})) and the spin projection 
$S^z$ are good quantum numbers independent of $\eta$ for the whole $\eta \geq 0$ range.

Consistently, the commutator $[\hat{S}^z,\hat{S}^{\pm}_{\eta}]$ has the known simple form,
\begin{equation}
[\hat{S}^z,\hat{S}^{\pm}_{\eta}] = \pm \hat{S}^{\pm}_{\eta} \, .
\label{commSzPhiDelta1}
\end{equation}

The expressions of the $q$-spin continuous $SU_q (2)$ symmetry
generators, Eqs. (\ref{OneSOphipm}) and (\ref{CasimirO}), continuously evolve into
those of the spin $SU(2)$ symmetry as $\eta$ is continuously decreased to zero. 
And the $q$-spin continuous $SU_q(2)$ symmetry itself also continuously evolves into the spin $SU(2)$ symmetry 
as the former symmetry parameter $q = \Delta + \sqrt{\Delta^2-1} = e^{\eta}$ tends to one and 
thus $\eta$ is adiabatically turned off to zero.

At $\eta =0$ and for $\eta >0$ the Bethe ansatz refers only to subspaces spanned either by the 
highest weight states (HWSs) or the lowest weight states (LWSs) of the $SU (2)$ and $SU_q(2)$ symmetries,
respectively \cite{Gaudin_71,Carmelo_22}. For such HWSs and LWSs one has that $S^z = S_q$ and $S^z = -S_q$, respectively.
In this paper we use a HWS Bethe ansatz. 

As confirmed below in Sec. \ref{SECIIB}, the irreducible representations of the continuous $SU_q(2)$ symmetry algebra
that involve the multiplet and singlet physical-spins configurations associated with the dimensions $(2S_q+1)$ and 
${\cal{N}}_{\rm singlet}  (S_q)$, respectively, of the overall dimension ${\cal{N}}(S_q) = (2S_q+1)\,{\cal{N}}_{\rm singlet}  (S_q)$ 
in Eq. (\ref{SumRuleSphi}) lead naturally to a representation for the $N$ spin-$1/2$ physical spins described by the Hamiltonian, Eq. (\ref{HD1}).
That representation is such that each irreducible representation refers to an energy eigenstate with $q$-spin in the 
range $0\leq S_q\leq N/2$ that is populated by physical spins in two types of configurations \cite{Carmelo_22}:\\

1) A number $M=M_{+1/2}+M_{-1/2} = 2S_q$ of {\it unpaired spins $1/2$} that participate in a multiplet configuration.
The number $M_{\pm 1/2}$ of such unpaired physical spins of projection $\pm 1/2$ of
an energy eigenstate is solely determined by its values of $S_q$ and $S^z$,
as it reads $M_{\pm 1/2} = S_q \pm S^z$, so that,
\begin{equation}
2S^z = M_{+1/2} - M_{-1/2} \hspace{0.20cm}{\rm and}\hspace{0.20cm}2S_q = M_{+1/2} + M_{-1/2} \, .
\label{2Sz2Sq}
\end{equation}

2) A complementary set of even number ${\cal{M}}=N-2S_q$ of {\it paired spins $1/2$}
that participate in singlet configurations.\\ 

This holds for {\it all} $2^N$ energy eigenstates of the Hamiltonian, Eq. (\ref{HD1}).
Within the corresponding representation in terms of the $N$ physical spins described by 
that Hamiltonian, the designation {\it $n$-pairs} refers both to {\it $1$-pairs} and {\it $n$-string-pairs} 
for $n>1$ \cite{Carmelo_23,Carmelo_22}:\\

- The internal degrees of freedom of a $1$-pair correspond to one unbound singlet
pair of physical spins. It is described by a $n=1$ single real Bethe rapidity. Its translational 
degrees of freedom refer to the $1$-band momentum $q_j \in [q_1^-,q_1^+]$ where $j = 1,...,L_1$ 
carried by each such a pair.\\

- The internal degrees of freedom of a 
$n$-string-pair refer to a number $n>1$ of singlet pairs of physical spins. They are
bound within a singlet configuration described by a corresponding complex Bethe $n$-string
given below in Eq. (\ref{LambdaIm}). Its translational degrees of freedom 
refer to the $n>1$ $n$-band momentum $q_j \in [q_n^-,q_n^+]$ where $j = 1,...,L_n$ carried by each such an $n$-pair.\\

For each corresponding $n$-band, the $j=1,...,L_n$ discrete set of $q_j$'s have for both $n=1$ and $n>1$ 
separation $q_{j+1}-q_j = {2\pi\over N}$. The Bethe-ansatz quantum numbers \cite{Gaudin_71} $I_j^n$ are
actually the discrete $n$-band momentum values $q_j = {2\pi\over N}I_j^n$ in units of ${2\pi\over N}$. They read,
\begin{eqnarray}
I_j^n & = & 0,\pm 1,...,\pm {L_n -1\over 2}
\hspace{0.20cm}{\rm for}\hspace{0.20cm}L_n\hspace{0.20cm}{\rm odd}
\nonumber \\
& = & \pm 1/2,\pm 3/2,...,\pm {L_n -1\over 2} \hspace{0.20cm}{\rm for}\hspace{0.20cm}L_n\hspace{0.20cm}{\rm even} \, ,
\label{Ij}
\end{eqnarray}
where,
\begin{eqnarray}
L_n & = & N_n + N^h_{n}\hspace{0.20cm}{\rm with}
\nonumber \\
N^h_{n} & = & 2S_q +\sum_{n'=n+1}^{\infty}2(n'-n)N_{n'} \, ,
\label{Nhn}
\end{eqnarray}
and $N_n$ is the number of occupied $q_j$'s and thus of $n$-pairs and 
$N^h_{n}$ that of unoccupied $q_j$'s and thus of $n$-holes.

The numbers $M=2S_q$ and ${\cal{M}} = N-2S_q$ of unpaired and paired physical spins $1/2$
of an energy eigenstate can then be exactly expressed in terms of those of the $n$-pairs as,
\begin{eqnarray}
M & = & 2S_q = N - \sum_{n=1}^{\infty}2n\,N_n 
\hspace{0.20cm}{\rm and}
\nonumber \\
{\cal{M}} & = & N-2S_q = \sum_{n=1}^{\infty}2n\,N_n  \, ,
\label{2Pi}
\end{eqnarray}
respectively.

The set of quantum numbers $\{I_j^n\}$, Eq. (\ref{Ij}), and thus the set $\{q_j= {2\pi\over N}I_j^n\}$ of 
$n$-band discrete momentum values have Pauli-like occupancies: The corresponding $n$-band momentum 
distributions read $N_n (q_j) = 1$ and $N_n (q_j) = 0$ for occupied and unoccupied $q_j$'s, respectively.
For each $n$-band the $q_j$'s have values in the interval $q_j \in [q_n^-,q_n^+]$ where,
\begin{equation}
q_n^{\pm} = \pm {\pi\over N}(L_n -1) + q_n^{\Delta}
\hspace{0.20cm}{\rm where}\hspace{0.20cm}
q_n^{\Delta} = {1\over 2}(q_n^- + q_n^+) \, ,
\label{qnpm}
\end{equation}
and the $n=1,...,\infty$ momentum values $q_n^{\Delta}$ can be expressed 
as $q_n^{\Delta}=-{\pi\over N}\delta L_n^{\eta}$ where $\delta L_n^{\eta}$ and its limiting values are given 
in Eqs. (12) and (13) of Ref. \onlinecite{Carmelo_22}, respectively.

The energy eigenstates are specified by the set of $n=1,...,\infty$ 
$n$-band momentum distributions $\{N_n (q_j)\}$ and described by a corresponding set of 
rapidities $\varphi_{n,l,j} = \varphi_{n,l} (q_j)$ that for $n=1$ are real and describe the $1$-pairs
and for $n>1$  are complex and describe the $n$-string-pairs. 

The corresponding $n>1$ Bethe
$n$-string structure depends on the system size. In the thermodynamic limit in which that structure simplifies 
the rapidities $\varphi_{n,l,j} = \varphi_{n,l} (q_j)$ can both for $n=1$ and $n>1$ be expressed as,
\begin{equation}
\varphi_{n,l} (q_j) = \varphi_{n} (q_j) + i(n + 1 -2l)\,\eta \hspace{0.20cm}{\rm where}\hspace{0.20cm}l=1,...,n \, .
\label{LambdaIm}
\end{equation}
Their real part refers to the $n$-band rapidity functions $\varphi_{n} (q_{j})$ 
defined by Bethe-ansatz equations given below in functional form. It is such that 
$\varphi_{n} (q_{j})\in [-\pi,\pi]$ where $\varphi_{n} (q_n^{\pm})=\pm\pi$.

The $n$-band discrete momentum values $q_j={2\pi\over N}I_j^n$'s such that $q_{j+1}-q_j = {2\pi\over N}$
can in the thermodynamic limit be described by continuous variables $q \in [q_n^-,q_n^+]$.
The corresponding $n$-band limiting momentum values $q_n^{\pm}$ in Eq. (\ref{qnpm}) can be expressed 
as $q_n^{\pm} = q_n (\pm\pi)$ in terms of the $n=1,...,\infty$ $n$-band momentum functions $q_n (\varphi)$ where $\varphi\in [-\pi,\pi]$ that
are the inverse functions of the $n=1,...,\infty$ rapidity functions $\varphi_n (q)$ 
where $q\in [q_n^-,q_n^+]$, so that $q_n^{\Delta} = {1\over 2}(q_n (\pi) + q_n (-\pi))$.

Alternatively to the set of $n=1,...,\infty$  $n$-band momentum distributions $\{N_n (q_j)\}$,
the energy eigenstates can be specified by a corresponding set of $n=1,...,\infty$ 
$n$-band rapidity-variable $\varphi\in [-\pi,\pi]$ distributions $\{\tilde{N}_n (\varphi)\}$ defined
by the relation,
\begin{equation}
\tilde{N}_n (\varphi_n (q_j)) = N_n (q_j) \, .
\label{tildeNN}
\end{equation}
The $n$-band momentum functions $q_n (\varphi)$ are then the unique solutions of the following set of $n=1,...,\infty$ of 
coupled Bethe-ansatz equations in functional form,
\begin{eqnarray}
&& q_n (\varphi) = 2\arctan \left(\coth \left({n\eta\over 2}\right)\tan\left({\varphi\over 2}\right)\right)
\nonumber \\
&& - {1\over 2\pi}\int_{-\pi}^{\pi}d\varphi'\,\tilde{N}_n (\varphi')\,2\pi\sigma_n (\varphi')
\nonumber \\
&& \times \Bigl[2\arctan \left(\coth (n\eta)\tan\left({\varphi-\varphi'\over 2}\right)\right)
\nonumber \\
&& + \sum_{l=1}^{n-1}4\arctan \left(\coth (l\eta)\tan\left({\varphi-\varphi'\over 2}\right)\right)\Bigr]
\nonumber \\
&& - {1\over 2\pi}\sum_{n' \neq\,n}\int_{-\pi}^{\pi}d\varphi'\,\tilde{N}_{n'} (\varphi')\,2\pi\sigma_{n'} (\varphi')
\nonumber \\
&& \times \Bigl[2\arctan \left(\coth \left({(n+n')\eta\over 2}\right)\tan\left({\varphi-\varphi'\over 2}\right)\right)
\nonumber \\
&& + 2\arctan \left(\coth \left({\vert n-n'\vert\eta\over 2}\right)\tan\left({\varphi-\varphi'\over 2}\right)\right) 
\nonumber \\
&& + \sum_{l=1}^{{n+n' - \vert n-n'\vert\over 2} -1}
\nonumber \\
&& \times
4\arctan\left(\coth\left({(\vert n-n'\vert + 2l)\eta\over 2}\right)\tan\left({\varphi-\varphi'\over 2}\right)\right)\Bigr] \, .
\nonumber \\
\label{BAqn}
\end{eqnarray}
The set of $n=1,...,\infty$ general distributions $2\pi\sigma_{n} (\varphi)$ appearing in this equation are given by,
\begin{equation}
2\pi\sigma_n (\varphi) = {d q_n (\varphi)\over d\varphi} \, .
\label{sigmanderivative}
\end{equation} 
The functional character of the coupled equations, Eq. (\ref{BAqn}), follows from their dependence on 
general $n$-band rapidity-variable distributions $\tilde{N}_n (\varphi)$, Eq. (\ref{tildeNN}), which have specific values
for each energy eigenstate.

Consistently with the $n$-band $q_j$'s being momentum values,
the momentum eigenvalues of the energy eigenstates of the Hamiltonian, Eq. (\ref{HD1}),
can be written in functional form as,
\begin{eqnarray}
P & = &  \pi \sum_{n=1}^{\infty}N_n + \sum_{n=1}^{\infty}\int_{-\pi}^{\pi}d\varphi\,\tilde{N}_n (\varphi)\,2\pi\sigma_n (\varphi)\,q_n (\varphi) 
\nonumber \\
& = &  \pi \sum_{n=1}^{\infty}N_n + \sum_{n=1}^{\infty}\int_{q_n^-}^{q_n^+}dq\,N_n (q)\,q
\nonumber \\
& = & \pi \sum_{n=1}^{\infty}N_n + \sum_{n=1}^{\infty}\sum_{j=1}^{L_n}N_{n} (q_{j})\,q_j \, .
\label{P0}
\end{eqnarray}

\subsection{Corresponding irreducible representations and states 
inside and outside the Bethe ansatz}
\label{SECIIB}

The Bethe ansatz describes both the translational and internal degrees of freedom
of each $n$-pair containing $2n$ physical spins in its $n$ singlet pairs. The
former refer to $N_n$ occupied $n$-band momentum values $q_j$ carried by
the corresponding $n$-pairs out of the $j = 1,...,L_n$ values $q_j \in [q_n^-,q_n^+]$ of that band.

Concerning the internal degrees of freedom of a $n>1$ $n$-string-pair, the set of $l=1,...,n$ imaginary terms $i(n + 1 -2l)\,\eta$ 
of the $n>1$ complex Bethe rapidity, Eq. (\ref{LambdaIm}), have within the physical-spins representation 
a precise physical meaning: They refer to the binding of the $l=1,...,n$ singlet pairs of physical 
spins within one $n$-string-pair. 

On the other hand, that imaginary part vanishes for $n=1$, consistently 
with the corresponding single real rapidity referring to one unbound singlet $1$-pair of such physical spins.

For each energy eigenstate, the physical-spins representation accounts both for a number 
${\cal{M}}=N-2S_q$ of paired physical spins and a number $M=2S_q$ of unpaired physical spins in its
singlet and multiplet configurations, respectively. 

Note though that only the translational degrees of freedom of the $M=2S_q$ unpaired physical spins 
are described within the Bethe ansatz, which involves a squeezed space construction \cite{Carmelo_22,Carmelo_24,Kruis_04}.
Such degrees of freedom are described by a number $M=2S_q$ of $n$-band momentum 
values, $q_j = {2\pi\over N}I_j^n$, out of the $N^h_{n}$, Eq. (\ref{Nhn}), unoccupied 
such values, {\it i.e.} $n$-holes, of each $n$-band with finite $N_n>0$ occupancy. 

On the other hand, the spin internal degrees of freedom of the $M=2S_q$ 
unpaired physical spins of an energy eigenstate is an issue beyond the Bethe ansatz. 
That ansatz refers only either to the HWSs or LWSs considered in Sec. \ref{SECIIA}.
For HWSs and LWSs all the $M=2S_q$ unpaired physical spins $1/2$ 
have the same projection $+1/2$ and $-1/2$, consistently with $S^z = S_q$ and $S^z = -S_q$, respectively.

The physical-spins representation though applies to the whole Hilbert space, as
it accounts for the spin internal degrees of freedom of the $M = 2S_q$ 
unpaired physical spins of all $2^N$ energy eigenstates.

Indeed, a number $n_z = 1,...,2S_q$ of continuous $SU_q(2)$ symmetry non-HWSs 
$\left\vert l_{\rm r}^{\eta},S_q,S^z\right\rangle = \left\vert l_{\rm r}^{\eta},S_q,S_q-n_z\right\rangle$ outside the Bethe-ansatz 
solution, which refer to different multiplet configurations of the
$M=(M_{+1/2}+M_{-1/2})=2S_q$ unpaired physical spins, are generated from a HWS 
$\left\vert l_{\rm r}^{\eta},S_q,S_q\right\rangle$ as,
\begin{equation} 
\left\vert l_{\rm r}^{\eta},S_q,S_q-n_z\right\rangle = 
{1\over \sqrt{{\cal{C}}_{\eta}}}({\hat{S}}^{-}_{\eta})^{n_z}\left\vert l_{\rm r}^{\eta},S_q,S_q\right\rangle \, .
\label{state}
\end{equation} 
Here $n_z\equiv S_q - S^z = 1,...,2S_q$, so that $S^z = S_q - n_z$, and,
\begin{equation}
{\cal{C}}_{\eta} = 
\prod_{l=1}^{n_z}{\sinh^2 (\eta\,(S_q+1/2)) - \sinh^2 (\eta\,(l - S_q - 1/2))\over\sinh^2 \eta} \, ,
\label{nonBAstatesDelta1}
\end{equation}
for $n_z= 1,...,2S_q$. Similarly to the $\Delta =1$ bare ladder spin operators $\hat{S}^{\pm}$, the 
action of the $\Delta = \cosh \eta >1$ $q$-spin ladder operators $\hat{S}^{\pm}_{\eta}$,
Eq. (\ref{OneSOphipm}), on $S_q > 0$ energy eigenstates indeed flips an {\it unpaired} physical spin projection,
as in Eq. (\ref{state}) for $\hat{S}^{-}_{\eta}$.

For the non-HWSs, Eq. (\ref{state}), the two sets of $n_z\equiv S_q - S^z = 1,...,2S_q = M_{-1/2}$ 
and $2S_q-n_z = S_q + S^z = M_{+1/2}$ unpaired physical spins have opposite $-1/2$
and $+1/2$ projections, respectively. Hence, the multiplet configurations that
involve the internal degrees of freedom of the $M = (M_{+1/2} + M_{-1/2}) = 2S_q$ unpaired physical spins
are generated as given in Eq. (\ref{state}). 

Within the present functional representation, the energy eigenvalues of any energy
eigenstate $\left\vert l_{\rm r}^{\eta},S_q,S^z\right\rangle$ of
the Hamiltonian, Eq. (\ref{HD1}), are in the thermodynamic limit given by,
\begin{eqnarray}
E (l_{\rm r}^{\eta},S_q,S^z) & = & - \sum_{n=1}^{\infty}\sum_{l=1}^{n}\sum_{j=1}^{L_n}
{J\sinh^2(\eta)\,N_{n} (q_{j})\over \cosh\eta - \cos\varphi_{n,l} (q_j)} 
\nonumber \\
& + & g\mu_B\,h\,{1\over 2}\left(M_{+1/2} - M_{-1/2}\right) \, .
\label{Energy0}
\end{eqnarray}
The $n$-string $l$-summation can be performed under the use of the expression,
Eq. (\ref{LambdaIm}), for the complex rapidity $\varphi_{n,l} (q_j)$. 
Under its use and accounting for symmetries of the $l$-summation, it is convenient 
to first replace it by a related $m$-summation such that,
\begin{eqnarray}
&& \sum_{l=1}^{n}{1\over \cosh\eta - \cos\varphi_{n,l} (q_j)} =
\nonumber \\
&& = 2\sum_{m=1}^{n/2}(\cosh\eta - \cosh ((n+1-2m)\eta)\cos\varphi_{n} (q_j))
\nonumber \\
&& \times \{(\cosh\eta - \cosh ((n+1-2m)\eta)\cos\varphi_{n} (q_j))^2 
\nonumber \\
&& + (\sinh ((n+1-2m)\eta)\sin\varphi_{n} (q_j))^{2}\}^{-1}
\nonumber \\
&& \hspace{2cm}{\rm for}\hspace{0.20cm}n\hspace{0.20cm} {\rm even}
\nonumber \\
&& = 2\sum_{m=1}^{(n-1)/2}
\nonumber \\
&& \times (\cosh\eta - \cosh ((n+1-2m)\eta)\cos\varphi_{n} (q_j))
\nonumber \\
&& \times \{(\cosh\eta - \cosh ((n+1-2m)\eta)\cos\varphi_{n} (q_j))^2 
\nonumber \\
&& + (\sinh ((n+1-2m)\eta)\sin\varphi_{n} (q_j))^{2}\}^{-1}
\nonumber \\
&& + {1\over \cosh\eta - \cos\varphi_{n} (q_j)}
\nonumber \\
&& \hspace{2cm}{\rm for}\hspace{0.20cm}n\hspace{0.20cm} {\rm odd} \, ,
\label{equalspectra}
\end{eqnarray}
where as reported in Sec. \ref{SECIIA} the rapidity function $\varphi_{n} (q_j)$ is the real part of the 
complex rapidity, Eq. (\ref{LambdaIm}). (here $m$ denotes a summation integer variable and not spin density.)
The $m$-summation can be exactly performed, which gives,
\begin{equation}
\sum_{l=1}^{n}{1\over \cosh\eta - \cos\varphi_{n,l} (q_j)} =
{\sinh^{-1}(\eta)\,\sinh (n\,\eta)\over \cosh (n\,\eta) - \cos\varphi_{n} (q_j)} \, ,
\label{equalspectrafinal}
\end{equation}
for both $n$ even and $n$ odd. Finally, the use of this result in Eq. (\ref{Energy0}) leads 
to the following simpler expression for the energy eigenvalues,
\begin{eqnarray}
&& E (l_{\rm r}^{\eta},S_q,S^z) = - J\sinh\eta\sum_{n=1}^{\infty}\sum_{j=1}^{L_n}
{\sinh (n\,\eta)\,N_{n} (q_{j})\over \cosh (n\,\eta) - \cos\varphi_{n} (q_{j})} 
\nonumber \\
&& \hspace{1cm} + g\mu_B\,h\,{1\over 2}\left(M_{+1/2} - M_{-1/2}\right)
\nonumber \\
&& = - {JL\sinh\eta\over 2\pi}\sum_{n=1}^{\infty}\int_{-\pi}^{\pi}d\varphi\,2\pi\sigma_n (\varphi) 
{\sinh (n\,\eta)\,\tilde{N}_n (\varphi)\over \cosh (n\,\eta) - \cos\varphi} 
\nonumber \\
&& \hspace{1cm} + g\mu_B\,h\,{1\over 2}\left(M_{+1/2} - M_{-1/2}\right) \, .
\label{Energy}
\end{eqnarray}

The continuous $SU_q (2)$ symmetry is behind the energy of the $n_z\equiv S_q - S^z = 1,...,2S_q$ non-HWSs, Eq. (\ref{state}),
outside the Bethe anstaz differing from that of the corresponding HWS only in the presence 
of a magnetic field $h$. As given in Eq. (\ref{Energy}), this difference refers to the values of the numbers of unpaired
physical spins of projection $\pm 1/2$, such that $M_{+1/2} - M_{-1/2} = 2S^z$, Eq. (\ref{2Sz2Sq}).
That symmetry indeed imposes that at $h=0$ the $2S_q +1$ states of the same $q$-spin tower
have exactly the same energy.

This is consistent with a HWS and all its non-HWSs of a $q$-spin tower with $2S_q+1$
states only differing in the values of the unpaired physical spins numbers
$M_{+1/2}=S_q + S^z$ and $M_{-1/2} = S_q  - S^z$. Otherwise, these $2S_q+1$
states have exactly the same $n$-pair occupancy configurations and corresponding 
$n$-pair numbers $\{N_n\}$, $n$-hole numbers $\{N_n^h\}$, and thus numbers $\{L_n\}$ where 
$L_n = N_n + N_n^h$, Eq. (\ref{Nhn}). They also have the same values for the set of 
$n=1,...,\infty$ $n$-band momentum distributions $\{N_n (q_j)\}$ and rapidity functions $\{\varphi_{n} (q_{j})\}$.

Finally, we confirm the consistency of the physical-spins representation by showing that
the number ${\cal{N}}_{\rm singlet} (S_q)$, Eq. (\ref{Nsinglet}), of 
paired physical-spins singlet configurations of a fixed-$S_q$ subspace exactly equals the number of
corresponding $n$-pairs independent occupancy configurations associated with the set of numbers 
$\{N_n\}$. This is a requirement that follows from {\it all} ${\cal{M}}=N-2S_q$ paired physical spins being contained in the set
of $n$-pairs of an energy eigenstate, each $n$-pair involving a number $n$ of {\it singlet} 
pairs of such physical spins.

We use the equality of the quantities given in Eq. (A.1) and in Appendix A's last equation
of Ref. \onlinecite{Takahashi_71}, respectively, for the isotropic case, which also applies to $\Delta >1$.
Indeed, the number relations in Eq. (\ref{Nhn}) are valid for $\Delta \geq 1$ and
under the unitary transformation, Eq. (\ref{Ophi}), $S$ is mapped into $S_q$ with $S_q =S$.

After some straightforward algebra that accounts for the HWS's number
of unpaired physical spins, which is given by $2S^z$, reading $2S_q$ for the non-HWSs
generated from it, Eq. (\ref{state}), one reaches the following more general 
{\it exact relation},
\begin{eqnarray}
{\cal{N}}_{\rm singlet} (S_q) & = & {N\choose N/2-S_q}-{N\choose N/2-S_q-1} 
\nonumber \\
& = & \sum_{\{N_{n}\}}\,\prod_{n =1}^{\infty} {L_n\choose N_n} \, .
\label{Nsinglet-MM}
\end{eqnarray}
The summation $\sum_{\{N_{n}\}}$ on its right-hand side is over all sets $\{N_{n}\}$ obeying
the exact fixed-$S_q$ sum rule $2S_q = N - \sum_{n=1}^{\infty}2n\,N_{n}$ in Eq. (\ref{2Pi}) and 
${L_n\choose N_n}$ gives the number of independent occupancy configurations of 
$N_n$ $n$-pairs over the $n$-band  $j =1,...,L_n$ discrete momentum values $q_j$
such that ${L_n\choose N_n}={L_n\choose N_n^h}$.

\section{Subspaces spanned by states generated from the ground state by a finite number of processes}
\label{SECIII}

The results reviewed in Sec. \ref{SECII} refer to the whole Hilbert space of the spin-$1/2$ $XXZ$ chain
in a longitudinal magnetic field. On the other hand, the expressions of some physical quantities given 
below in Secs. \ref{SECIV}, \ref{SECV}, and \ref{SECVI} involve basic quantities that refer to subspaces 
spanned by energy eigenstates generated from the ground state by a finite number $N_{\rm ex}$ of elementary processes. The
specificity of these subspaces follows from $N_{\rm ex}/N\rightarrow 0$ in the present thermodynamic limit, $N\rightarrow\infty$. 
Here we introduce the corresponding basic quantities needed for our studies.

For such subspaces one has that $q_n^{\Delta}=0$ in Eq. (\ref{qnpm}),
the $n$-band limiting values, Eq. (\ref{qnpm}), having simple expressions
$q_1^{\pm} = \pm k_{F\uparrow}$ and $q_n^{\pm} = \pm (k_{F\uparrow}-k_{F\downarrow})$ for $n>1$
where $k_{F\uparrow} = {\pi\over 2}(1+m)$ and $k_{F\downarrow} = {\pi\over 2}(1-m)$.
As in Sec. \ref{SECIIA}, within the thermodynamic limit the $n$-band momentum values $q_j={2\pi\over N}I_j^n$'s 
can be described by $n$-band continuous momentum variables $q \in [q_n^-,q_n^+]$, which here simply read
$q\in [-k_{F\uparrow},k_{F\uparrow}]$ for $n=1$ and $q\in [-(k_{F\uparrow}-k_{F\downarrow}),(k_{F\uparrow}-k_{F\downarrow})]$
for $n>1$. 

Ground states refer to a $1$-band occupied $1$-pair Fermi sea for $q \in [-k_{F\downarrow},k_{F\downarrow}]$ 
with $1$-holes for $\vert q\vert\in [k_{F\downarrow},k_{F\uparrow}]$ and empty $n$-bands for $n>1$
and thus with $n$-holes for the whole interval $q\in [-(k_{F\uparrow}-k_{F\downarrow}),(k_{F\uparrow}-k_{F\downarrow})]$
of these bands.

The finite number $N_{\rm ex}$ of elementary processes that generate the excited states that span
the subspaces considered here from the ground state refer to changes in the occupancies of the discrete 
$n$-band momentum values $q_j$ whose spacing is $q_{j+1}-q_j = {2\pi\over N}$.
We thus expand the Bethe-ansatz equations, Eq. (\ref{BAqn}), and 
energy eigenvalues, Eq. (\ref{Energy}), in the small $n$-band momentum distribution deviations
$\delta N_n (q_j) = N_n (q_j) - N_n^0 (q_j)$ and corresponding
$n$-band rapidity-variable distribution deviations 
$\delta \tilde{N}_n (\varphi) = \tilde{N}_n (\varphi) - \tilde{N}_n^0 (\varphi)$,
which in the thermodynamic limit refer to small $n$-band continuous momentum distribution deviations,
\begin{equation}
\delta N_n (q) = N_n (q) - N_n^0 (q) \, .
\label{deltaNnq}
\end{equation}
Here $\tilde{N}_n^0 (\varphi)$ and $N_n^0 (q)=N_n^0 (q_j)$ refer to the ground state.
Such distribution deviations obey the relation $\delta \tilde{N}_n (\varphi_n (q)) = \delta N_n (q)$, Eq. (\ref{tildeNN}).
\begin{figure}
\begin{center}
\subfigure{\includegraphics[width=5.65cm]{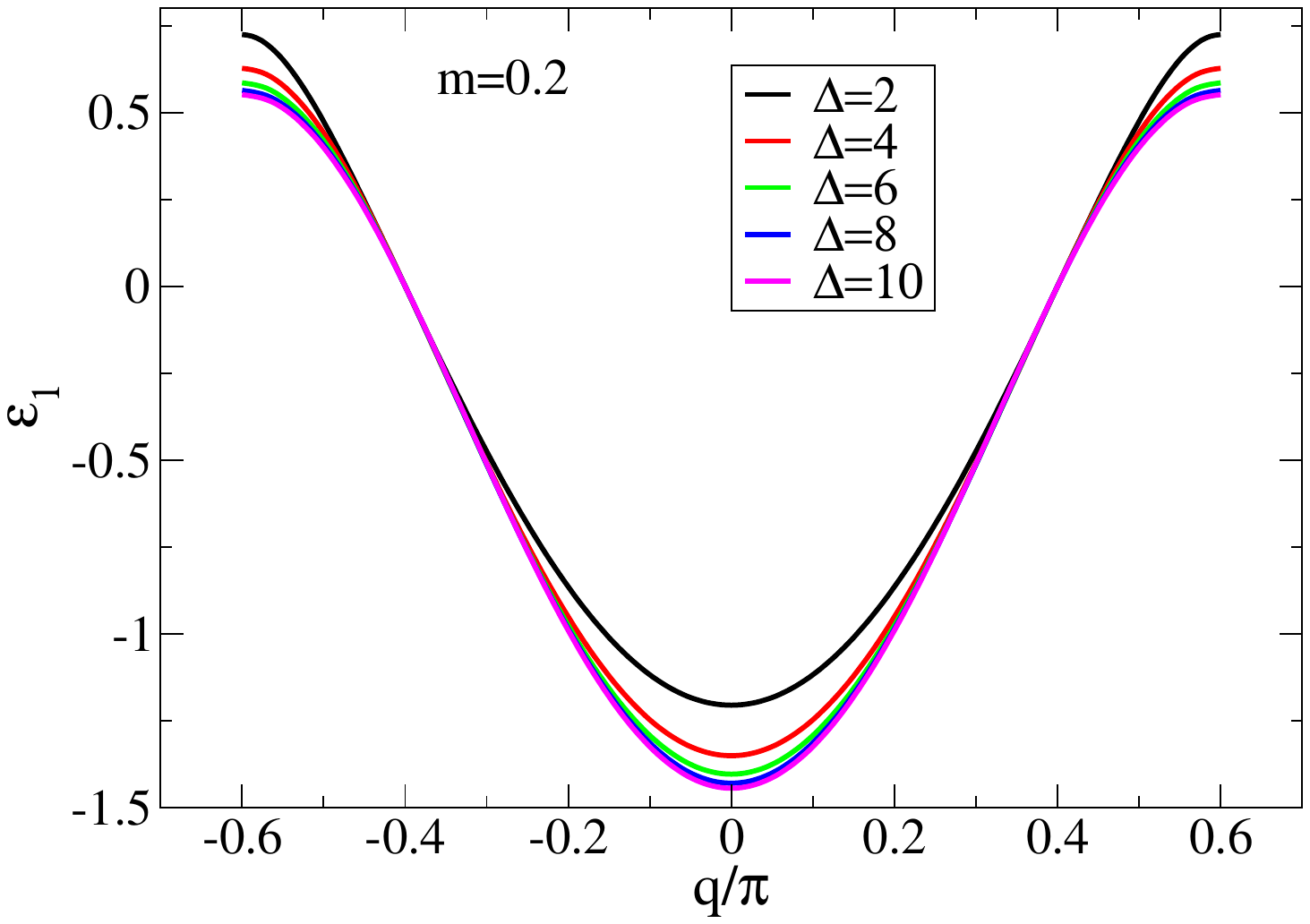}}
\hspace{0.25cm}
\subfigure{\includegraphics[width=5.65cm]{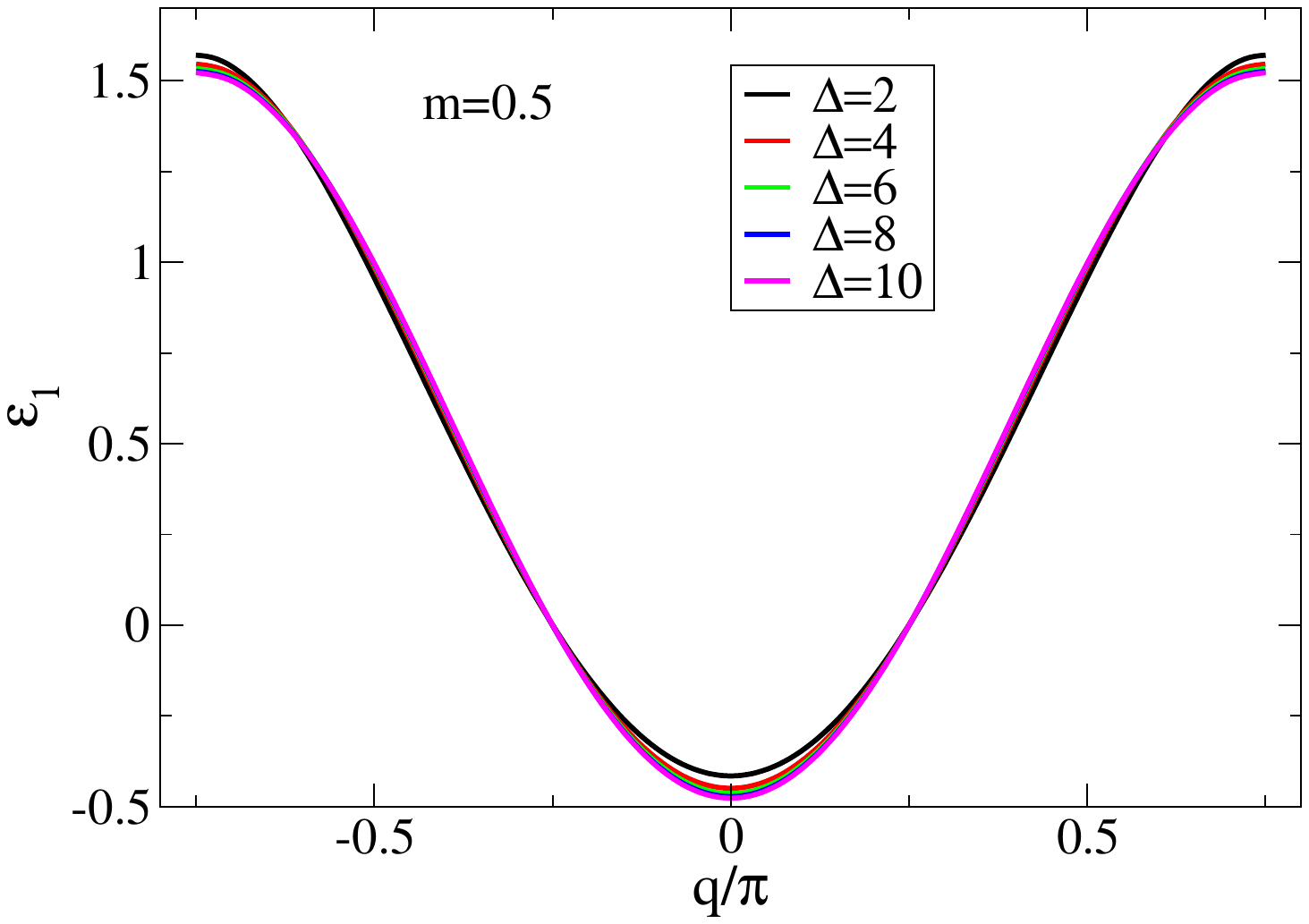}}
\hspace{0.25cm}
\subfigure{\includegraphics[width=5.65cm]{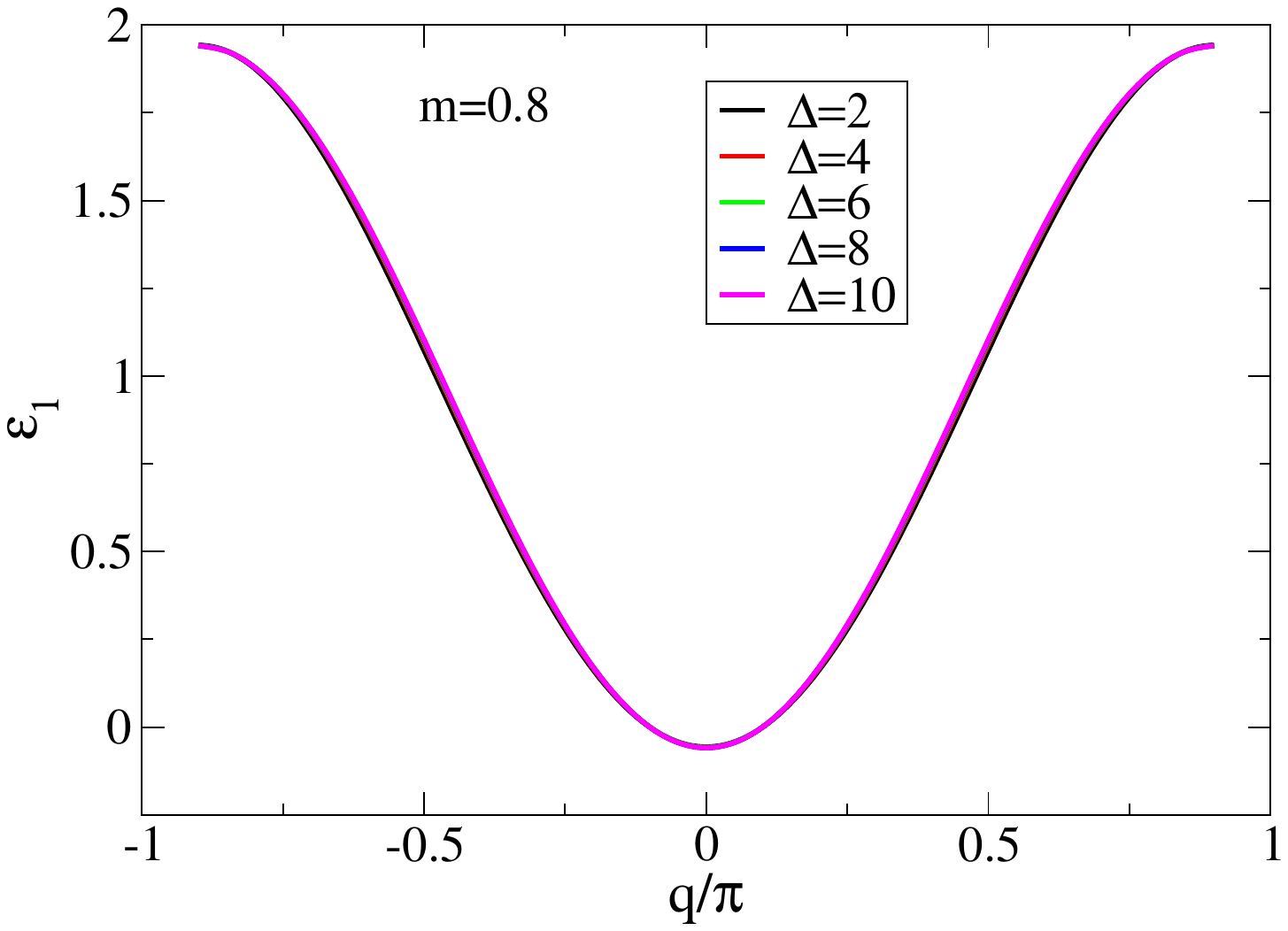}}
\caption{The $1$-pair energy dispersion $\varepsilon_{1} (q)$ in units of $J$
defined by Eqs. (\ref{equA4n}) and (\ref{equA4n10}) for $n=1$ plotted as
a function of $q/\pi$ for $1$-band momentum $q \in [-k_{F\uparrow},k_{F\uparrow}]$, spin densities
$m=0.2$, $m=0.5$, and $m=0.8$, and spin anisotropies $\Delta =2,4,6,8,10$.
From Ref. \onlinecite{Carmelo_22}.}
\label{FigChaos1}
\end{center}
\end{figure}

For ground states, one has that $\tilde{N}_n^0 (\varphi) =0$ and $N_n^0 (q) =0$
for $n>1$ in the above deviations. The ground-state $1$-band
momentum function $q_1 (\varphi)$ is the inverse of the $1$-band rapidity function
$\varphi_{1}(q)\in [-\pi,\pi]$. In the thermodynamic limit it is a solution of Bethe-ansatz equation, Eq. (\ref{BAqn})
with $\tilde{N}_1^0 (\varphi)=1$ for $\varphi\in [-B,B]$, $\tilde{N}_1^0 (\varphi)=0$ 
for $\vert\varphi\vert \in [B,\pi]$, and $\tilde{N}_n^0 (\varphi)=0$ for $n>1$
where the parameter $B$ is such that,
\begin{eqnarray}
B & = & \varphi_{1} (k_{F\downarrow}) \hspace{0.20cm}{\rm with}
\nonumber \\
\lim_{m\rightarrow 0} B & = & \pi\hspace{0.20cm}{\rm and}\hspace{0.20cm}
\lim_{m\rightarrow 1} B = 0 \hspace{0.20cm}{\rm for}\hspace{0.20cm}\Delta > 1 \, .
\label{QB-r0rs}
\end{eqnarray}
Indeed, one has in the thermodynamic limit in terms of $n$-band continuous momentum distributions
that $N_1^0 (q)=1$ for $q\in [-k_{F\downarrow},k_{F\downarrow}]$, 
$N_1^0 (q)=0$ for $\vert q\vert \in [k_{F\downarrow},k_{F\uparrow}]$, and $N_n^0 (q)=0$ for $n>1$.

The ground-state rapidity functions $\varphi_{n}(q)$ have simple analytical expressions
at spin density values $m=0$ and $m=1$.
In the spin-insulating quantum phase for fields $h\in [0,h_{c1}]$ and $m=0$,
the interval $q\in [-(k_{F\uparrow} - k_{F\downarrow}),(k_{F\uparrow} - k_{F\downarrow})]$ 
of the $n>1$ rapidity functions $\varphi_{n}(q)\in [-\pi,\pi]$ argument collapses to $q=0$. 

On the other hand, for $h\in [0,h_{c1}]$ and $m=0$ the $1$-band rapidity function $\varphi_1 (q)$ reads,
\begin{equation}
\varphi_1 (q) = \pi {F (q,u_{\eta})\over K (u_{\eta})} \hspace{0.20cm}{\rm for}\hspace{0.20cm}
q\in [-\pi/2,\pi/2] \, ,
\label{varphi1}
\end{equation}
where $K (u_{\eta})$ is the complete elliptic integral given in Eq. (\ref{elliptic}) and 
the elliptic integral $F (q,u_{\eta})$ is given by,
\begin{equation}
F (q,u_{\eta}) = \int_0^{q}d\theta {1\over \sqrt{1 - u_{\eta}^2\sin^2\theta}} \, .
\label{Felliptic}
\end{equation}  
In the opposite limit of $h=h_{c2}$ and $m=1$, the rapidity function $\varphi_n (q)$
has the following closed-form expression valid for $n\geq 1$,
\begin{equation}
\varphi_n (q) = 2\arctan\left(\tanh\left({n\,\eta\over 2}\right)\tan\left({q\over 2}\right)\right)
\hspace{0.10cm}{\rm for}\hspace{0.20cm} q \in [-\pi,\pi] \, .
\label{varphinqm1}
\end{equation}

Expanding the excitation energy $\delta E = E (l_{\rm r}^{\eta},S_q,S^z) - E_{GS}$ 
of excited energy eigenstates generated from the ground state by a finite number $N_{\rm ex}$ of 
elementary processes up to second order in the small $n$-band distribution deviations, Eq. (\ref{deltaNnq}),
leads in the thermodynamic limit, $N\rightarrow\infty$, to the following expression,
\begin{eqnarray}
\delta E & = & \sum_{n=1}^{\infty}\sum_{j=1}^{L_n}\varepsilon_n (q_j)\,\delta N_n (q_j) 
\nonumber \\
& + & {1\over N}\sum_{n,n'=1}^{\infty}
\sum_{j=1}^{L_n}\sum_{j'=1}^{L_{n'}}{1\over 2}\,f_{n\,n'} (q_j,q_{j'})\,\delta N_n (q_j)\delta N_{n'} (q_{j'}) 
\nonumber \\
& = & {N\over 2\pi}\sum_{n=1}^{\infty}\int_{q_n^-}^{q_n^+}dq\,\varepsilon_n (q)\,\delta N_n (q) 
\nonumber \\
& + & {N\over 4\pi^2}\sum_{n,n'=1}^{\infty}\int_{q_n^-}^{q_n^+}dq\int_{q_{n'}^-}^{q_{n'}^+}dq'
{1\over 2}\,f_{n\,n'} (q,q')
\nonumber \\
& \times & \delta N_n (q)\delta N_{n'} (q') \, .
\label{DEnHm}
\end{eqnarray}
The $n$-pairs energy dispersion $\varepsilon_n (q)$ and the $f$-functions $f_{n\,n'} (q,q')$
appearing here are for $n\geq 1$ given by \cite{Carmelo_23,Carmelo_22},
\begin{eqnarray}
\varepsilon_{n} (q) & = & {\bar{\varepsilon}_{n}} (\varphi_n (q)) \hspace{0.20cm}{\rm and}\hspace{0.20cm}
\varepsilon_{n}^0 (q)={\bar{\varepsilon}_{n}^0} (\varphi_n (q))\hspace{0.20cm}{\rm where}
\nonumber \\
{\bar{\varepsilon}_{n}} (\varphi) & = & {\bar{\varepsilon}_{n}^0} (\varphi) + \Bigl(n - \delta_{n,1}{1\over 2}\Bigr)\, g\mu_B\,h
\hspace{0.20cm}{\rm for}\hspace{0.20cm}h\in [0,h_{c1}]
\nonumber \\
{\bar{\varepsilon}_{n}} (\varphi) & = &
{\bar{\varepsilon}_{n}^0} (\varphi) + n\,g\mu_B\,h
\hspace{0.20cm}{\rm for}\hspace{0.20cm}h\in ]h_{c1},h_{c2}] \, ,
\label{equA4n}
\end{eqnarray}
and \cite{Carmelo_18},
\begin{eqnarray}
&& f_{n\,n'} (q,q') = v_n (q)\,2\pi\Phi_{n\,n'} (q,q') + v_{n'} (q')\,2\pi\Phi_{n'\,n} (q',q) 
\nonumber \\
&& \hspace{0.5cm} + 
{v_ 1 (k_{F\downarrow})\over 2\pi}\sum_{\iota = \pm}2\pi\Phi_{1\,n} (\iota k_{F\downarrow},q)\,2\pi\Phi_{1\,n'} (\iota k_{F\downarrow},q') \, ,
\label{ffnHmGEN}
\end{eqnarray}
respectively. Here $\varepsilon_{n} (q)$ and $\varepsilon_{n}^0 (q)$ are the $n$-band energy dispersions with
zero-energy level measured relative to the ground state for a fixed magnetic field in the range $h \in [0,h_{c2}]$ and
to the $h=0$ ground state, respectively.

The rapidity-variable dispersions ${\bar{\varepsilon}_{n}^0} (\varphi)$ appearing in Eq. (\ref{equA4n}) 
are given below, the critical magnetic fields $h_{c1}$ and $h_{c2}$ in that equation are defined in Eq. (\ref{criticalfields}), 
and the $n$-band group velocity in Eq. (\ref{ffnHmGEN}) reads,
\begin{equation}
v_n (q) = {\partial\varepsilon_n (q)\over\partial q} \, .
\label{vnq}
\end{equation}

Only the $1$-pair phase shifts $2\pi\Phi_{1\,n}(q,q')$ where $n\geq 1$ of those 
appearing in Eq. (\ref{ffnHmGEN}) contribute to the physical quantities given in the ensuing
sections. Within the physical-spins representation,
$2\pi\Phi_{1\,n} (q,q')$ (and $-2\pi\Phi_{1\,1} (q,q')$) is the phase shift acquired by one $1$-pair scatterer 
of $1$-band momentum $q \in [-k_{F\downarrow},k_{F\downarrow}]$ due to creation of one $n$-pair (and $1$-hole) scattering center 
at $n$-band (and $1$-band) momentum $q'\in [-k_{F\uparrow},-k_{F\downarrow}]$ or
$q'\in [k_{F\downarrow},k_{F\uparrow}]$ for $n=1$ and
$q' \in [-(k_{F\uparrow} - k_{F\downarrow}),(k_{F\uparrow} - k_{F\downarrow})]$ for $n>1$
(and $q' \in [-k_{F\downarrow},k_{F\downarrow}]$) under a transition from the ground state to an excited state.
Such $1$-pair phase shifts are given by,
\begin{equation}
2\pi\Phi_{1\,n}(q,q') = 2\pi\bar{\Phi }_{1\,n} \left(\varphi_1 (q),\varphi_n(q')\right) 
\hspace{0.20cm}{\rm for}\hspace{0.20cm}n\geq 1 \, ,
\label{Phi-barPhi}
\end{equation}
where the rapidity-variable dependent phase shifts 
$2\pi\bar{\Phi }_{1\,n} \left(\varphi,\varphi'\right)$ are 
defined by the following integral equations,
\begin{eqnarray}
2\pi\bar{\Phi }_{1\,1} \left(\varphi,\varphi'\right) & = & 
2\arctan\left(\coth \eta\tan\left({\varphi - \varphi'\over 2}\right)\right)
\nonumber  \\
& + & \int_{-B}^{B} d\varphi''\,G_1 (\varphi - \varphi'')\,2\pi\bar{\Phi }_{1\,1} \left(\varphi'',\varphi'\right) \, ,
\nonumber  \\
{\rm and} &&
\nonumber  \\
2\pi\bar{\Phi }_{1\,n} \left(\varphi,\varphi'\right) & = & 
2\sum_{\iota=\pm 1}
\nonumber  \\
&& \arctan\left(\coth\left({(n + \iota)\,\eta\over 2}\right)\tan\left({\varphi - \varphi'\over 2}\right)\right)
\nonumber  \\
& + & \int_{-B}^{B} d\varphi''\,G_1 (\varphi - \varphi'')\,2\pi\bar{\Phi }_{1\,n} \left(\varphi'',\varphi'\right) \, ,
\label{Phis1n}
\end{eqnarray}
for $n>1$. The kernel reads
$G_1 (\varphi) = - {1\over{2\pi}}{\sinh (2\eta)\over \cosh (2\eta) - \cos (\varphi)}$.

The rapidity-variable dependent dispersions ${\bar{\varepsilon}_{n}^0} (\varphi)$ in Eq. (\ref{equA4n}) are defined by the equations,
\begin{eqnarray}
{\bar{\varepsilon}_{n}^0} (\varphi) & = & \int_{0}^{\varphi }d\varphi^{\prime}2J\gamma_{n} (\varphi^{\prime}) + A_n^{0}
\hspace{0.20cm}{\rm where}
\nonumber \\
A_1^{0} & = & - J(1 + \cosh \eta) 
\nonumber \\
& + & {1\over\pi}\int_{-B}^{B}d\varphi^{\prime}\,2J\gamma_{1} (\varphi^{\prime})
\arctan\left(\coth \eta\tan\left({\varphi^{\prime}\over 2}\right)\right)
\nonumber \\
&& {\rm and}
\nonumber \\
A_n^{0} & = & -J {\sinh \eta\over\sinh (n\,\eta)}\left(1 + \cosh (n\,\eta)\right) 
\nonumber \\
& + & {1\over\pi}\sum_{\iota=\pm 1}\int_{-B}^{B}d\varphi^{\prime}\,2J\gamma_{1} (\varphi^{\prime})
\nonumber \\
& \times & \arctan\left(\coth \left({(n + \iota)\,\eta\over 2}\right)\tan\left({\varphi^{\prime}\over 2}\right)\right) \, ,
\label{equA4n10}
\end{eqnarray}
for $n> 1$. The energy-related distribution $2J\gamma_{n} (\varphi)$ obeys the following equation for $n\geq 1$,
which is an integral equation at $n=1$,
\begin{eqnarray}
2J\gamma_{n} (\varphi) & = & J\,{\sinh \eta\,\sinh (n\,\eta)\sin (\varphi)\over (\cosh (n\,\eta) - \cos (\varphi))^2} 
\nonumber \\
& + & \int_{-B}^{B}d\varphi^{\prime}\,G_n (\varphi - \varphi^{\prime})\,2J\gamma_{1} (\varphi^{\prime})  \, ,
\label{equA6}
\end{eqnarray}
where $G_n (\varphi) = - {1\over{2\pi}}\sum_{\iota=\pm 1}{\sinh ((n+\iota)\,\eta)\over \cosh ((n+\iota)\,\eta) - \cos (\varphi)}$.

\begin{figure}
\begin{center}
\centerline{\includegraphics[width=8.5cm]{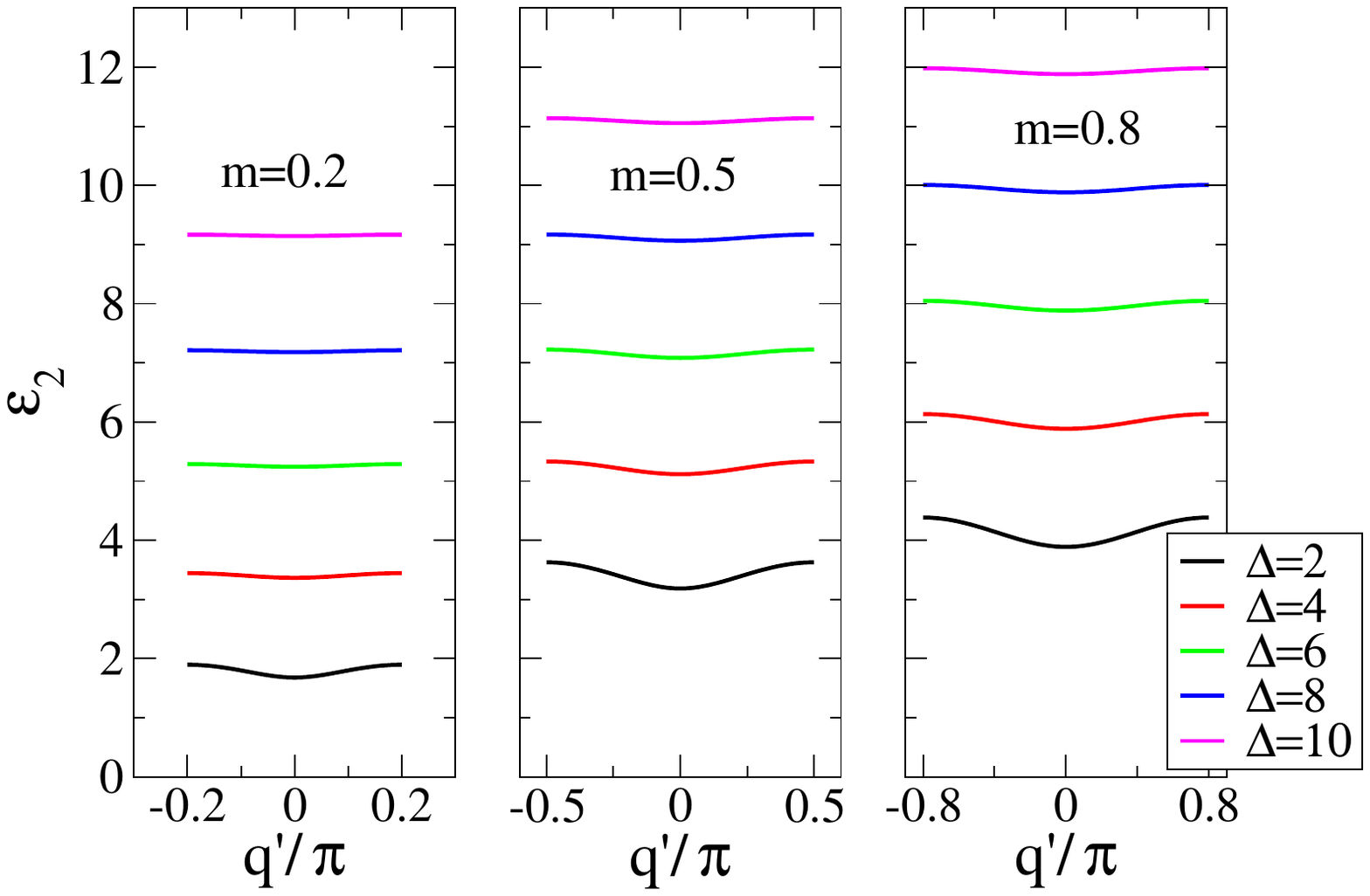}}
\caption{The $2$-string-pair energy dispersion $\varepsilon_{2} (q)$ 
in units of $J$ is plotted as a function of $q/\pi$ for $2$-band momentum $q\in [-(k_{F\uparrow} - k_{F\downarrow}),(k_{F\uparrow} - k_{F\downarrow})]$
(with $q$ denoted in Ref.  \onlinecite{Carmelo_23} by $q'$ for $n>1$), spin densities $m=0.2$, $m=0.5$, $m=0.8$, and spin anisotropies $\Delta =2,4,6,8,10$.
It is associated with a Bethe string of length two. From Ref. \onlinecite{Carmelo_23}.}
\label{FigChaos2}
\end{center}
\end{figure}

\begin{figure}
\begin{center}
\centerline{\includegraphics[width=8.5cm]{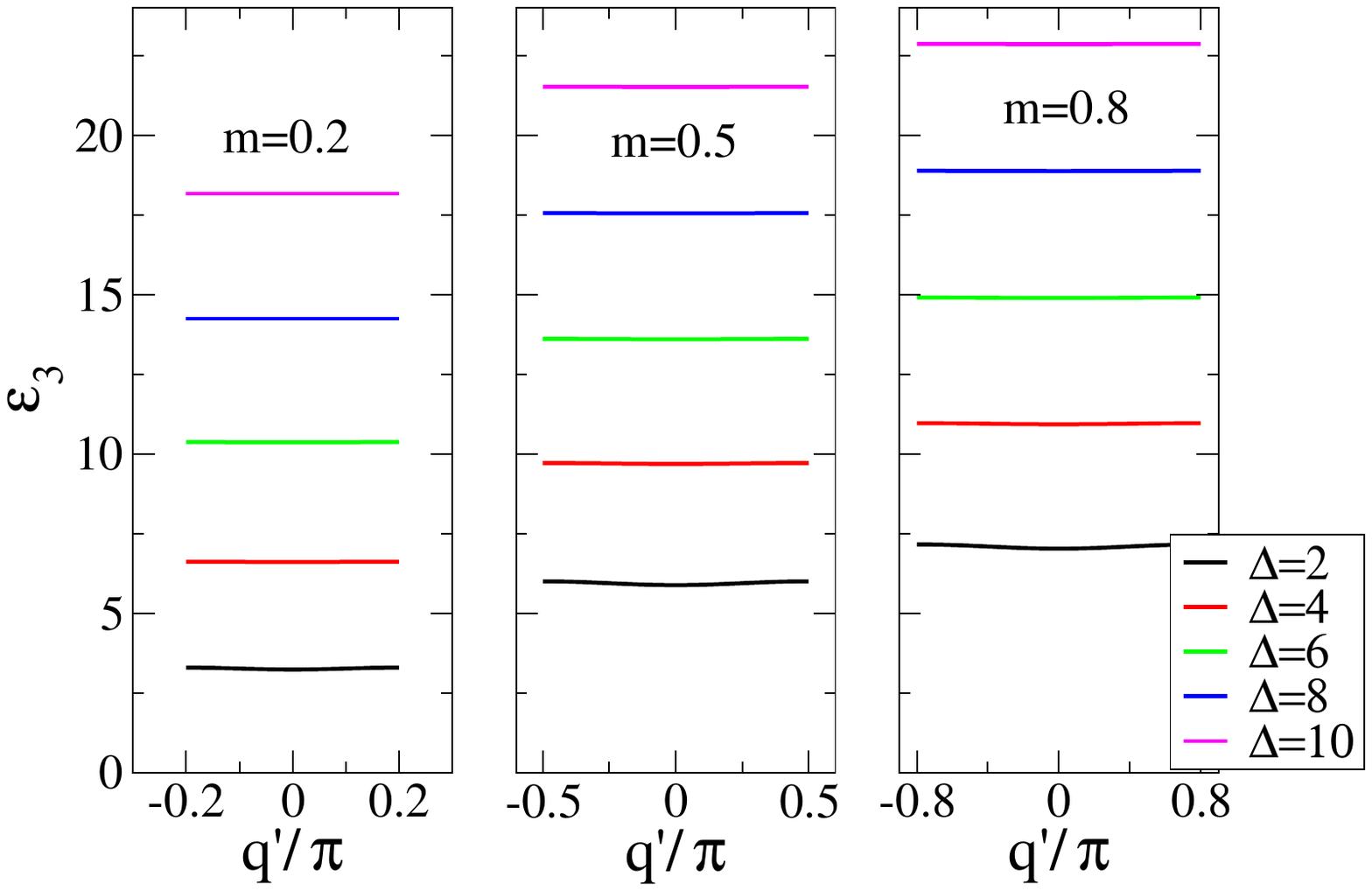}}
\caption{The same as in Fig. \ref{FigChaos2} for the
$3$-string-pair energy dispersion $\varepsilon_{3} (q)$ 
associated with a Bethe string of length three. From Ref. \onlinecite{Carmelo_23}.}
\label{FigChaos3}
\end{center}
\end{figure}

The $n$-pair energy dispersions $\varepsilon_{n} (q)$, Eqs. (\ref{equA4n}) and (\ref{equA4n10}),
are plotted in units of $J$ in Figs. \ref{FigChaos1}, \ref{FigChaos2}, and \ref{FigChaos3} for $n=1$, $n=2$, and
$n=3$, respectively, as a function of $q/\pi$ for spin densities $m=0.2$, $m=0.5$, $m=0.8$ and
anisotropies $\Delta =2,4,6,8,10$. For $n=2$ and $n=3$ such dispersions are gapped.

For $h \in [0,h_{c1}]$ and $m=0$ and for $h=h_{c2}$ and $m=1$, the energy dispersions
$\varepsilon_{n} (q)$ and $\varepsilon_{n}^0 (q)$ have the following simple analytical expressions,
\begin{eqnarray}
\varepsilon_{1} (q) & = & \varepsilon_{1}^0 (q) + {1\over 2}\, g\mu_B\,h
\nonumber \\
\varepsilon_{1}^0 (q) & = & - {J\over\pi}\sinh\eta\,K (u_{\eta}) \sqrt{1 - u_{\eta}^2\sin^2 q} 
\nonumber  \\
&& {\rm for}\hspace{0.20cm}q \in [-\pi/2,\pi/2]\hspace{0.20cm}{\rm and}\hspace{0.20cm}h \in [0,h_{c1}] 
\nonumber \\
&& {\rm and}
\nonumber  \\
\varepsilon_{1} (q) & = & \varepsilon_{n}^0 (q) + J (1+ \Delta) = J (1 - \cos q)
\nonumber \\
\varepsilon_{1}^0 (q) & = & - J (\Delta + \cos q)
\nonumber  \\
&& {\rm for}\hspace{0.20cm}q \in [-\pi,\pi]\hspace{0.20cm}{\rm and}\hspace{0.20cm}h = h_{c2} \, ,
\label{vareband1m0m1}
\end{eqnarray}
at $n=1$ and,
\begin{eqnarray}
\varepsilon_{n} (q) & = & \varepsilon_{n}^0 (q) + n\,g\mu_B\,h = (n-1)\,g\mu_B\,h
\nonumber  \\
\varepsilon_{n}^0 (q) & = & - g\mu_B\,h 
\nonumber \\
&& {\rm for}\hspace{0.20cm} q' = 0 \hspace{0.20cm}{\rm and}\hspace{0.20cm}h \in [0,h_{c1}] 
\nonumber \\
&& {\rm and}
\nonumber \\
\varepsilon_{n} (q) & = & \varepsilon_{n}^0 (q) + n\,J (1+ \Delta) 
\nonumber \\
\varepsilon_{n}^0 (q) & = & - J\,{\sinh \eta\over\sinh (n\,\eta)}(\cosh (n\,\eta) + \cos q) 
\nonumber  \\
&& {\rm for}\hspace{0.20cm} q \in [-\pi,\pi] \hspace{0.20cm}{\rm and}\hspace{0.20cm}h = h_{c2} \, ,
\label{qdepvarepsilonm0m1}
\end{eqnarray}
for $n>1$. The complete elliptic integral $K (u_{\eta})$ appearing in Eq. (\ref{vareband1m0m1}) 
is given in Eq. (\ref{elliptic}).

The $1$-band group velocity, 
\begin{equation}
v_1(k_{F\downarrow})\hspace{0.20cm}{\rm where}\hspace{0.20cm}
v_1 (q) = {\partial\varepsilon_1 (q)\over\partial q} \, ,
\label{v1q}
\end{equation}
and $v_1 (q)$ is the $n$-band group velocity given in Eq. (\ref{vnq}) at $n=1$ appears in 
several expressions of physical quantities given in the ensuing sections. 

For anisotropies $\Delta >1$ and $\Delta = 1$, spin density $m=0$, and 
magnetic field $0\leq h\leq h_{c1}$, where $h_{c1}=0$ for $\Delta =1$, the
general $1$-band group velocity $v_1 (q)$ reads,
%CORR factor 1/2
\begin{eqnarray}
v_{1} (q) & = & {J\over 2\pi}\sinh \eta\,K (u_{\eta}){u_{\eta}^2\sin 2q\over \sqrt{1 - u_{\eta}^2\sin^2 q}} 
\hspace{0.20cm}{\rm for}\hspace{0.20cm}\Delta > 1
\nonumber \\
v_{1} (q) & = & J{\pi\over 2} \sin q \hspace{0.20cm}{\rm for}\hspace{0.20cm}\Delta = 1 \, .
\label{v1qm0}
\end{eqnarray}
On the other hand, for $\Delta \geq 1$, $m=1$, and $h= h_{c2}$, where $h_{c2}=h_c$ for $\Delta =1$
it is given by,
\begin{equation}
v_1 (q) = J\sin q \hspace{0.20cm}{\rm for}\hspace{0.20cm}\Delta \geq 1 \, .
\label{v1qm1}
\end{equation}

The corresponding $1$-band group velocity at $q=k_{F\downarrow}$, Eq. (\ref{v1q}), is thus such that, 
\begin{eqnarray}
v_{1} (k_{F\downarrow}) & = & v_{1} \left({\pi\over 2}(1-m)\right)
\nonumber \\
& = & v_{1} \left({\pi\over 2}\right) = 0 \hspace{0.20cm}{\rm for}\hspace{0.20cm}\Delta > 1 \hspace{0.20cm}{\rm and}\hspace{0.20cm}m=0
\nonumber \\
& = & v_{1} \left({\pi\over 2}\right) = J{\pi\over 2} \hspace{0.20cm}{\rm for}\hspace{0.20cm}\Delta = 1 
\hspace{0.20cm}{\rm and}\hspace{0.20cm}m=0
\nonumber \\
& = & v_{1} (0) = 0 \hspace{0.20cm}{\rm for}\hspace{0.20cm}\Delta \geq 1 \hspace{0.20cm}{\rm and}\hspace{0.20cm}m=1 \, .
\label{v1qm0kF}
\end{eqnarray}
Hence for $m=0$ and $0\leq h\leq h_{c1}$ it vanishes for $\Delta >1$ yet is finite at $\Delta =1$ when $h_{c1}=0$.

Similarly to the group velocity $v_{1} (k_{F\downarrow})$, the following phase-shift parameter $\xi$ appears
in several expressions of physical quantities given in the ensuing sections,
\begin{equation}
\xi = 1 + {1\over 2\pi}\sum_{\iota=\pm 1} (\iota)\,2\pi\Phi_{1\,1}\left(k_{F\downarrow},\iota k_{F\downarrow}\right) \, .
\label{x-aaPM}
\end{equation}
It has values in the intervals $\xi \in [1/2,1]$ for $\Delta >1$ and $\xi \in [1/\sqrt{2},1]$ for $\Delta = 1$, its
limiting values being,
\begin{eqnarray}
\xi & = & {1\over 2}\hspace{0.20cm}{\rm for} \hspace{0.20cm}m\rightarrow 0\hspace{0.20cm}{\rm and} \hspace{0.20cm}\Delta > 1
\nonumber \\
\xi & = & {1\over\sqrt{2}}\hspace{0.20cm}{\rm for} \hspace{0.20cm}m\rightarrow 0\hspace{0.20cm}{\rm and} \hspace{0.20cm}\Delta = 1
\nonumber \\
\xi & = & 1\hspace{0.20cm}{\rm for} \hspace{0.20cm}m\rightarrow 1\hspace{0.20cm}{\rm and} \hspace{0.20cm}\Delta \geq 1 \, .
\label{xilimits}
\end{eqnarray}
This parameter involves the $1$-pair phase shift $2\pi\Phi_{1,1} (q,q')$, 
Eq. (\ref{Phi-barPhi}), as given in Eq. (\ref{x-aaPM}), where in $2\pi\Phi_{1,1} (k_{F\downarrow},k_{F\downarrow})$ the two 
momentum values differ by $2\pi/N$. Hence within the physical-spins representation the parameter $\xi$ 
is determined by physical-spins $1$-pair - $1$-pair scattering. 

This actually reveals that the usual Tomonaga-Luttinger liquid (TLL) parameter $K$ \cite{Horvatic_20}
is determined by physical-spins $1$-pair - $1$-pair scattering. Indeed, it is directly related to the phase-shift parameter 
$\xi$, Eq. (\ref{x-aaPM}), as $K = \xi^2$, so that in terms of phase shifts it reads
$K = (1 + {1\over 2\pi}\sum_{\iota = \pm1}(\iota)2\pi\Phi_{1,1} (k_{F\downarrow},\iota\,k_{F\downarrow}))^2$.

\section{Low-temperature dynamical properties for fields $h_{c1}<h<h_{c2}$}
\label{SECIV}

Magnetization experimental results for spin chains in the zigzag materials BaCo$_2$V$_2$O$_8$ 
and SrCo$_2$V$_2$O$_8$ are explained well in terms of a 
spin-$1/2$ $XXZ$ chain in a longitudinal magnetic field with
anisotropy $\Delta\approx 2$ \cite{Kimura_07,Okunishi_07,Kimura_08,Han_21}. 

In addition, for their low-temperature spin-conducting phases,
the magnetic-field dependencies of the energies of the sharp peaks in the transverse 
components $S^{\mp\pm} (k,\omega)$ of the spin dynamic structure factor 
observed by optical experiments \cite{Wang_18,Wang_19} have been quantitatively described 
by that purely 1D spin chain. In the case of the longitudinal component $S^{zz} (k,\omega)$ observed by neutron
scattering \cite{Bera_20}, this applies to the energies of the sharp peaks, yet their spectral weights are enhanced
relative to those predicted by the 1D physics, as discussed below in Sec. \ref{SECVI}.

Such spin-conducting phases occur for longitudinal magnetic fields $h_{c1}<h<h_{c2}$,
where $h_{c1}\approx 3.8$\,T and $h_{c2}\approx 22.9$\,T for BaCo$_2$V$_2$O$_8$ and
$h_{c1}\approx 3.8$\,T and $h_{c2}\approx 28.7$\,T for SrCo$_2$V$_2$O$_8$.
The 1D physics of these zigzag materials also includes the experimental identification of finite-energy sharp peaks in
the transverse component $S^{+-} (k,\omega)$ associated with excited states containing
exotic complex Bethe strings of length two and three \cite{Wang_18,Wang_19,Bera_20} described by the 
exact Bethe-ansatz solution \cite{Gaudin_71,Carmelo_22,Gaudin_14,Takahashi_71,Takahashi_99}
of the spin-$1/2$ $XXZ$ chain in a longitudinal magnetic field. 

A more accurate description of the quasi-1D materials BaCo$_2$V$_2$O$_8$ and SrCo$_2$V$_2$O$_8$
should involve a three-dimensional (3D) system of chains, each described by the
spin-$1/2$ $XXZ$ chain Hamiltonian, Eq. (\ref{HD1}), plus a small interchain term 
containing an effective interchain coupling $J'$ such that $J'\ll J$. 

The 1D physics then emerges at very low temperatures just above a small critical transition 
temperature $T_c (h)$, below which the 3D physics associated with the interchain coupling $J'$ applies. 
That very small critical transition temperature is at each value of the magnetic field $h$ the largest of two critical 
transition temperatures $T_c^{z} (h)$ and $T_c^{x} (h)$ associated with longitudinal and transverse orders, respectively.

One can calculate the expression of these two critical transition temperatures within interchain mean-field theory \cite{Okunishi_07}. 
Except for the effective interchain coupling $J'$, such expressions involve physical quantities given in Sec. \ref{SECIII}, which are
associated with the 1D physics emerging just above $T_c (h)$. 
The small critical transition temperature $T_c (h)$ and corresponding two critical transition temperatures 
$T_c^{z} (h)$ and $T_c^{x} (h)$ are given by \cite{Okunishi_07},
\begin{eqnarray}
T_c (h) & = & {\rm max}\{T_c^{z} (h),T_c^{x} (h)\}\hspace{0.20cm}{\rm where} 
\nonumber \\
T_c^{z} (h) & = & {v_1(k_{F\downarrow})\over 2\pi}
\nonumber \\
& \times & \left(\Delta\,4J' A_1^z{\sin (\pi \xi^2)\over v_1 (k_{F\downarrow})}
B^2 \left({\xi^2\over 2}, 1 - \xi^2\right)\right)^{1\over 2 (1-\xi^2)}
\nonumber \\
T_c^{x} (h) & = & {v_1(k_{F\downarrow})\over 2\pi}
\nonumber \\
& \times & \left(4J' A_0^x{\sin \left({\pi\over 4\xi^2}\right)\over v_1 (k_{F\downarrow})}
B^2 \left({1\over 8\xi^2}, 1 - {1\over 4\xi^2}\right)\right)^{2\xi^2\over 4\xi^2 - 1} \, .
\nonumber \\
\label{TcTc}
\end{eqnarray}
Here $4$ in $4J'$ is the coordination number for $3D$, $v_1(k_{F\downarrow})$ is 
the $1$-band group velocity at $q=k_{F\downarrow}$, Eqs. (\ref{v1q}) and (\ref{v1qm0kF}), 
$\xi$ is the phase-shift parameter, Eqs. (\ref{x-aaPM}) and (\ref{xilimits}), and
the Euler beta function $B (x,y)$ can be expressed in
terms of the gamma function as,
\begin{equation}
B (x,y)= {\Gamma (x)\Gamma (y)\over \Gamma (x+y)} \, .
\label{Bxy}
\end{equation}
The non-universal TLL pre-factors $A_0^x$ and $A_1^z$ of the static spin correlation functions 
also appearing in Eq. (\ref{TcTc}) can be numerically computed \cite{Hikihara_04} and are
plotted below in Sec. \ref{SECIVC}, as they appear in the expressions of other physical
quantities studied in that section.
\begin{figure*}
\includegraphics[width=0.45\textwidth]{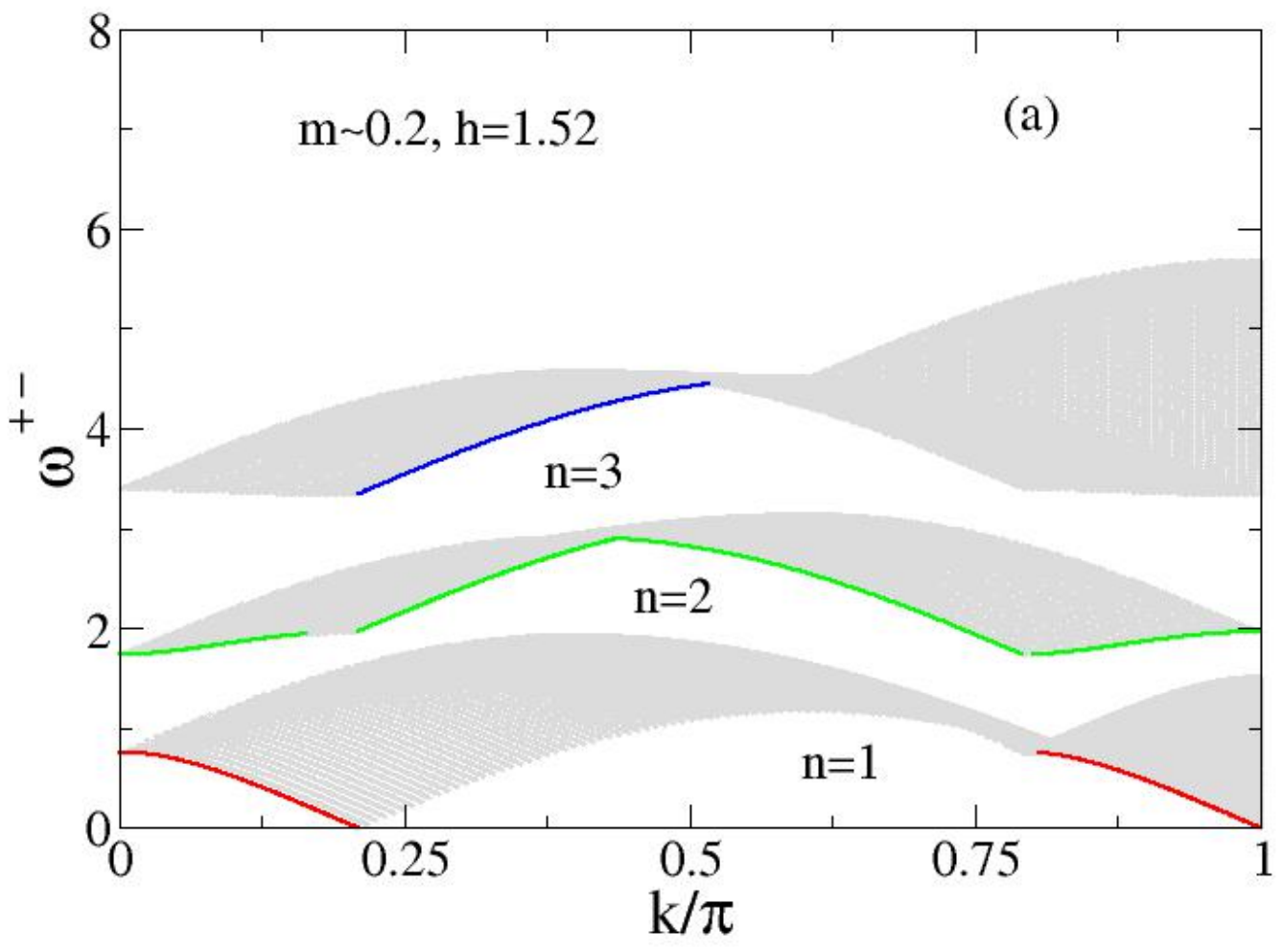}
\includegraphics[width=0.45\textwidth]{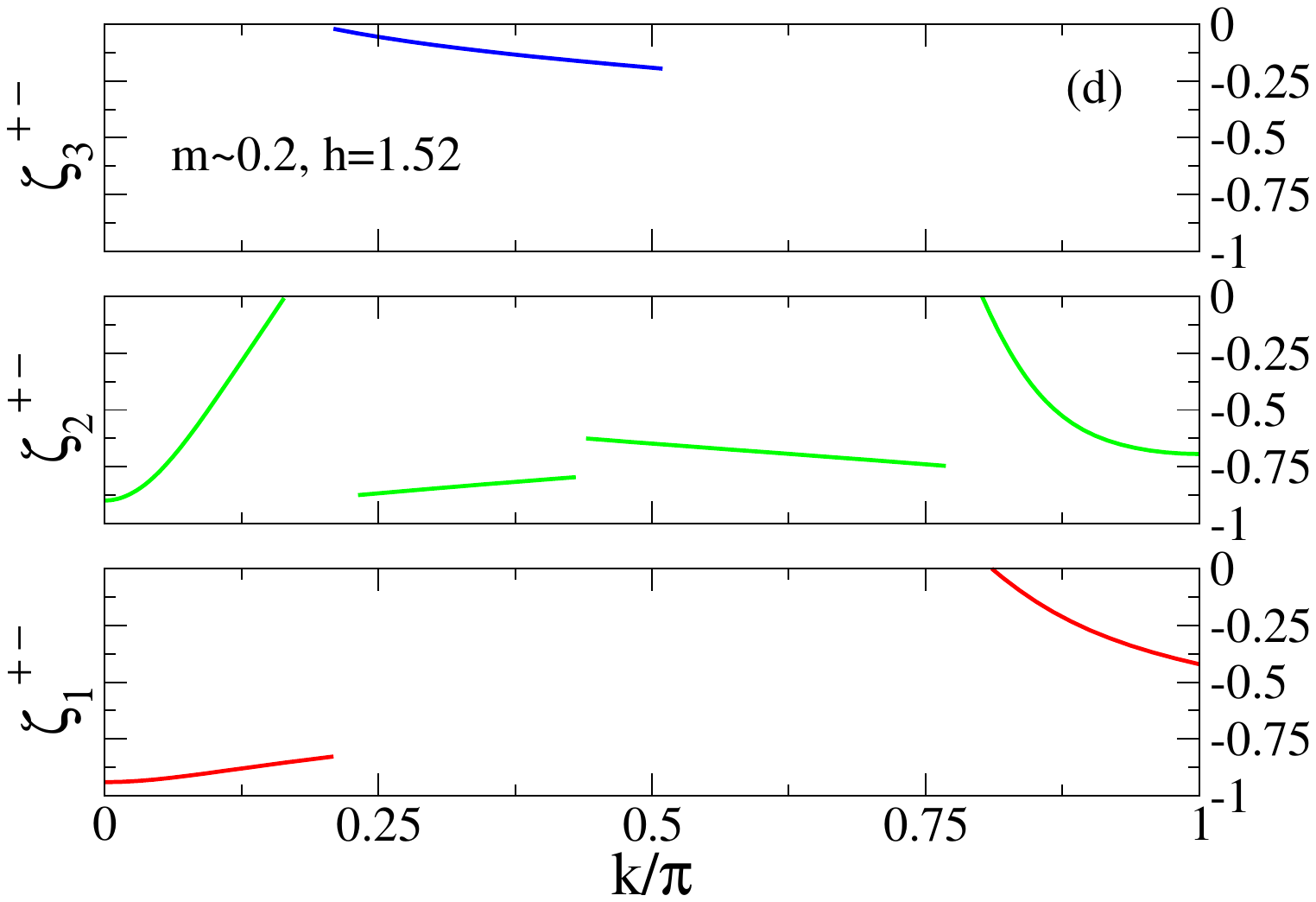}
\includegraphics[width=0.45\textwidth]{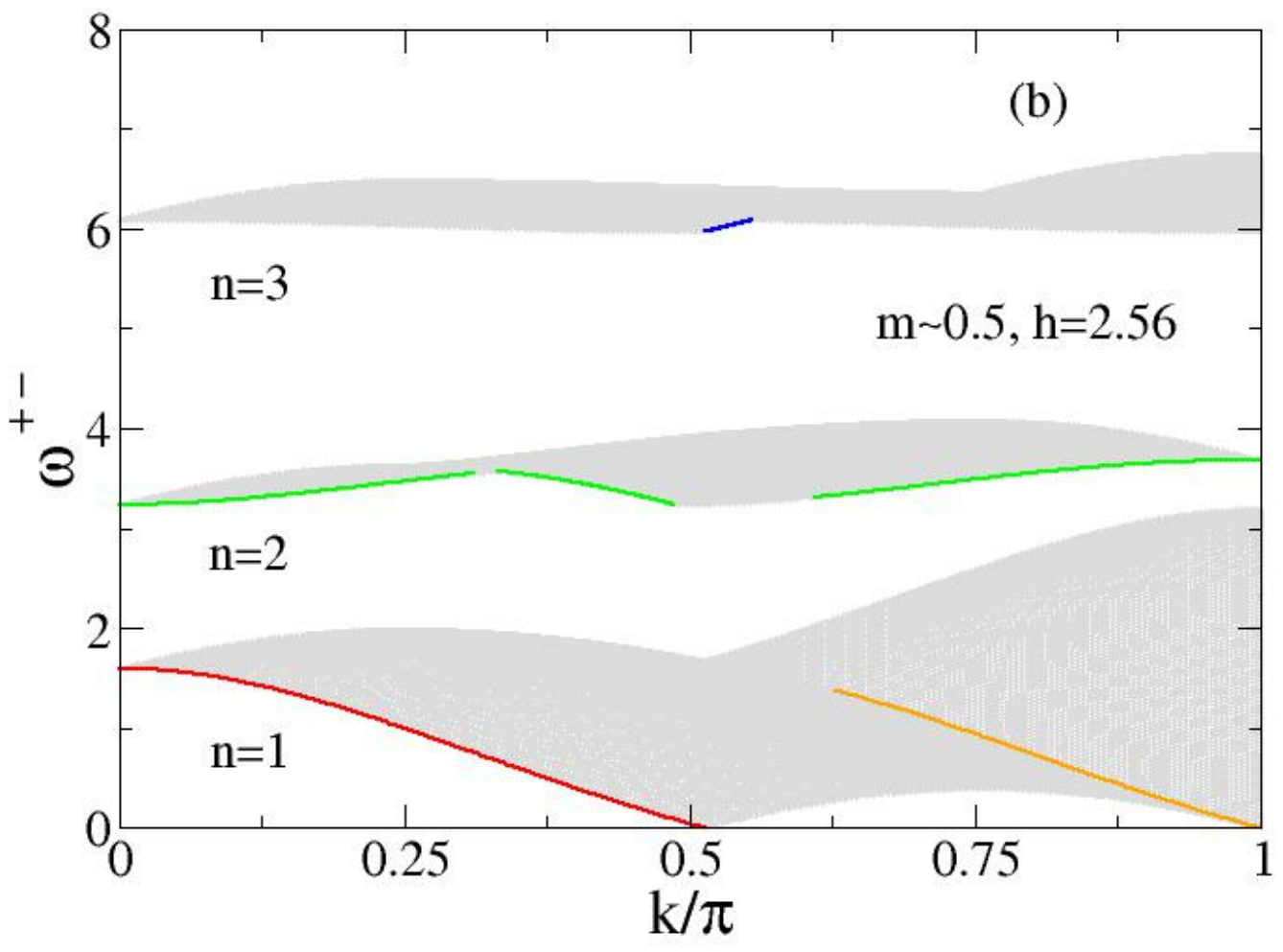}
\includegraphics[width=0.45\textwidth]{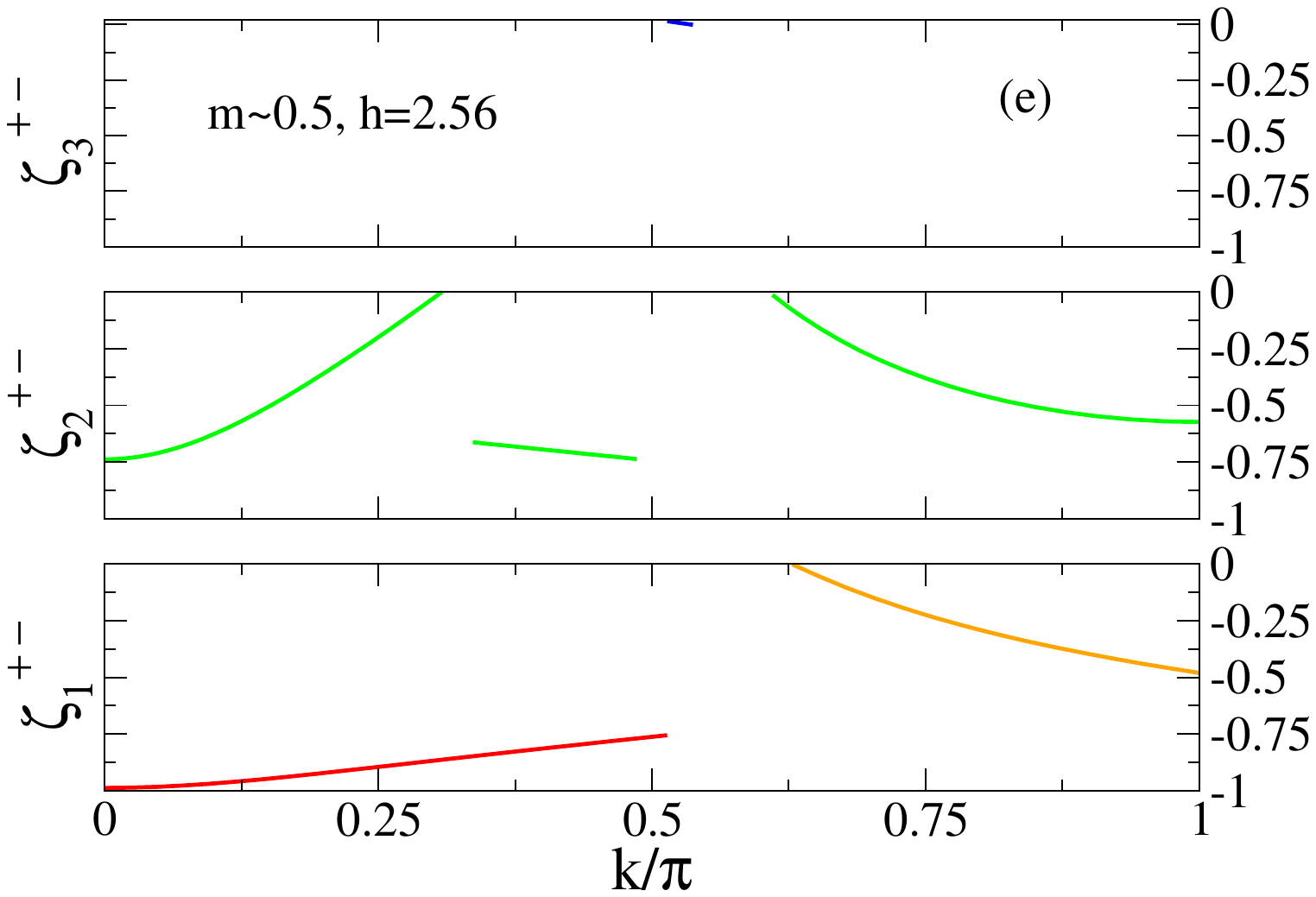}
\includegraphics[width=0.45\textwidth]{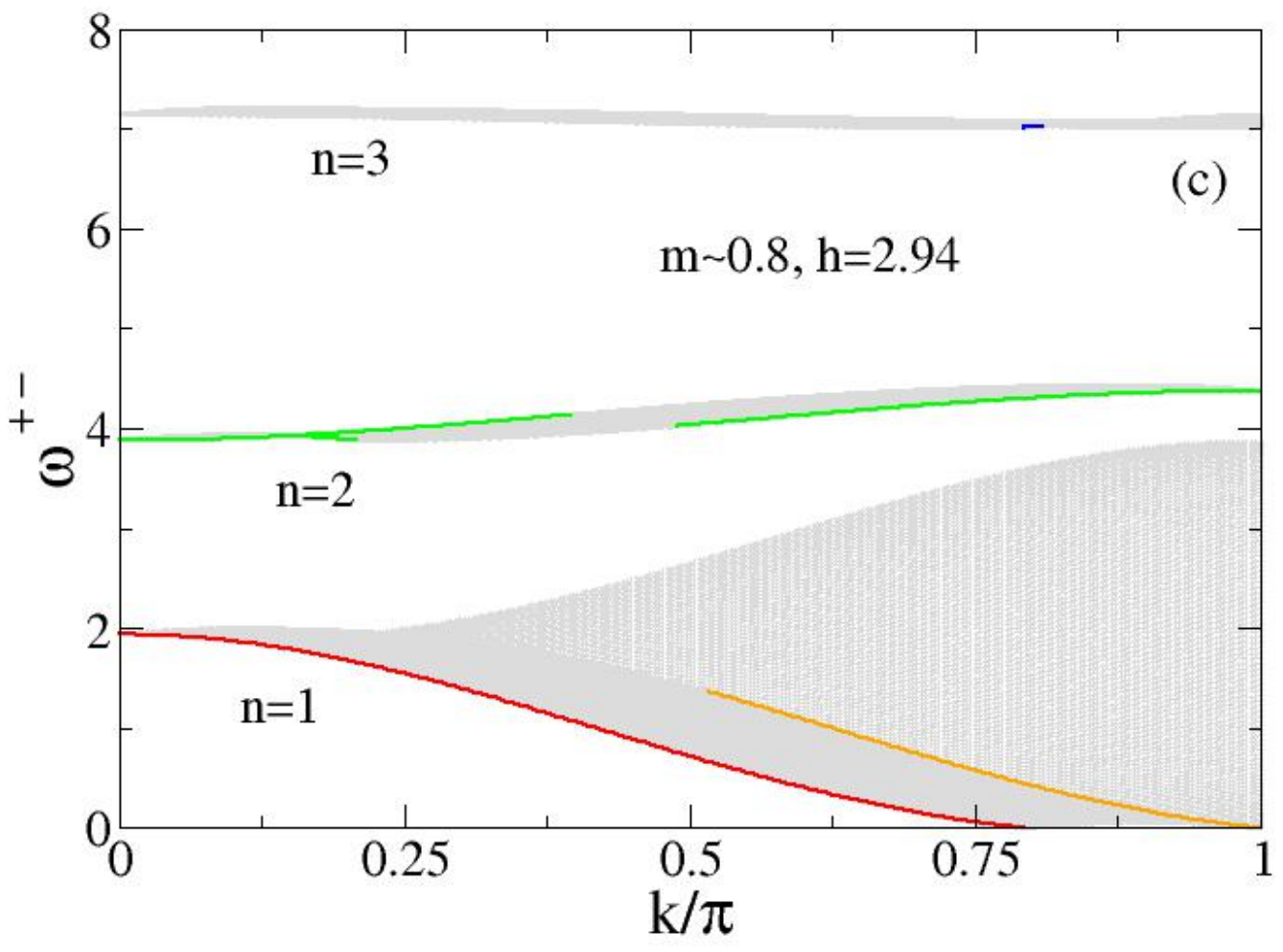}
\includegraphics[width=0.45\textwidth]{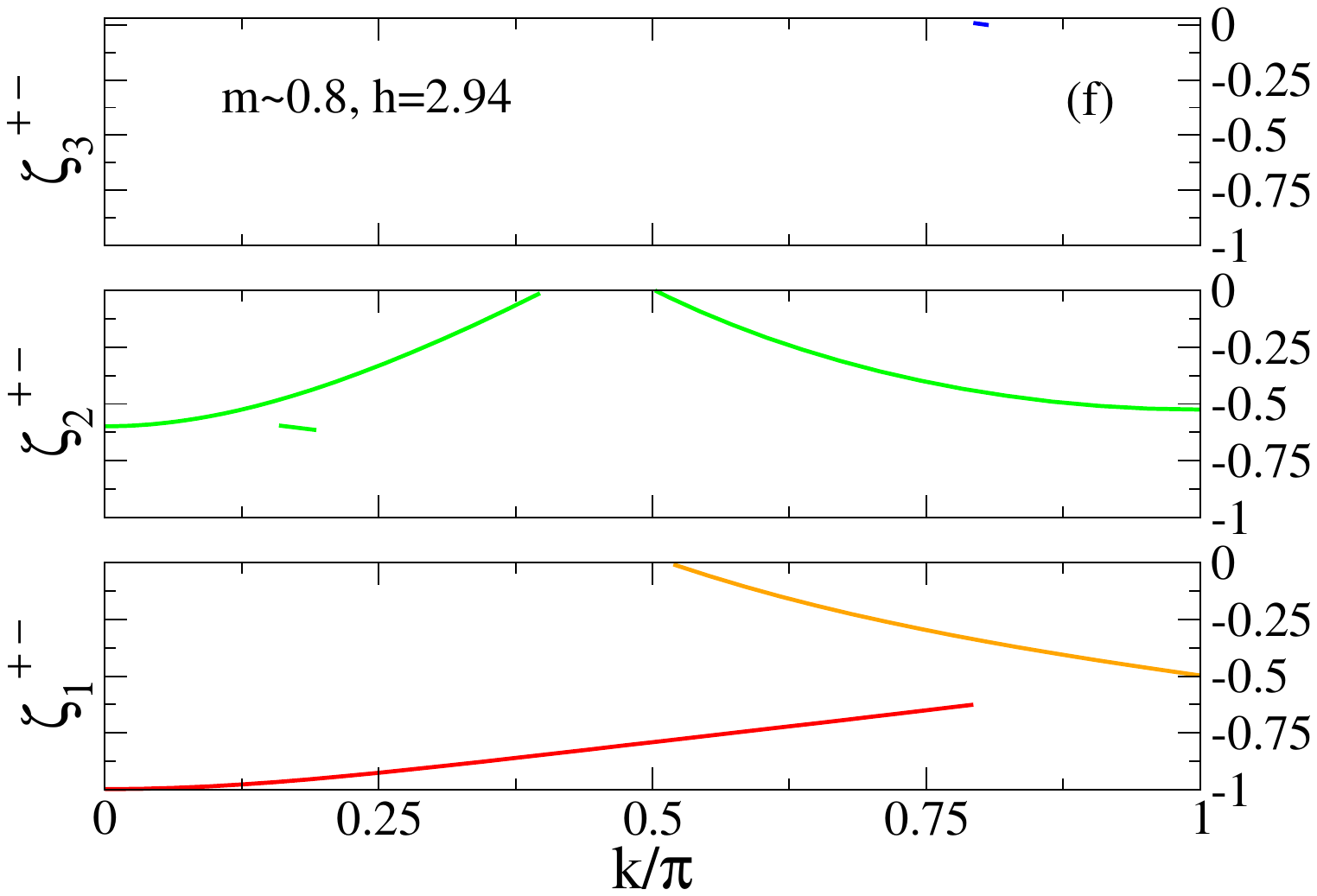}
\caption{\label{FigChaos4}
The $(k,\omega)$-plane (a)-(c) $n=1$, $n=2$, and $n=3$ $n$-continua where in the thermodynamic limit there is significant
spectral weight in $S^{+-} (k,\omega)$ for the spin-$1/2$ $XXZ$ chain in a longitudinal magnetic field with anisotropy $\Delta =2$. 
The corresponding negative $k$ dependent exponents that
control the line shape $S^{+-} (k,\omega)\propto (\omega - E^{+-}_n (k))^{\zeta^{+-}_n (k)}$ 
in the $k$ intervals near the lower thresholds of such continua (d)-(f). The spin densities 
in (a), (b), and (c) are $m=0.209\approx 0.2$, $m=0.514\approx 0.5$, and $m=0.793\approx 0.8$, respectively. The corresponding
$h$ values are given in units of $J/(g\mu_B)$. The exponents are negative
in the $k$ intervals of the $n$-continua lower thresholds marked in the spectra (a)-(c)
and near the branch line running through the $1$-continuum in (b) and (c). On the marked lines in the 
$(k,\omega)$-plane $S^{+-} (k,\omega)$ displays sharp peaks. From Ref. \onlinecite{Carmelo_23}.}
\end{figure*} 

As shown in Fig. 2 of Ref. \onlinecite{Okunishi_07} where the indices $(ic)$ and $(s)$ refer to longitudinal
$(z)$ and transverse $(x)$, respectively, one has that $T_c (h) = T_c^{z} (h)$ and  $T_c (h) = T_c^{x} (h)$
for two subdomains $h\in [h_{c1},h_*]$ and $h\in [h_*,h_{c2}]$, respectively, where the
magnetic field $h_*$ such that $T_c^{z} (h_*)=T_c^{x} (h_*)$
increases upon increasing the anisotropy $\Delta$. 

\subsection{The spin dynamical structure factor}
\label{SECIVA}

The spin dynamical structure factor components $S^{xx} (k,\omega)$, $S^{yy} (k,\omega)$. and $S^{zz} (k,\omega)$ 
are for the spin-$1/2$ $XXZ$ chain given by,
\begin{eqnarray}
S^{aa} (k,\omega) & = & \sum_{j=1}^N e^{-ik j}\int_{-\infty}^{\infty}dt\,e^{-i\omega t}\langle GS\vert\hat{S}^{a}_j (t)\hat{S}^{a}_j (0)\vert GS\rangle 
\nonumber \\
& = & \sum_{\nu}\vert\langle \nu\vert\hat{S}^{a}_k\vert GS\rangle\vert^2
\delta (\omega - \omega^{aa}_{\nu} (k)) 
\nonumber \\
&& {\rm for}\hspace{0.40cm}a =x,y,z \,  .
\label{SDSF}
\end{eqnarray}
Here the spectra read $\omega^{aa}_{\nu} (k) = (E_{\nu}^{aa} - E_{GS})$, $E_{\nu}^{aa}$ refers to
the energies of the excited energy eigenstates that contribute to the 
dynamical structure factors components $aa= xx,yy,zz$, $\sum_{\nu}$ is the sum over such states,
$E_{GS}$ is the initial ground state energy, and $\hat{S}^{a}_k$ are for $a =x,y,z$ the 
Fourier transforms of the usual local $a =x,y,z$ spin operators $\hat{S}^{a}_j$, respectively. 
\begin{figure*}
\includegraphics[width=0.45\textwidth]{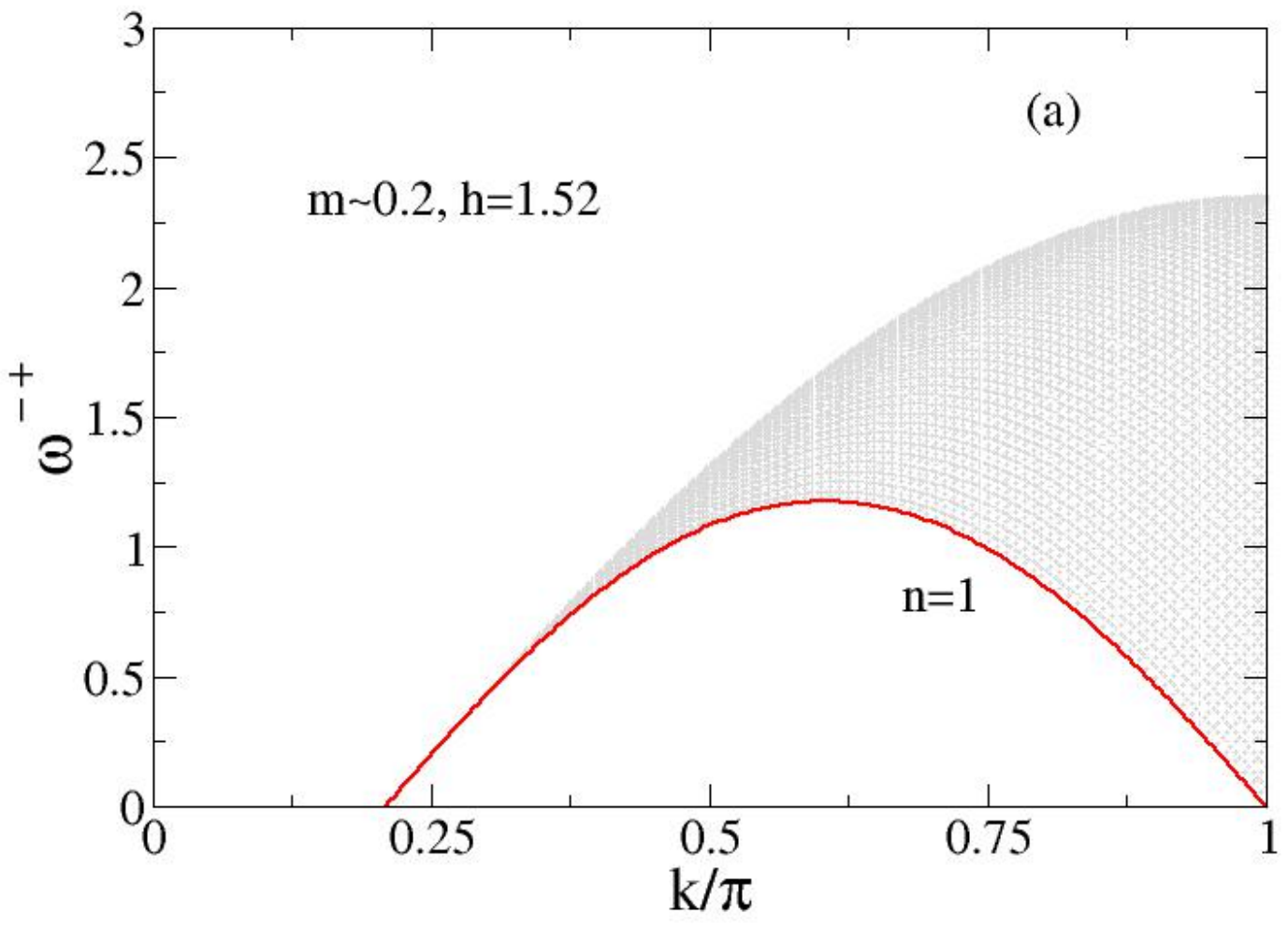}
\includegraphics[width=0.45\textwidth]{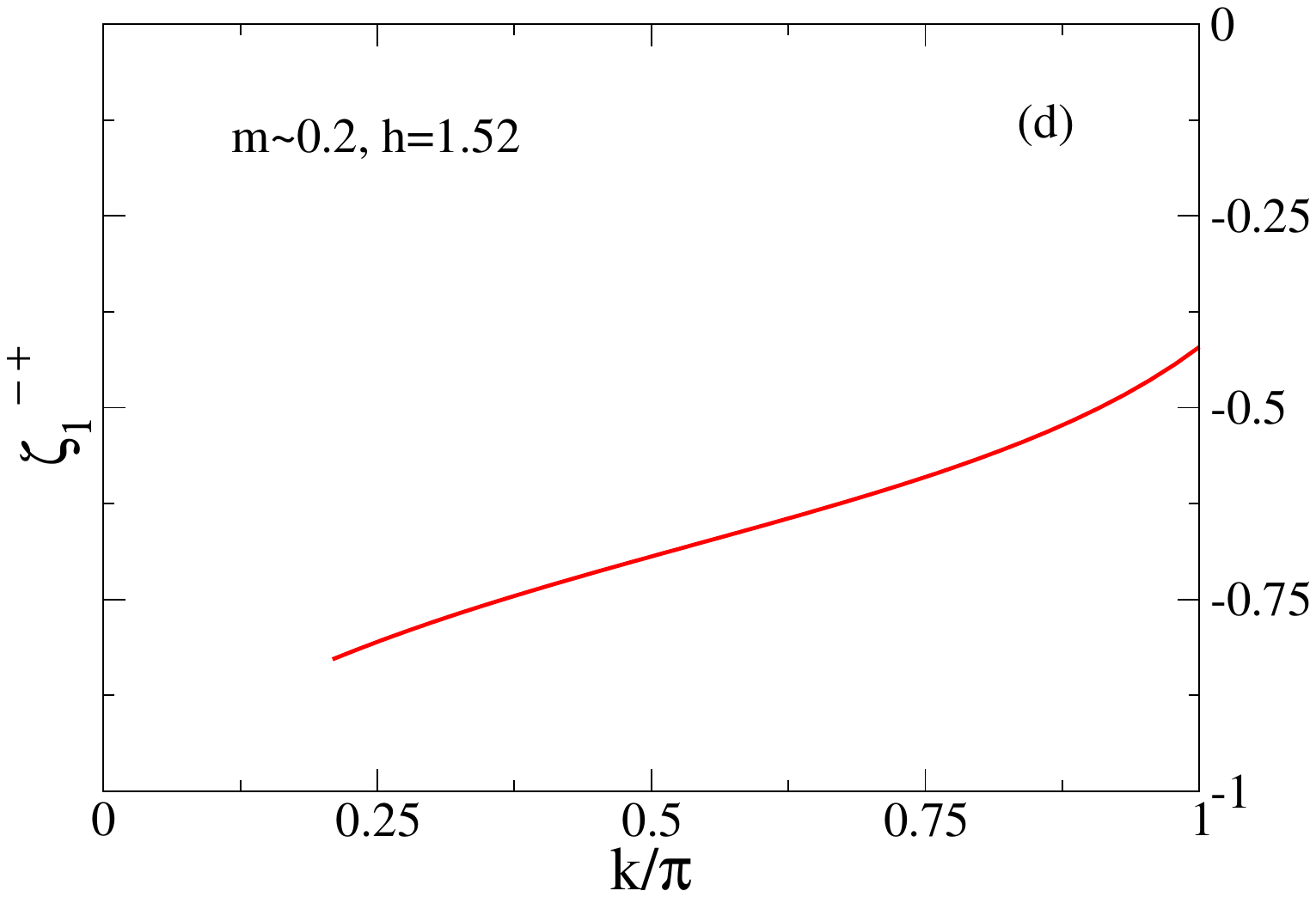}
\includegraphics[width=0.45\textwidth]{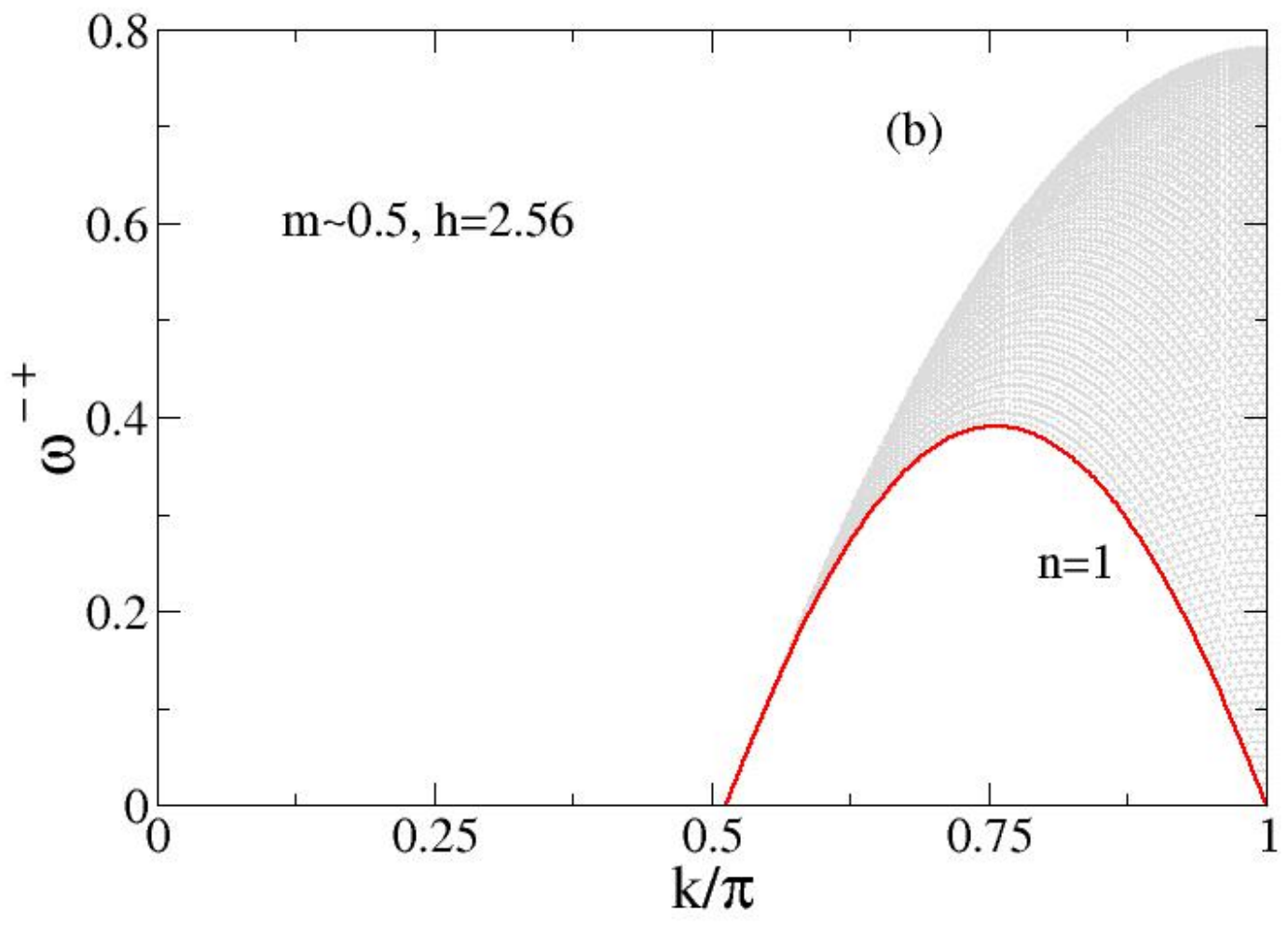}
\includegraphics[width=0.45\textwidth]{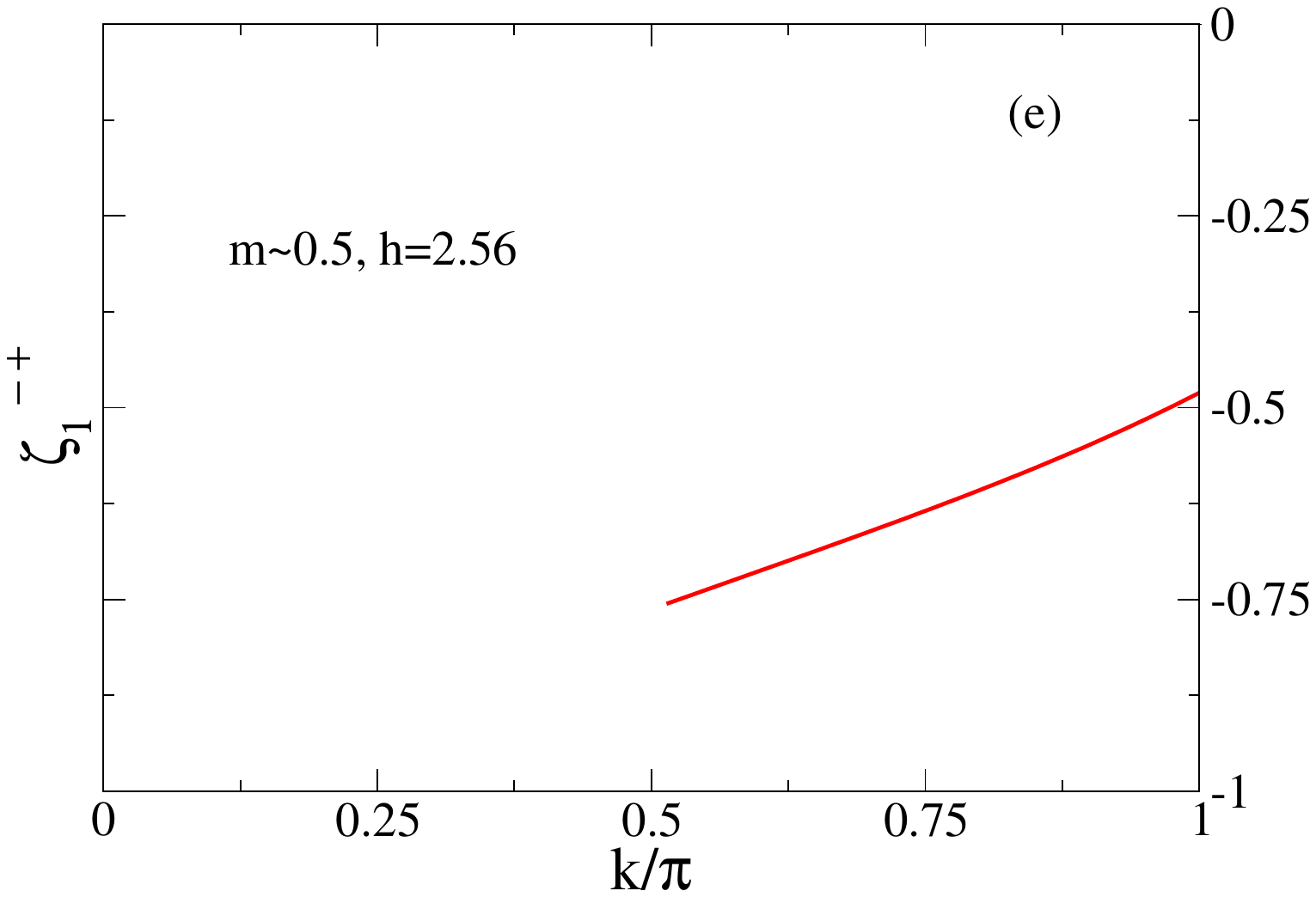}
\includegraphics[width=0.45\textwidth]{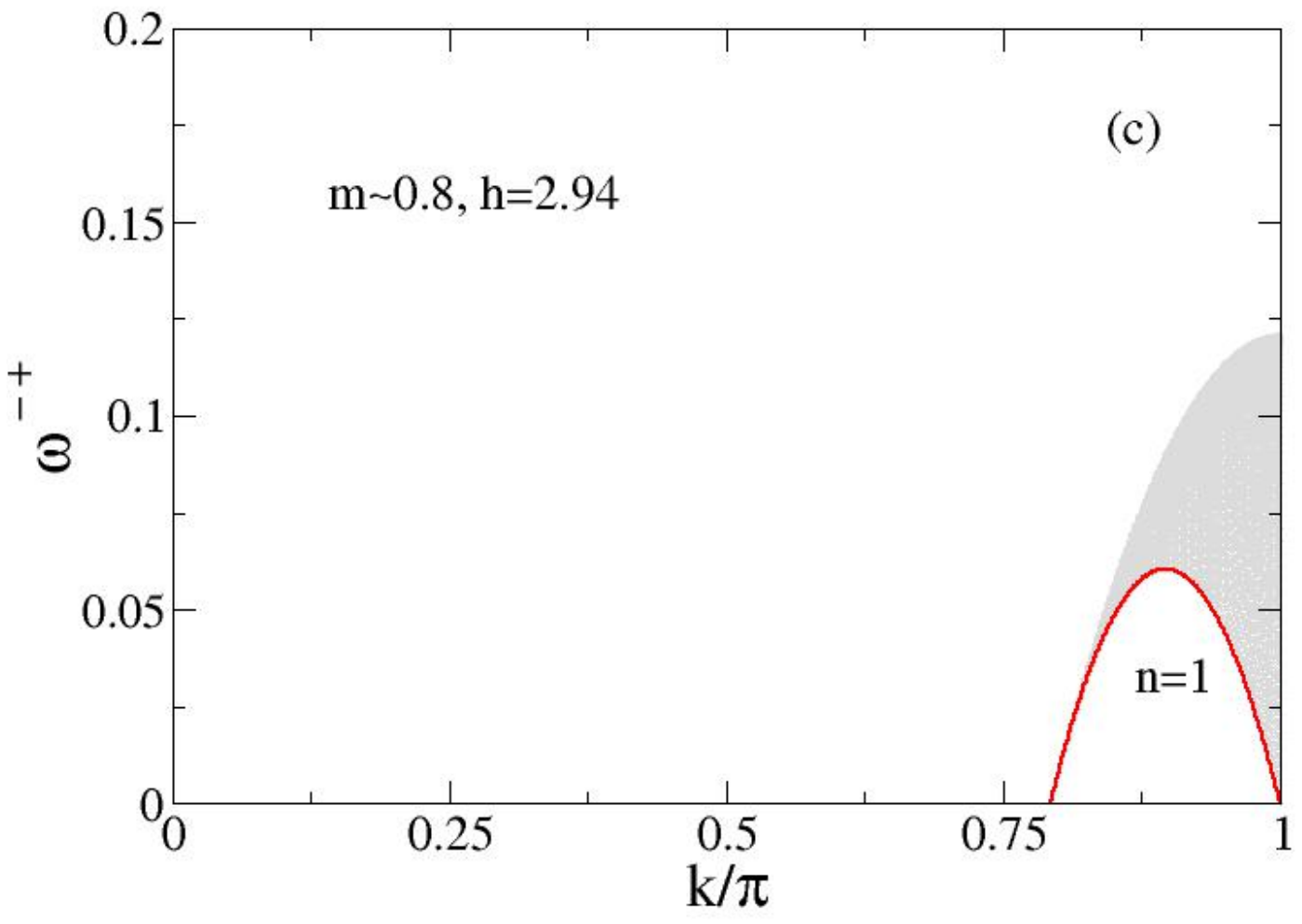}
\includegraphics[width=0.45\textwidth]{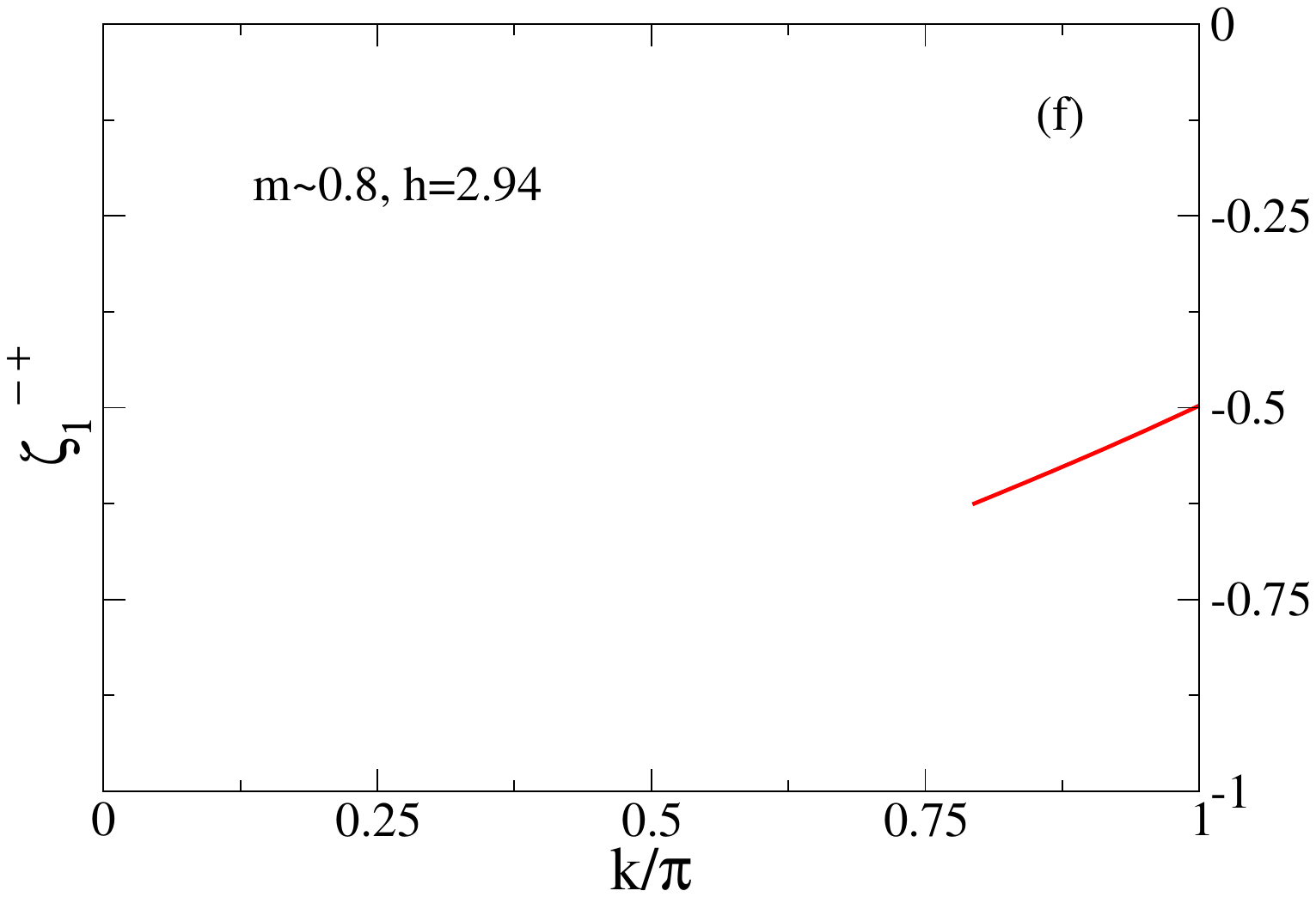}
\caption{\label{FigChaos5}
The $(k,\omega)$-plane $1$-continuum where in the thermodynamic limit there is significant spectral weight in 
$S^{-+} (k,\omega)$ for the spin-$1/2$ $XXZ$ chain in a longitudinal magnetic field with anisotropy $\Delta =2$ 
(a)-(c). The corresponding negative $k$-dependent exponent for its whole $k$ interval (d)-(f).
The spin densities in (a), (b), and (c) are the same as in Fig. \ref{FigChaos4}.
The corresponding $h$ values are given in units of $J/(g\mu_B)$. On this $1$-continuum lower threshold 
$S^{-+} (k,\omega)$ displays sharp peaks. From Ref. \onlinecite{Carmelo_23}.}
\end{figure*} 

The components $S^{+-} (k,\omega)$ and $S^{-+} (k,\omega)$ of the spin dynamical structure factor are directly related to 
the two transverse components $S^{xx} (k,\omega)$ and $S^{yy} (k,\omega)$, Eq. (\ref{SDSF})
for $aa=xx,yy$, which are identical,  $S^{xx} (k,\omega)=S^{yy} (k,\omega)$, as follows,
\begin{equation}
S^{xx} (k,\omega) = S^{yy} (k,\omega) = {1\over 4}\left(S^{+-} (k,\omega)+S^{-+} (k,\omega)\right) \, .
\label{xxPMMP}
\end{equation}

Concerning the spin dynamical structure factor components in Eqs. (\ref{SDSF}) and (\ref{xxPMMP}),
we are mainly interested in energies just above the $n$-continua lower-threshold spectra of the excited states.
The reason is that the sharp peaks studied below in Sec. \ref{SECIVB} 
are located at those lower thresholds. Such $(k,\omega)$-plane $n$-continua where $n=1$, $n=2$, and $n=3$ are classified
according to the corresponding excited states having no $n>1$ Bethe $n$-strings, one
single $2$-string, and one single $3$-string, respectively. Within the physical-spins representation,
this corresponds to such states not being populated by $n$-string-pairs, being populated by one
$2$-string-pair, and being populated by one $3$-string-pair, respectively.
The $2$-continuum and the $3$-continuum are gapped. 

The $n=1,2,3$ $n$-continua of the component $S^{+-} (k,\omega)$
are shown in Figs. \ref{FigChaos4} (a)-(c) and the $n=1$ $n$-continuum of
$S^{-+} (k,\omega)$ in Figs. \ref{FigChaos5} (a)-(c). 
They refer to the regions in the $(k,\omega)$-plane where there is significant spectral weight in 
the components $S^{+-} (k,\omega)$ and $S^{-+} (k,\omega)$, respectively.
(For the $n=1,2$ $n$-continua where there is significant spectral weight in the longitudinal 
component $S^{zz} (k,\omega)$, see Fig. 4 (a)-(c) of Ref. \onlinecite{Carmelo_23}.)

The panels (a),(b),(c) of Figs. \ref{FigChaos4} and \ref{FigChaos5} 
refer to anisotropy $\Delta =2$ and spin densities $m=0.209\approx 0.2$, $m=0.514\approx 0.5$, 
and $m=0.793\approx 0.8$, respectively. The field $h$ values corresponding to the above spin densities given in
these figures are in units of $J/(g\mu_B)$. In these units the critical fields,
Eq. (\ref{criticalfields}), and the magnetic field $h_{1/2}$ that refers to spin density $m=1/2$
read $h_{c1} = 0.39$, $h_{1/2} = 2.53$, and $h_{c2} = 3.00$ for the anisotropy $\Delta =2$
used for SrCo$_2$V$_2$O$_8$ and $h_{c1} = 0.52$, $h_{1/2} = 2.69$, and $h_{c2} = 3.17$ for 
the anisotropy $\Delta =2.17$ used for BaCo$_2$V$_2$O$_8$ \cite{Carmelo_23}.

At fixed excitation momentum $k$ and small values of the energy deviation
$(\omega - E^{ab}_{n} (k))\geq 0$, the dynamical theory used in Refs. \onlinecite{Carmelo_23} and
 \onlinecite{Carmelo_22} provides the following exact power-law form
for the spin dynamic structure factor $ab = +-,-+,zz$ components,
\begin{equation}
S^{ab} (k,\omega) = C_{ab}^n (k)
\left({\omega - E^{ab}_{n} (k)\over 4\pi\,B_1^{ab}\,v_1 (k_{F\downarrow})}\right)^{\zeta_{n}^{ab} (k)} \, .
\label{MPSs}
\end{equation}
Here $E^{ab}_{n} (k)$ denotes the $n$-continua lower-threshold spectra of the excited states.
The momentum $k$ dependent exponents $\zeta_{n}^{ab} (k)$ and pre-factor functions 
$C_{ab}^n (k)$ in Eq. (\ref{MPSs}) are given by \cite{Carmelo_23,Carmelo_22},
\begin{equation}
\zeta_{n}^{ab} (k) = - 1  + \sum_{\iota = \pm 1}\Phi_{\iota}^2 (k) \, ,
\label{zetaabk}
\end{equation}
and
\begin{eqnarray}
C_{ab}^n (k) & = & {1\over \vert\zeta_{n}^{ab} (k)\vert}
\nonumber \\
& \times & \prod_{\iota =\pm 1}
{e^{-f_0^{ab} + f_2^{ab}\left(2{\tilde{\Phi}}_{\iota}\right)^2 - f_4^{ab}\left(2{\tilde{\Phi}}_{\iota}\right)^4} \over \Gamma (\Phi_{\iota}^2 (k))} \, ,
 \label{Cabn}
\end{eqnarray}
respectively. Here ${\tilde{\Phi}}_{\iota}$ is the scattering part of the general functional $\Phi_{\iota}$
given by \cite{Carmelo_23,Carmelo_22},
\begin{eqnarray}
\Phi_{\iota} & = & \iota{\delta N_1^F\over 2} + \delta J_1^F + {\tilde{\Phi}}_{\iota} =
\iota\,\delta N_{1,\iota}^F + {\tilde{\Phi}}_{\iota} \hspace{0.20cm}{\rm where}
\nonumber \\
{\tilde{\Phi}}_{\iota} & = & {1\over 2\pi}\sum_{n=1}^{3}\,\sum_{j=1}^{L_{n}}\delta N_{n}(q_{j})\,2\pi\Phi_{1,n}(\iota k_{F\downarrow},q_{j}) \, ,
\label{functional}
\end{eqnarray}
the index $n=1,2,3$ refers the $(k,\omega)$-plane $n$-continua shown in 
Fig. \ref{FigChaos4} (a)-(c), and the $l=0,2,4$ coefficients $0<f_l^{ab}<1$ depend on $\eta$ 
and are different for each spin dynamic structure factor component.
(Fig. \ref{FigChaos5} (a)-(c) only displays the $1$-continuum
because there is almost no spectral-weight contributions to the component $S^{-+} (k,\omega)$
from states populated by $n>1$ $n$-string pairs.)

The dependence of the functional $\Phi_{\iota}$, Eq. (\ref{functional}), on the excitation momentum $k$ occurs through its direct
relation to the $n$-band momentum values $q_j$ in the phase shifts $2\pi\Phi_{1,n}(\iota k_{F\downarrow},q_{j})$ in the expression of 
${\tilde{\Phi}}_{\iota}$. The index $\iota =\pm 1$ in $\iota k_{F\downarrow}$ refers to the left $(\iota = -1)$ and right $(\iota = +1)$ 
$1$-band Fermi points and $\delta N_{1}^F = \sum_{\iota = \pm 1}\delta N_{1,\iota}^F$ and
$\delta J_{1}^F =  {1\over 2}\sum_{\iota = \pm 1}\iota\,\delta N_{1,\iota}^F$ are deviations under the ground-state - excited 
state transitions. Here $\delta N_{1,\iota}^F$ is the deviation in the number of $1$-pairs at and very near such $\iota = \pm 1$
$1$-band Fermi points.

The line shape just above the $(k,\omega)$-plane $n$-continua lower thresholds $k$ intervals 
for $n=1,2,3$ of $S^{+-} (k,\omega)$ and above the $(k,\omega)$-plane $1$-continuum lower threshold 
$k$ interval of $S^{-+} (k,\omega)$ where there are sharp peaks
is controlled by exponents $\zeta_{n}^{+-} (k)$ and $\zeta_{1}^{-+} (k)$, respectively,
whose general expression is given in Eqs. (\ref{zetaabk}) and (\ref{functional}).
They are negative in the lower thresholds $k$ intervals marked in Figs. \ref{FigChaos4} (a)-(c) and
\ref{FigChaos5} (a)-(c). 

The $k$ dependence in the intervals where they are negative is shown 
in Figs. \ref{FigChaos4} (d)-(f) and \ref{FigChaos5} (d)-(f)
for the components $S^{+-} (k,\omega)$ and $S^{-+} (k,\omega)$, respectively. 
(For the $k$ dependence in the intervals where the $n=1,2$ exponents $\zeta_{n}^{zz} (k)$
of the longitudinal component $S^{zz} (k,\omega)$ are negative for
the spin-$1/2$ $XXZ$ chain in a longitudinal magnetic field, see Fig. 4 (d)-(f) of 
Ref. \onlinecite{Carmelo_23}.)

In the following we review results on the description of sharp peaks in the dynamical structure factor's
transverse components $S^{\pm\mp} (k,\omega)$ of the spin-$1/2$ $XXZ$ chain in a longitudinal magnetic 
field. This includes those that were experimentally observed in the quasi-1D material SrCo$_2$V$_2$O$_8$ \cite{Wang_18,Bera_20},
which are shown to be described by that spin-chain model with the parameter sets $\Delta =2.00$, 
$J=3.55$ meV, and $g =6.2$ suitable to that material \cite{Carmelo_23}. 

For a similar description of the sharp peaks in the longitudinal component $S^{z} (k,\omega)$ experimentally 
observed in that material and those in $S^{\pm} (k,\omega)$ experimentally observed in 
BaCo$_2$V$_2$O$_8$ \cite{Wang_19}, see Ref. \onlinecite{Carmelo_23}.

\subsection{Selected sharp peaks at fixed momentum values $k=0,\pi/2,\pi$ in the $(h,\omega)$-plane}
\label{SECIVB}

Besides momentum dependencies, our study includes extracting the longitudinal
magnetic field $h$ dependencies in the thermodynamic limit of the negative exponents that control the line shape at and near the 
sharp peaks in $S^{+-} (k,\omega)$ and $S^{-+} (k,\omega)$
at the momentum values $k=0$, $k=\pi/2$, and $k=\pi$ 
at which they were experimentally observed in SrCo$_2$V$_2$O$_8$ \cite{Wang_18,Bera_20}. 

The momentum values $k=0$, $k=\pi/2$, and $k=\pi$ 
of the sharp peaks observed experimentally 
belong to the marked $k$ intervals of the $n$-continua lower thresholds 
shown in Figs. \ref{FigChaos4} (a)-(c) and \ref{FigChaos5} (a)-(c).
When at such momentum values the corresponding lower threshold is not marked,
the exponent is not negative and there is no sharp peak.
\begin{figure}
\begin{center}
\subfigure{\includegraphics[width=8.50cm]{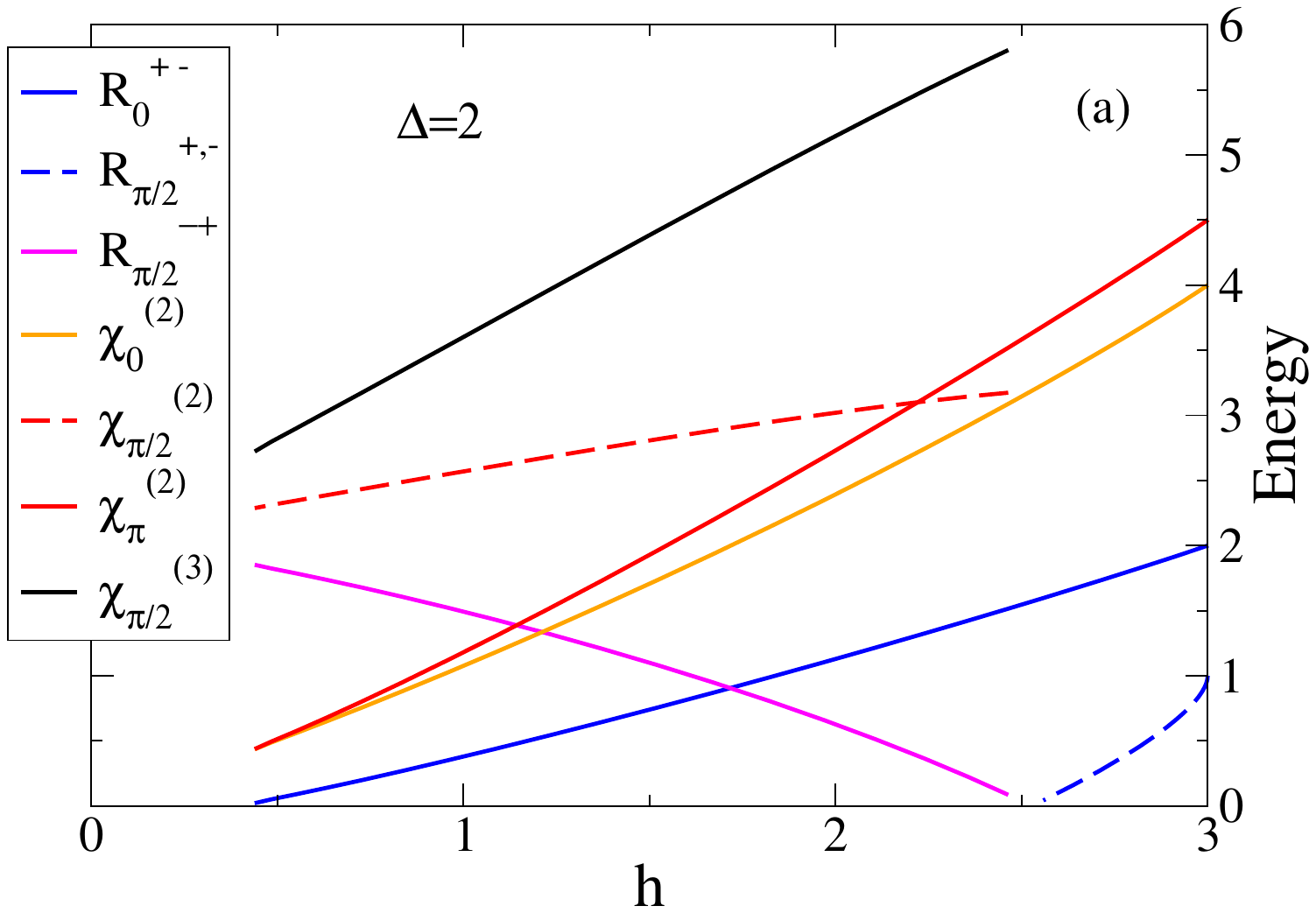}}
\hspace{0.50cm}
\subfigure{\includegraphics[width=8.50cm]{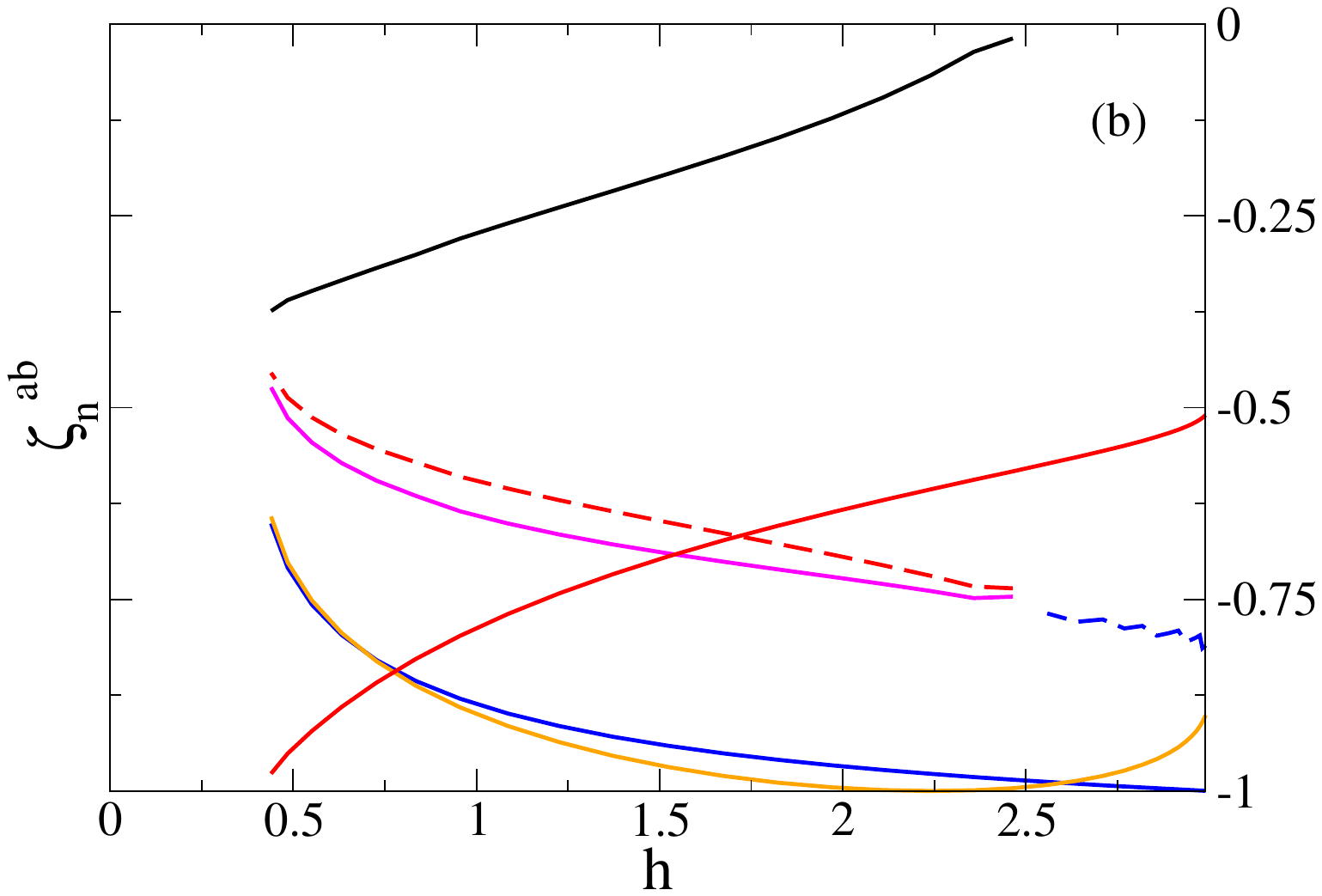}}
\caption{The energies in units of $J$ of the sharp peaks $R^{+-}_{0}$, $R^{+-}_{\pi/2}$, $R^{-+}_{\pi/2}$, $\chi^{(2)}_0$,
$\chi^{(2)}_{\pi/2}$, $\chi^{(2)}_{\pi}$, and  $\chi^{(3)}_{\pi/2}$ in the transverse components $S^{+-} (k,\omega)$ and $S^{-+} (k,\omega)$
versus the magnetic field $h$ for $h\in [h_{c1},h_{c2}]$ in units of $J/(g\mu_B)$ 
for the spin-$1/2$ $XXZ$ chain in a longitudinal magnetic field with $\Delta =2.00$;
The corresponding magnetic field $h$ dependencies of the negative exponents 
that control the line shape near such sharp peaks (b). 
The expressions of these energies and exponents are given in Eqs. 
(\ref{EPM0})-(\ref{EPM3PI2}). From Ref. \onlinecite{Carmelo_23}.}
\label{FigChaos6}
\end{center}
\end{figure}

The following thermodynamic-limit results are for the spin-$1/2$ $XXZ$ chain in 
a longitudinal field $h_{c1}<h<h_{c2}$ with anisotropy $\Delta =2$ 
representative of the 1D physics of SrCo$_2$V$_2$O$_8$.
At and near the sharp peaks denoted by $R^{+-}_{0}$, $R^{+-}_{\pi/2}$, $R^{-+}_{\pi/2}$, $R^{zz}_{\pi}$, 
$\chi^{(2)}_0$, $\chi^{(2)}_{\pi/2}$, $\chi^{(2)}_{\pi}$, and 
$\chi^{(3)}_{\pi/2}$ in Ref. \onlinecite{Carmelo_23}, the dynamical theory
used in the studies of Refs. \onlinecite{Carmelo_23} and \onlinecite{Carmelo_22}
gives for small values of the energy deviation $(\omega - E^{ab}_{n} (k,h))\geq 0$
from the $ab=-+,+-$ $n$-continuum lower-threshold energy $E^{ab}_{n} (k,h)$
at momentum $k$ and field $h$ a line shape of power-law form \cite{Carmelo_23,Carmelo_22},
\begin{eqnarray}
R_{k}^{ab} & = & S^{ab} (k,\omega) = \bar{C}^{ab}_1 (k) \,
\Bigl(\omega - E^{ab}_{1} (k,h)\Bigr)^{\zeta_{1}^{ab} (k,h)}  
\nonumber \\
\chi^{(n)}_k & = & S^{+-} (k,\omega) = \bar{C}^{+-}_n (k)
\Bigl(\omega - E^{+-}_{n} (k,h)\Bigr)^{\zeta_{n}^{+-} (k,h)} 
\nonumber \\
&& {\rm where}
\nonumber \\
\bar{C}^{ab}_ n (k) & = & {C_{ab}^n (k)\over (4\pi\,B_1^{ab}\,v_1 (k_{F\downarrow}))^{\zeta_{n}^{ab} (k,h)}} 
\hspace{0.20cm}{\rm for}\hspace{0.20cm}n = 1,2,3  \, .
\label{Rkab}
\end{eqnarray}
Here the momentum $k$ dependent exponents $\zeta_{n}^{ab} (k)$ and pre-factor functions 
$C_{ab}^n (k)$ are given in Eqs. (\ref{zetaabk}) and (\ref{Cabn}), respectively, and the
remaining quantities are those also appearing in Eq. (\ref{MPSs}).

The sharp-peak line shapes given in Eq. (\ref{Rkab}) refer to zero temperature. Hence
we expect that the sharp modes observed in experiments 
\cite{Wang_18,Bera_20} at low temperatures just above the very
small critical transition temperature $T_c (h)$, Eq. (\ref{TcTc}),
to be a bit smeared by thermal fluctuations and coupling to phonons. 
\begin{figure}
\begin{center}
\subfigure{\includegraphics[width=8.50cm]{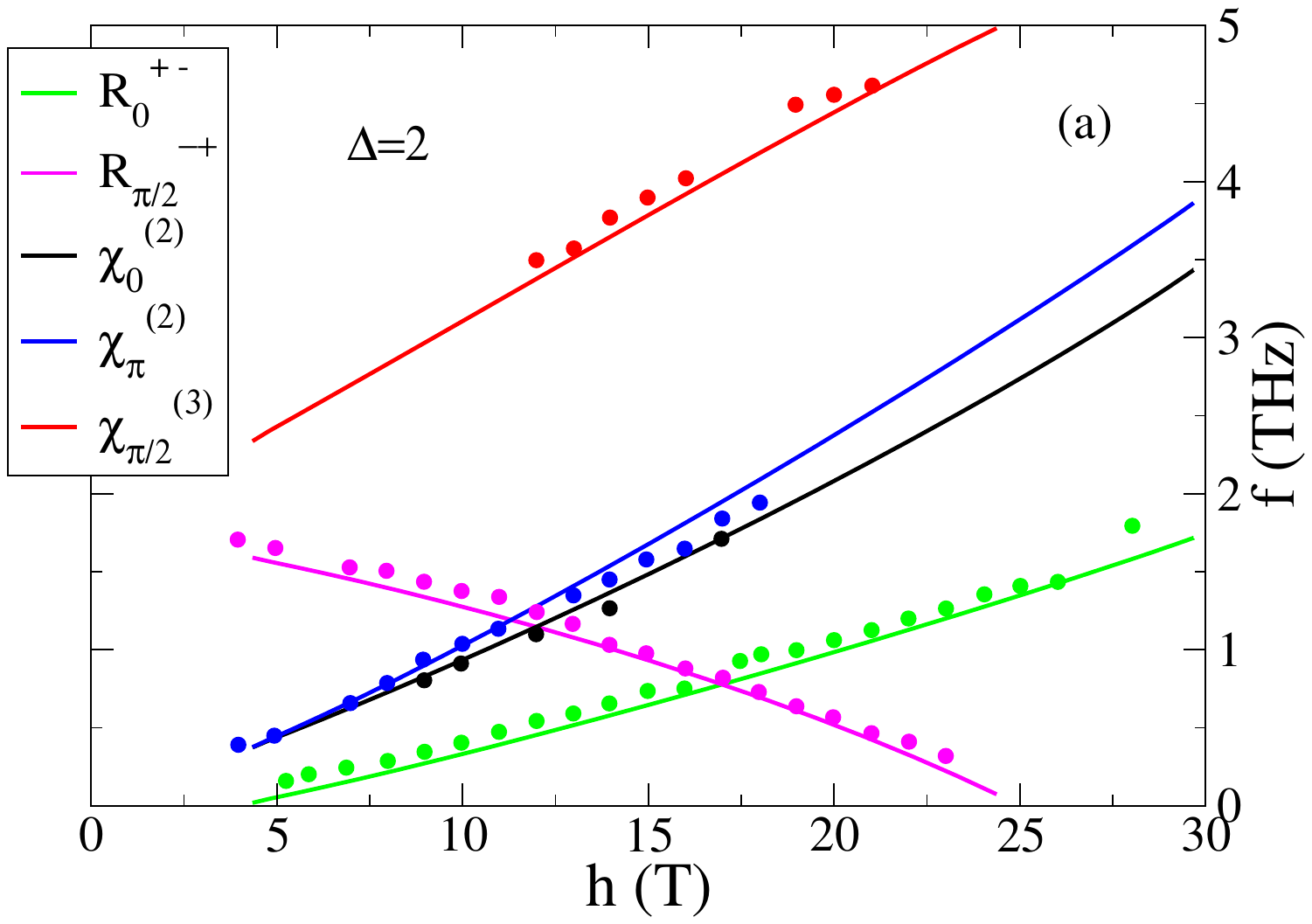}}
\hspace{0.50cm}
\subfigure{\includegraphics[width=8.50cm]{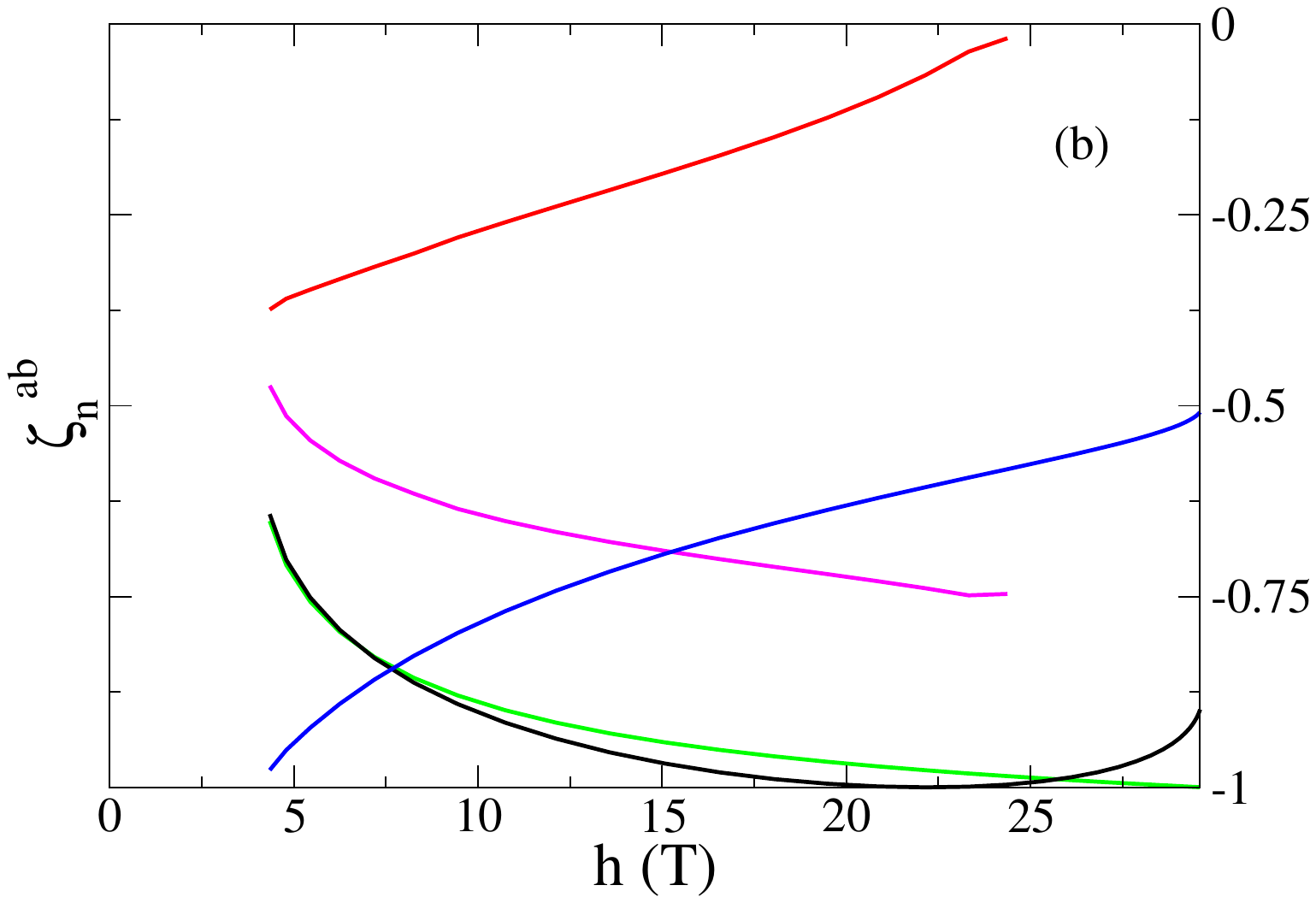}}
\caption{The dependencies on the magnetic field $h\in [h_{c1},h_{c2}]$ 
in Tesla of the frequencies in units of THz associated with the 
energies of the transverse sharp peaks $R_{0}^{+-}$, $R_{\pi/2}^{-+}$, 
$\chi^{(2)}_{0}$, $\chi^{(2)}_{\pi}$, and $\chi^{(3)}_{\pi/2}$ experimentally observed in SrCo$_2$V$_2$O$_8$
for the spin-$1/2$ $XXZ$ chain in a longitudinal magnetic field with
$\Delta =2.00$, $J=3.55$ meV, and $g =6.2$ (a); The corresponding negative exponents 
for the line shape, Eq. (\ref{Rkab}), obtained for that spin-chain model (b). 
Expressions both of the energies corresponding to these frequencies and of the latter 
exponents are given in Eqs. (\ref{EPM0}), (\ref{EMPPI2}), (\ref{EPM20}), (\ref{EPM2pi}), and 
(\ref{EPM3PI2}). Such theoretical frequency dependencies on $h\in [h_{c1},h_{c2}]$ 
are to be compared with those of the corresponding sharp peaks points experimentally 
observed in SrCo$_2$V$_2$O$_8$ also shown in (a), which are
those displayed in Fig. 4 of Ref. \onlinecite{Wang_18} with $h_{c1}=B_c$ and $h_{c2}=B_s$. From Ref. \onlinecite{Carmelo_23}.}
\label{FigChaos7}
\end{center}
\end{figure}

The expressions of the $n=1,2,3$ lower threshold energies $E^{+-}_{n} (k,h)$ and $n=1$ lower threshold energy $E^{-+}_{1} (k,h)$
given in the following involve the $n$-pair energy dispersions $\varepsilon_{n} (q)$, Eqs. (\ref{equA4n}) and (\ref{equA4n10}),
that are plotted in units of $J$ in Figs. \ref{FigChaos1}, \ref{FigChaos2}, and \ref{FigChaos3} for $n=1$, $n=2$, and
$n=3$, respectively. The exponents $\zeta_{n}^{ab} (k,h)$, Eqs. (\ref{zetaabk}) and (\ref{functional}), whose expressions
are also given in the following involve the phase shifts $2\pi\Phi_{1\,n}(\iota\,k_{F\downarrow},q) = 
2\pi\bar{\Phi }_{1\,n} \left(\iota\,B,\varphi_n (q)\right)$ for $n=1,2,3$, the
parameter $\xi$ defined in Eq. (\ref{x-aaPM}), and the also phase-shift related parameter,
\begin{eqnarray}
\xi_{1\,n}^0 = {1\over\pi}\times 2\pi\Phi_{1,n}\left(k_{F\downarrow},0\right)\hspace{0.20cm}{\rm for}\hspace{0.20cm}n=2,3  \, .
\label{x1n}
\end{eqnarray}

Such lower threshold energies and momentum-dependent exponents appearing in the expressions, Eq. (\ref{Rkab}), of the line shape 
at and near the sharp peaks $R^{+-}_{0}$, $R^{+-}_{\pi/2}$, $R^{-+}_{\pi/2}$, $R^{zz}_{\pi}$, 
$\chi^{(2)}_0$, $\chi^{(2)}_{\pi/2}$, $\chi^{(2)}_{\pi}$, and $\chi^{(3)}_{\pi/2}$ are for anisotropy $\Delta =2$ given by,
\begin{eqnarray}
E^{+-}_1 (0,h) & = & \varepsilon_1 (k_{F\uparrow})\in [0,2J]
\nonumber \\
\zeta^{+-}_1 (0,h) & = & -1 + \sum_{\iota =\pm 1}\Bigl(- {\xi\over 2} 
+ \Phi_{1,1}(\iota k_{F\downarrow},-k_{F\uparrow})\Bigr)^2
\nonumber \\
&& {\rm for}\hspace{0.20cm}h \in [h_{c1},h_{c2}] \, ,
\label{EPM0}
\end{eqnarray}
for $R^{+-}_{0}$, 
\begin{eqnarray}
E^{-+}_1 (\pi/2,h) & = & - \varepsilon_1 \Bigl({(k_{F\uparrow}-k_{F\downarrow})\over 2}\Bigr) \in [0,1.876 J]
\nonumber \\
\zeta^{-+}_1 (\pi/2,h) & = & -1 
\nonumber \\
& + & \sum_{\iota =\pm 1}\Bigl(- {\xi\over 2} 
- \Phi_{1,1}\Bigl(\iota k_{F\downarrow},{(k_{F\uparrow}-k_{F\downarrow})\over 2}\Bigr)\Bigr)^2 
\nonumber \\
&& {\rm for}\hspace{0.20cm}h \in [h_{c1},h_{1/2}] \, ,
\label{EMPPI2}
\end{eqnarray}
for $R^{+-}_{\pi/2}$, 
\begin{eqnarray}
E^{+-}_1 (\pi/2,h) & = & \varepsilon_1 \Bigl({(k_{F\uparrow}-k_{F\downarrow})\over 2}\Bigr)\in [0,J]
\nonumber \\
\zeta^{+-}_1 (\pi/2,h) & = & -1\nonumber \\
& + & \sum_{\iota =\pm 1} \Bigl(-{\xi\over 2} 
+ \Phi_{1,1}\Bigl(\iota k_{F\downarrow},- {(k_{F\uparrow}-k_{F\downarrow})\over 2}\Bigr)\Bigr)^2 
\nonumber \\
&& {\rm for}\hspace{0.20cm}h \in [h_{1/2},h_{c2}] \, ,
\label{EPMPI2}
\end{eqnarray}
for $R^{-+}_{\pi/2}$, 
\begin{eqnarray}
E^{zz}_1 (\pi,h) & = & \varepsilon_1 (k_{F\uparrow})\in [0,2J]
\nonumber \\
\zeta_{1}^{zz} (\pi,h) & = & -1 
\nonumber \\
& + & \sum_{\iota =\pm 1}\left(- {\iota\over 2\xi_{1\,1}} + {\xi_{1\,1}\over 2} 
+ \Phi_{1,1}(\iota k_{F\downarrow},k_{F\uparrow})\right)^2 
\nonumber \\
&& {\rm for}\hspace{0.20cm}h \in [h_{c1},h_{c2}] \, ,
\label{EZZPI}
\end{eqnarray}
for $R^{zz}_{\pi}$, 
\begin{eqnarray}
E^{+-}_{2} (0,h) & = & \varepsilon_2 (0)\in [0.389 J,4J] 
\nonumber \\
\zeta^{+-}_{2} (0,h) & = & -1 + {1\over 2}\left({1\over 2\xi} - \xi_{1\,2}^0\right)^2 
\nonumber \\
&& {\rm for}\hspace{0.20cm}h \in [h_{c1},h_{c2}] \, ,
\label{EPM20}
\end{eqnarray}
for $\chi^{(2)}_0$, 
\begin{eqnarray}
&& E^{+-}_{2} (\pi/2,h) = \varepsilon_ 2 (0) - \varepsilon_1 \Bigl({(k_{F\uparrow}-k_{F\downarrow})\over 2}\Bigr)
\nonumber \\
&& \hspace{2cm}\in [2.265 J,3.190 J]
\nonumber \\
&& \zeta^{+-}_{2} (\pi/2,h) = -1 
\nonumber \\
&& + \sum_{\iota=\pm 1} \Bigl(\iota {\xi_{1\,2}^0\over 2} + {\xi\over 2}
- \Phi_{1,1}\Bigl(\iota k_{F\downarrow},- {(k_{F\uparrow}-k_{F\downarrow})\over 2}\Bigr)\Bigr)^2 
\nonumber \\
&& \hspace{2cm}{\rm for}\hspace{0.20cm}h \in [h_{c1},h_{1/2}] \, ,
\label{EPM2PI2}
\end{eqnarray}
for $\chi^{(2)}_{\pi/2}$, 
\begin{eqnarray}
E^{+-}_{2} (\pi,h) & = & \varepsilon_ 2 (k_{F\uparrow}-k_{F\downarrow})\in [0.389 J,4.5J]
\nonumber \\
\zeta_{2}^{+-} (\pi) & = & -1 
\nonumber \\
& + & \sum_{\iota=\pm 1}
\Bigl(- {\iota\over 2\xi} + \xi 
+ \Phi_{1,2}(\iota k_{F\downarrow},k_{F\uparrow}-k_{F\downarrow})\Bigr)^2 
\nonumber \\
&& {\rm for}\hspace{0.20cm}h \in [h_{c1},h_{c2}] \, ,
\label{EPM2pi}
\end{eqnarray}
for $\chi^{(2)}_{\pi}$ and,
\begin{eqnarray}
&& E^{+-}_{3} (\pi/2,h) = \varepsilon_ 3 (0) - \varepsilon_1 \Bigl({(k_{F\uparrow}-k_{F\downarrow})\over 2}\Bigr)
\nonumber \\
&& \hspace{2cm}\in [2.654 J,5.891 J]
\nonumber \\
&& \zeta^{+-}_{3} (\pi/2,h) = -1 
\nonumber \\
&& + \sum_{\iota=\pm 1}\Bigl(- {\iota\over 2\xi} + \iota {\xi_{1\,3}^0\over 2} + {\xi\over 2}
- \Phi_{1,1}\Bigl(\iota k_{F\downarrow},{(k_{F\uparrow}-k_{F\downarrow})\over 2}\Bigr)\Bigr)^2 
\nonumber \\
&& \hspace{2cm}{\rm for}\hspace{0.20cm}h \in [h_{c1},h_{1/2}] \, ,
\label{EPM3PI2}
\end{eqnarray}
for $\chi^{(3)}_{\pi/2}$.

As given in Eqs. (\ref{EPM0})-(\ref{EPM3PI2}), depending on which specific sharp peaks, they occur for four
ranges of magnetic fields: $h\in [h_{c1},h_{c2}]$, $h\in [h_{c1},h_{1/2}]$, and $h\in [h_{1/2},h_{c2}]$.
The theoretical dependencies on the magnetic field $h$ in units of $J/(g\mu_B)$
of the energies in units of $J$ and of the corresponding exponents given in Eqs. (\ref{EPM0})-(\ref{EPM3PI2}) 
of such transverse sharp peaks are plotted in Figs. \ref{FigChaos6} (a) and (b), respectively, for anisotropy $\Delta =2$.
The specific energy lines $h$ ranges in these figures are those given above for which in the thermodynamic limit the 
corresponding exponents are negative. Only for such ranges there are sharp peaks.

Our thermodynamic-limit theoretical dependencies on the magnetic 
field $h$ in Tesla for the ranges of the frequencies in THz corresponding 
to the lower-threshold energies given in Eqs. (\ref{EPM0})-(\ref{EPM3PI2})  
of the subset of sharp peaks $R^{+-}_{0}$, $R^{-+}_{\pi/2}$, $\chi^{(2)}_0$, $\chi^{(2)}_{\pi}$, and 
$\chi^{(3)}_{\pi/2}$ experimentally observed in SrCo$_2$V$_2$O$_8$ by optical experiments 
are plotted in Fig. \ref{FigChaos7} (a). 
The corresponding experimental points in the $(h,\omega)$ plane
that describe the $h$ dependencies of the frequencies displayed in Fig. 4 of Ref. \onlinecite{Wang_18} for SrCo$_2$V$_2$O$_8$
are also shown in that figure. 

The negative exponents that control the line shape near such peaks whose expressions are given 
in Eqs. (\ref{EPM0})-(\ref{EPM3PI2}) are plotted as a function of the magnetic field $h$ in Fig. \ref{FigChaos7} (b). 
Such exponents general expression, Eqs. (\ref{zetaabk}) and (\ref{functional}), was obtained in Ref. \onlinecite{Carmelo_23}.

Comparison with the experimental dependence on $h\in [h_{c1},h_{c2}]$ of the frequencies of the
sharp peaks displayed in Fig. 4 of Ref. \onlinecite{Wang_18} for
SrCo$_2$V$_2$O$_8$ with those plotted in Fig. \ref{FigChaos7} (a) for
the spin-$1/2$ chain with $\Delta =2$ and $J=3.55$\,meV confirms the excellent quantitative agreement.
Our theoretical results refer to the thermodynamic limit. They agree with the
finite-size results used in Ref. \onlinecite{Wang_18} to describe the same experimental data.
\begin{figure}
\begin{center}
\subfigure{\includegraphics[width=4.25cm]{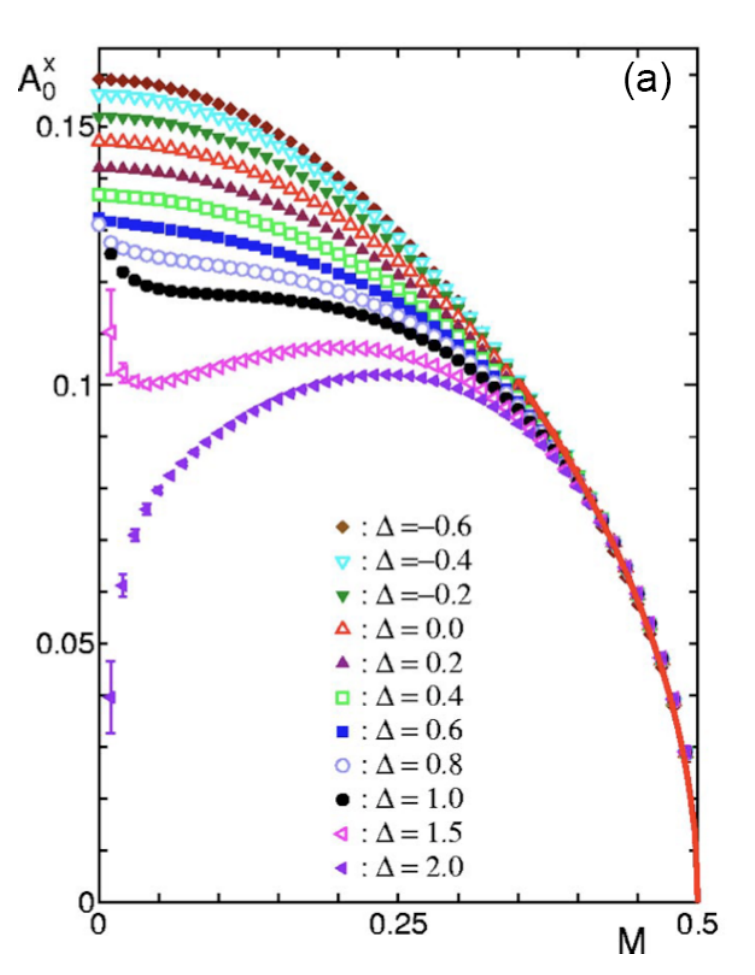}}
\subfigure{\includegraphics[width=4.25cm]{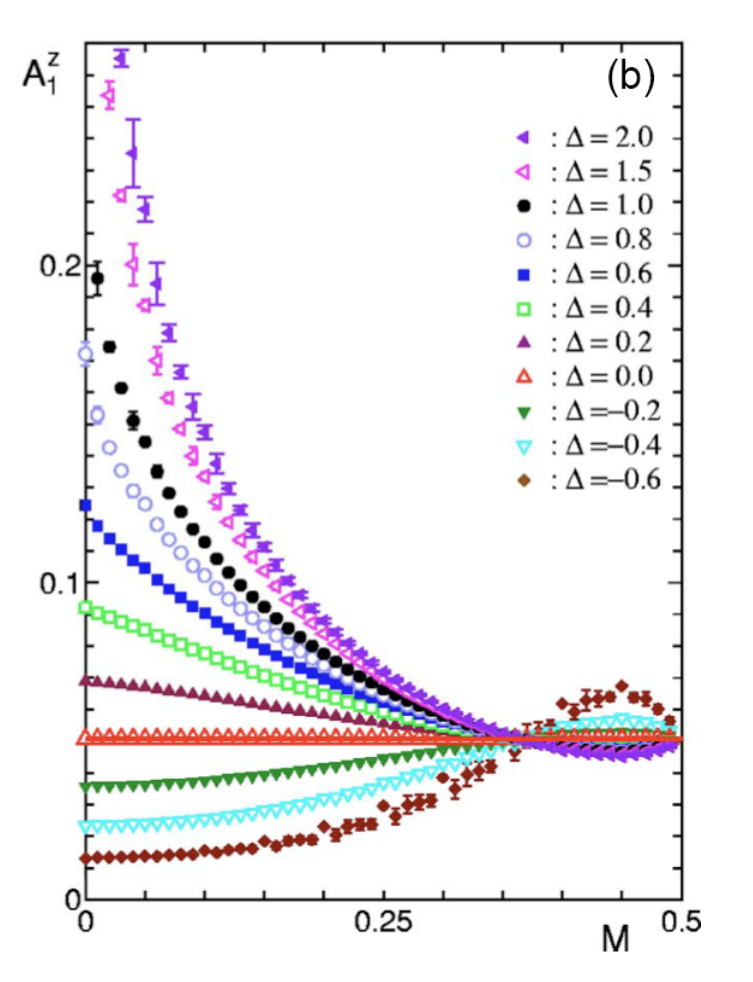}}
\caption{The dependence of the pre-factors (a) $A_x^0$ and (b) $A_z^1$ of the static spin correlation functions 
appearing in Eqs. (\ref{TcTc}) and (\ref{T1-1limit-1D})
on $M = m^z = m/2$ for spin density $m\in [0,1]$ at different anisotropy 
values $\Delta$ for the spin-$1/2$ $XXZ$ chain in a longitudinal magnetic field
as numerically computed in Ref. \onlinecite{Hikihara_04}.
The lines of importance for BaCo$_2$V$_2$O$_8$ and SrCo$_2$V$_2$O$_8$ 
refer to anisotropy $\Delta = 2$.
From Refs. \onlinecite{Carmelo_23} and \onlinecite{Hikihara_04}.}
\label{FigChaos8}
\end{center}
\end{figure}

\subsection{Quantities related to the dynamical structure factor}
\label{SECIVC}

The spin dynamical structure factor components in Eqs. (\ref{SDSF}) and (\ref{xxPMMP}) appear in expressions of several
physical quantities. An example further discussed below Sec. \ref{SECVI} is the NMR relaxation rate, which can 
be expressed as \cite{Dupont_16},
\begin{eqnarray}
{1\over T_1} & = & {1\over T_1^{\parallel}} + {1\over T_1^{\perp}}\hspace{0.20cm}{\rm where}
\nonumber \\
{1\over T_1^{\parallel}} & = & {\gamma^2\over 2}\sum_k\vert A_{\parallel} (k)\vert^2 
S^{zz} (k,\omega_0)
\hspace{0.20cm}{\rm and} 
\nonumber \\
{1\over T_1^{\perp}}  & = & {\gamma^2 \over 2}\sum_k\vert A_{\perp} (k)\vert^2 (S^{+-} (k,\omega_0)+S^{-+} (k,\omega_0)) \, .
\nonumber \\
\label{T1-1}
\end{eqnarray}
Here $\omega_0$ is the NMR frequency, $\gamma$ is the gyromagnetic ratio, and $A_{\parallel} (k)$ and $A_{\perp} (k)$
are the longitudinal and transverse hyperfine form factors, respectively.

In the present case of the spin-$1/2$ $XXZ$ chain in a longitudinal magnetic field, these two
hyperfine form factors are peaked at $k=2k_{F\downarrow}$ and $k=\pi$, respectively \cite{Carmelo_23}.
For low-energy $\omega/(k_B T)\ll 1$, the longitudinal and transverse components of
the NMR relaxation rate in Eq. (\ref{T1-1}) can then be written as \cite{Dupont_16},
\begin{eqnarray}
{1\over T_1^{\parallel}} & = & {\gamma^2\,\vert A_{\parallel} (2k_{F\downarrow})\vert^2\over 2}\,{A_1^z\cos (\pi \xi^2)\over v_1 (k_{F\downarrow})}
B (\xi^2, 1 - 2\xi^2)
\nonumber \\
& \times & \left({2\pi\,T\over v_1 (k_{F\downarrow})}\right)^{\zeta_{\parallel}}\hspace{0.20cm}{\rm and} 
\nonumber \\
{1\over T_1^{\perp}} & = & {\gamma^2\,\vert A_{\perp} (\pi)\vert^2\over 2}\,{A_{0}^x\cos\left({\pi\over 4\xi^2}\right)\over v_1 (k_{F\downarrow})}
B \left({1\over 4\xi^2}, 1 - {1\over 2\xi^2}\right)
\nonumber \\
& \times & \left({2\pi\,T\over v_1 (k_{F\downarrow})}\right)^{\zeta_{\perp}} \, .
\label{T1-1limit-1D}
\end{eqnarray}
The expressions of the exponents $\zeta_{\parallel}$ and $\zeta_{\perp}$ appearing here
involve only the the phase-shift parameter $\xi$ defined in Eqs. (\ref{x-aaPM}) and are given by,
\begin{equation}
\zeta_{\parallel} = 2\xi^2 - 1\hspace{0.20cm}{\rm and}\hspace{0.20cm}
\zeta_{\perp} = {1\over 2\xi^2} - 1 \, .
\label{exppnNMR}
\end{equation}
The other quantities appearing in Eq. (\ref{T1-1limit-1D}) are the same as in Eq. (\ref{TcTc}). 
\begin{figure}
\begin{center}
\centerline{\includegraphics[width=8.5cm]{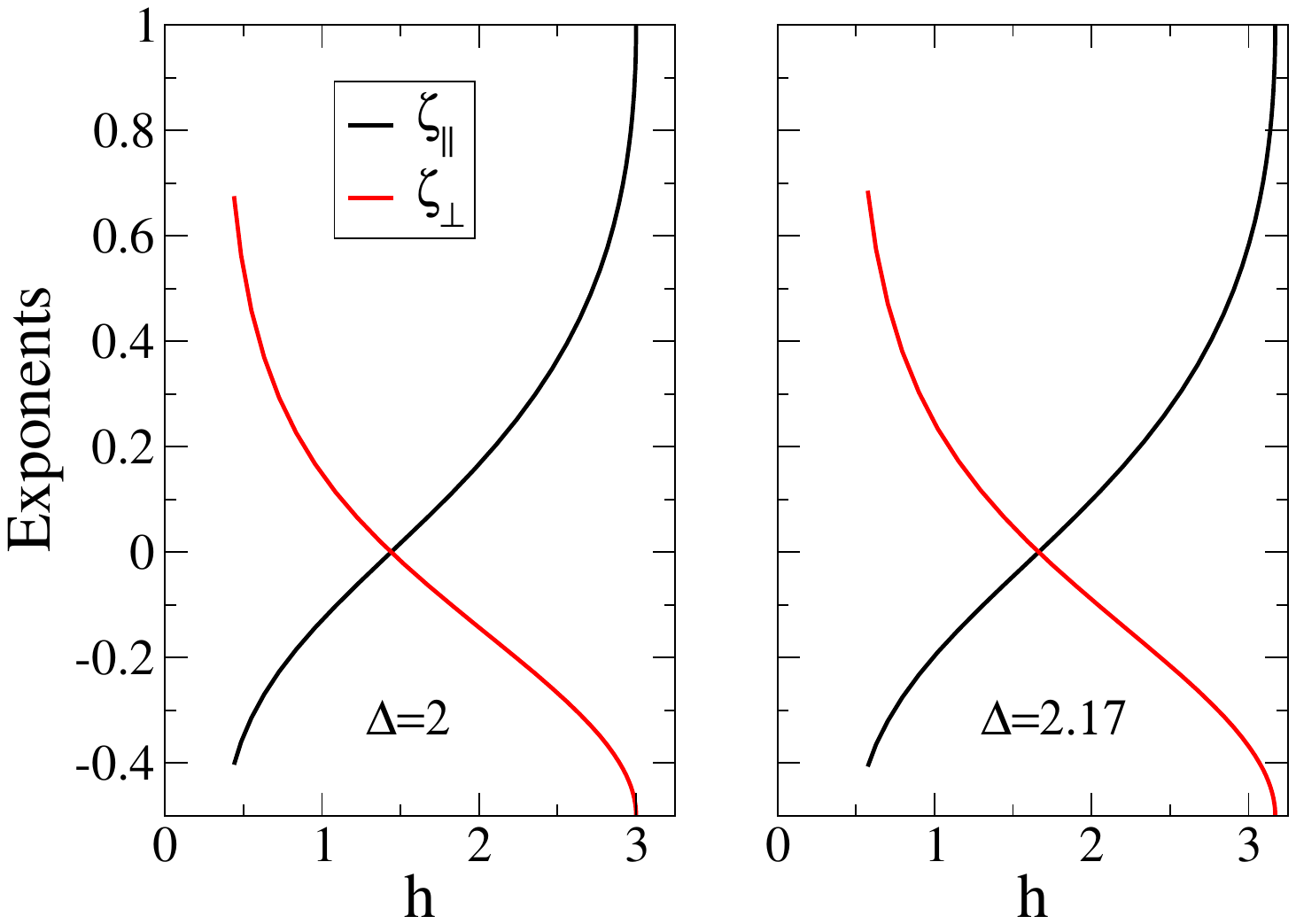}}
\caption{The exponents $\zeta_{\parallel}$ and $\zeta_{\perp}$, Eq. (\ref{exppnNMR}), 
for the purely 1D spin-$1/2$ $XXZ$ chain in a longitudinal magnetic field 
plotted as a function of that field for $h\in [h_{c1},h_{c2}]$ at anisotropies 
$\Delta =2$ and $\Delta =2.17$ suitable to the zigzag materials SrCo$_2$V$_2$O$_8$
and BaCo$_2$V$_2$O$_8$, respectively. From Ref. \onlinecite{Carmelo_23}.}
\label{FigChaos9}
\end{center}
\end{figure}

As mentioned above, the non-universal TLL pre-factors $A_0^x$ and $A_1^z$ of the static spin correlation functions 
appearing both in Eqs. (\ref{TcTc}) and (\ref{T1-1limit-1D}) can be numerically computed \cite{Hikihara_04}. They
are plotted in Fig. \ref{FigChaos8} as a function of $M = m^z = m/2$ for spin density $m\in [0,1]$ and
several $\Delta$ values. The exponents, Eq. (\ref{exppnNMR}), are plotted in Fig. \ref{FigChaos9} as a function 
of the magnetic field $h\in [h_{c1},h_{c2}]$ for anisotropies $\Delta = 2$ and $\Delta = 2.17$ used in Sec. \ref{SECVI} for 
SrCo$_2$V$_2$O$_8$ and BaCo$_2$V$_2$O$_8$, respectively. Their limiting behaviors are,
\begin{eqnarray}
\zeta_{\parallel} & = & -1/2\hspace{0.20cm}{\rm and}\hspace{0.20cm}\zeta_{\perp} = 1
\hspace{0.20cm}{\rm for}\hspace{0.20cm}h\rightarrow h_{c1} 
\nonumber \\
\zeta_{\parallel} & = & \zeta_{\perp} = 0
\hspace{0.20cm}{\rm for}\hspace{0.20cm}h = h_* 
\nonumber \\
\zeta_{\parallel} & = & 1\hspace{0.20cm}{\rm and}\hspace{0.20cm}\zeta_{\perp} = -1/2
\hspace{0.20cm}{\rm for}\hspace{0.20cm}h\rightarrow h_{c2}  \, ,
\label{zetaLimits}
\end{eqnarray}
where the magnetic field $h_*$ is that at which the lines for 
the exponents $\zeta_{\parallel}$ and $\zeta_{\perp}$ cross each other in Fig. \ref{FigChaos9},
at which they read $\zeta_{\parallel}=\zeta_{\perp} =0$. 

\begin{figure}
\begin{center}
\centerline{\includegraphics[width=8.5cm]{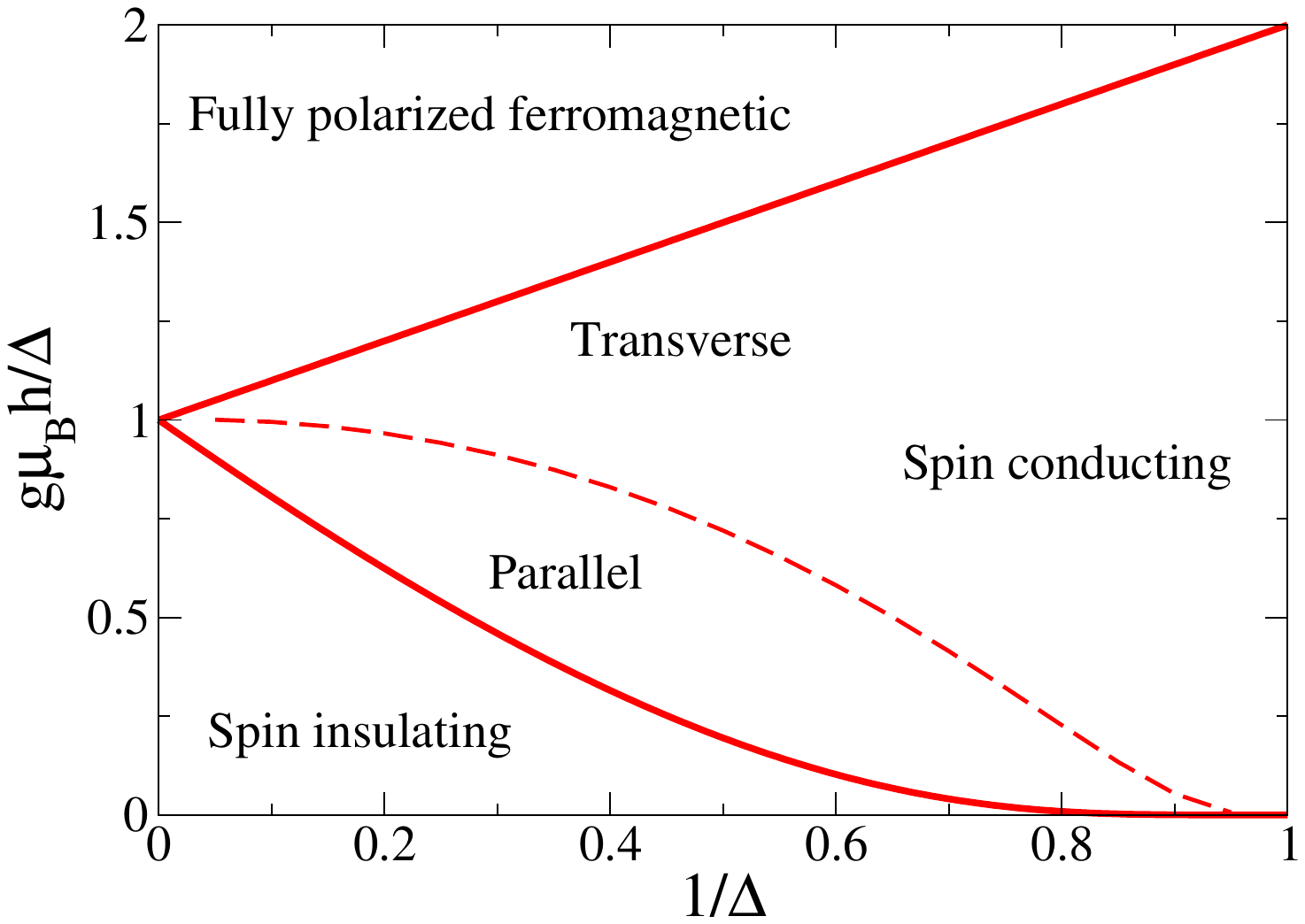}}
\caption{The spin-$1/2$ $XXZ$ chain phase diagram of the magnetic energy over anisotropy, $g\mu_B h/\Delta$, 
in units of $J$ versus inverse anisotropy $\epsilon = 1/\Delta\in [0,1]$.
The energy absolute value over anisotropy lines (a) 
$\lim_{h\rightarrow h_{c1}}g\mu_B h/\Delta$,
(b) $g\mu_B h_*/\Delta$, and (c) $\lim_{h\rightarrow h_{c2}}g\mu_B h/\Delta$ 
separate (a) the spin-insulating phase from the spin-conducting phase with dominant longitudinal
relaxation-rate fluctuations, (b) the latter from the spin-conducting phase with dominant transverse
relaxation-rate fluctuations, and (c) the latter from the fully-polarized ferromagnetic phase, 
respectively. The intermediate magnetic field $h_*$ is that at which $\zeta_{\parallel}=\zeta_{\perp} =0$
where the exponents $\zeta_{\parallel}$ and $\zeta_{\perp}$ are given in Eq. (\ref{exppnNMR})
and plotted in Fig. \ref{FigChaos9}. From Ref. \onlinecite{Carmelo_23}.}
\label{FigChaos10}
\end{center}
\end{figure}

The middle dashed line in the diagram of Fig. \ref{FigChaos10} refers to $g\mu_B h_*/\Delta$
and separates the field regions $h_{c1}<h<h_*$ and $h_{*}<h<h_{c2}$ where the 
longitudinal and and transverse term of the NMR relaxation rate $1/T_1$ in Eq. (\ref{T1-1limit-1D}) dominates, respectively. 
As also illustrated in Fig. \ref{FigChaos10}, $\lim_{h\rightarrow h_{c1}}g\mu_B h/\Delta$
and $\lim_{h\rightarrow h_{c2}}g\mu_B h/\Delta$
fully control the spin-$1/2$ $XXZ$ chain phase diagram of the magnetic energy 
over anisotropy, $g\mu_B h/\Delta$, versus $\epsilon = 1/\Delta\in [0,1]$. 

Note though that the intermediate magnetic field $h_*$ at which $\zeta_{\parallel}=\zeta_{\perp} =0$
as defined here for the purely 1D spin-$1/2$ $XXZ$ chain in a longitudinal magnetic field and that
at which $T_c^{z} (h_*)=T_c^{x} (h_*)$ for a 3D system of weakly 
%CORR the the REPLACED BY the
coupled such chains where the two critical transition temperatures associated 
with longitudinal and transverse orders are given in Eq. (\ref{TcTc}) 
have in general different values. 

For the spin-$1/2$ $XXZ$ chain in a longitudinal magnetic field, analysis of the expressions, Eq. (\ref{T1-1limit-1D}),
with the exponents $\zeta_{\parallel}$ and $\zeta_{\perp}$ given in Eq. (\ref{exppnNMR}) and plotted in Fig. \ref{FigChaos9} 
reveals that the NMR spin-lattice relaxation rate $1/T_1 = 1/T_1^{\parallel} + 1/T_1^{\perp}$, Eq. (\ref{T1-1}), 
is dominated by its divergent longitudinal term $1/T_1^{\parallel}$ for fields $h\in [h_{c1},h_*]$ when $\zeta_{\parallel}<0$ and 
$\zeta_{\perp} >0$ and by its divergent transverse term $1/T_1^{\perp}$ for $h\in [h_*h_{c2}]$ when $\zeta_{\parallel}>0$ 
and $\zeta_{\perp} <0$. 

In Sec. \ref{SECV} we discuss issues related to the spin-transport properties of
the spin-$1/2$ $XXZ$ chain. We return to the relation of that model to quasi-1D
materials in Sec. \ref{SECVI}, where we discuss and justify deviations from the 1D physics associated
with the NMR relaxation rate expressions, Eqs. (\ref{T1-1}) and (\ref{T1-1limit-1D}), experimentally observed in
the zigzag materials BaCo$_2$V$_2$O$_8$ and SrCo$_2$V$_2$O$_8$.

\section{The spin carriers, their spin currents, and the contributions to ballistic spin transport}
\label{SECV}

\subsection{The spin-transport carriers and their elementary spin currents}
\label{SECVA}

In the presence of a uniform vector potential (twisted boundary conditions), 
the Hamiltonian $\hat{H}=\hat{H} (\Phi/L)$ where $\Phi = \Phi_{\uparrow} = -\Phi_{\downarrow}$,
remains solvable by the Bethe ansatz \cite{Shastry_90,Zotos_99}.
After some straightforward algebra using the corresponding $\Phi\neq 0$ Bethe-ansatz equations \cite{Carmelo_24}, 
which refer to Eq. (\ref{BAqn}) for $n=1,...,\infty$ with $q_n (\varphi)$ replaced by
$q_n (\varphi) -2n{\Phi\over N}$, one finds that for HWSs the momentum eigenvalues
are in the thermodynamic limit for finite $\Phi$ given by,
\begin{equation}
P_{\Phi} = P + {\Phi\over N}\,(N - \sum_{n=1}^{\infty}2n\,N_{n}) \, ,
\label{PPhi}
\end{equation}
where $P$ is given in Eq. (\ref{P0}) and refers to $\Phi =0$.
The number of physical spins that couple to the vector potential is 
given by the factor that multiplies ${\Phi\over N}$ in Eq. (\ref{PPhi}).
From the use of the exact sum rule
$M = 2S_q = N - \sum_{n=1}^{\infty}2n\,N_n$, Eq. (\ref{2Pi}), one finds 
that such a number equals that of unpaired physical spins.

The term ${\Phi\over N}\,N$ in ${\Phi\over N}\,2S_q = {\Phi\over N}\,(N - \sum_{n=1}^{\infty}2n\,N_{n})$
refers to {\it all} $N$ physical spins coupling to the vector potential
in the absence of physical-spins singlet pairing. Indeed, the negative coupling counter terms $-\sum_{n=1}^{\infty}2n\,N_n$ 
refer to the number $2n$ of paired physical spins in each $n$-pair. This applies both
to $1$-pairs and to $n$-string-pairs for which $n>1$. These counter terms {\it exactly cancel} the positive coupling of the 
corresponding $2n$ paired physical spins in each $n$-pair. 
As a result of such counter terms, only the $M = 2S_q = N - \sum_{n=1}^{\infty}2n\,N_n$ 
unpaired physical spins couple to the vector potential and carry spin current.

A similar analysis for non-HWSs, Eq. (\ref{state}), gives 
\begin{equation}
P_{\Phi} = P + {\Phi\over N}(M_{+1/2} - M_{-1/2}) \, ,
\label{PPhinonHWS}
\end{equation}
where $P$ is given in Eq. (\ref{P0}) and $M_{\pm 1/2} =  N/2 - \sum_{n=1}^{\infty}n\,N_{n} \pm S^z$ is the number $M_{\pm 1/2} = S_q \pm S^z$ 
of unpaired physical spins of projection $\pm 1/2$. Hence $(M_{+1/2} - M_{-1/2}) = 2S^z$ in Eq. (\ref{PPhinonHWS}),
as given in Eq. (\ref{2Sz2Sq}). This confirms that only the $M_{\pm 1/2}$ unpaired physical spins 
of projection $\pm 1/2$ couple to a uniform vector potential. Hence they are indeed the spin carriers.
Only them carry spin current. 

Consistently with the opposite sign of the coupling to a uniform vector potential
of the unpaired physical spins of projection $+1/2$ and $-1/2$, respectively, revealed by Eq. (\ref{PPhinonHWS}),
it is found below that they also carry elementary spin currents $j^z_{\pm 1/2}$ with opposite sign,
$j^z_{+1/2}= - j^z_{- 1/2}$.

To evaluate the elementary spin currents $j^z_{\pm 1/2}$, it is useful to revisit some operator 
and expectation values relations \cite{Carmelo_15A}.
The $z$ component of the spin current operator $\hat{\vec{J}}$,
\begin{equation}
\hat{J}^z = -i\,J\sum_{j=1}^{N}(\hat{S}_j^+\hat{S}_{j+1}^- - \hat{S}_{j+1}^+\hat{S}_j^-)  \, ,
\label{c-s-currents}
\end{equation}
plays a key role in our study. The commutators, 
\begin{eqnarray}
\left[\hat{J}^z,\hat{S}^{\pm}\right] & = &
\left[\hat{S}^{z},\hat{J}^{\pm}\right] = \pm \hat{J}^{\pm} 
\, ; \hspace{0.25cm}
\left[\hat{J}^{\pm},\hat{S}^{\mp}\right] = \pm 2\hat{J}^z
\nonumber \\
\left[\hat{J}^z,\hat{S}^{z}\right] & = & 0
\, ; \hspace{0.25cm}
\left[\hat{J}^z,(\hat{\vec{S}})^2\right]  = \hat{J}^{+}\hat{S}^{-} - \hat{S}^{+}\hat{J}^{-} \, ,
\label{comm-currents}
\end{eqnarray}
are also useful for it. Here, in addition to the $z$-component spin-current operator $\hat{J}^z$, Eq. (\ref{c-s-currents}), the other two 
operator components $\hat{J}^{\pm}$ of the spin-current operator read,
\begin{equation}
\hat{J}^{+} = (\hat{J}^{-})^{\dag} = 2i\,J\sum_{j=1}^{N}(\hat{S}_j^+\hat{S}_{j+1}^z - \hat{S}_{j+1}^+\hat{S}_j^z) \, .
\label{Jpm}
\end{equation}

To start with we consider the $\Delta =1$ isotropic point associated with the spin-$1/2$ $XXX$ chain.
The $S>0$ HWSs $\vert l_{\rm r}^0,S,S\rangle$ and the $S=S^{z}=0$ states (which are simultaneously 
HWSs and LWSs $\vert l_{\rm r}^0,0,0\rangle$) used
in our operator algebra manipulations obey the well-known transformation laws
$\hat{S}^{+}\vert l_{\rm r} ,S,S\rangle = 0$ and
$\hat{S}^{+}\vert l_{\rm r},0,0\rangle = \hat{S}^{-}\vert l_{\rm r},0,0\rangle = 0$,
which follow straightforwardly from the corresponding $SU(2)$ symmetry operator algebra. 

We consider a class of spin-current operator matrix elements $\langle l_{\rm r}^0,S,S^z\vert\hat{J}^z\vert l_{\rm r}^{0'},S,S^{z}\rangle$
between states with the same arbitrary spin $S\geq 1$ (and $S\geq 1/2$ in the case
of $N$ being an odd integer) and $S^{z}$ values. The following general result is valid for $S\geq 1$ (and $S \geq 1/2$),
\begin{equation}
\langle  l_{\rm r}^0,S,S^z\vert\hat{J}^z\vert  l_{\rm r}^{0'},S,S^z\rangle = {S^z\over S}
\langle  l_{\rm r}^0 ,S,S\vert\hat{J}^z\vert  l_{\rm r}^{0'},S,S\rangle \, , 
\label{currents-gen}
\end{equation}
where $S^z = S - n_z$ and $n_z = 1,...,2S$. It is obtained by
combining the systematic use of the commutators given in Eq. (\ref{comm-currents}) with the
above state transformation laws.
The calculations to reach Eq. (\ref{currents-gen}) are relatively easy for non-HWSs whose generation 
from HWSs involves small $n_z=S-S^z$ values. The calculations become lengthy as the $n_z$ 
value increases, but they remain straightforward. 

Analysis of the matrix elements, Eq. (\ref{currents-gen}), reveals that the 
$l_{\rm r}^0 = l_{\rm r}^{0'}$ and $S^z = 0$ spin-current expectation values 
$\langle  l_{\rm r} ,S,0\vert\hat{J}^z\vert  l_{\rm r},S,0\rangle$ all vanish for 
$S\geq 1$ (and $S\geq 1/2$). However, we also need such spin-current expectation values for
$S=S^z=0$. Those refer to the particular case, $l_{\rm r}^0 = l_{\rm r}^{0'}$, of the
general matrix elements $\langle l_{\rm r}^0,0,0\vert\hat{J}^z\vert l_{\rm r}^{0'},0,0\rangle$, 
which in the following are shown to vanish. Such  matrix elements connect the energy eigenstates 
$\vert  l_{\rm r}^0 ,0,0\rangle$ and $\vert  l_{\rm r}^{0'},0,0\rangle$, which are both LWSs and HWSs. 
It follows from Eq. (\ref{comm-currents}) that the spin-current operator $\hat{J}^z$, Eq. (\ref{c-s-currents}), may be written 
as the commutator $\hat{J}^z = {1\over 2}[\hat{J}^{+},\hat{S}^{-}]$.
Thus the spin-current matrix elements $\langle l_{\rm r}^0,0,0\vert\hat{J}^z\vert l_{\rm r}^{0'},0,0\rangle$ can be written as,
\begin{eqnarray}
\langle l_{\rm r}^0,0,0\vert\hat{J}^z\vert l_{\rm r}^{0'},0,0\rangle & = &
{1\over 2}(\langle l_{\rm r}^0 ,0,0\vert\hat{J}^{+}\hat{S}^{-}\vert l_{\rm r}^{0'},0,0\rangle
\nonumber \\
& - & \langle l_{\rm r}^0,0,0\vert\hat{S}^{-}\hat{J}^{+}\vert l_{\rm r}^{0'},0,0\rangle) .
\label{matrixele00}
\end{eqnarray}

That this expression vanishes is readily confirmed by applying the above-state
transformation laws. A similar result holds for all 
matrix elements of the form $\langle  l_{\rm r}^0,S,0\vert\hat{J}^z\vert  l_{\rm r}^{0'},S+\delta S,0\rangle$
where $S \geq 0$ and $S' = S + \delta S\geq 0$, which are found to vanish unless $\delta S = \pm 1$.
Hence all $S^z = 0$ spin-current expectation values $\langle  l_{\rm r}^0,S,0\vert\hat{J}^z\vert  l_{\rm r}^0,S,0\rangle$ vanish for 
$S\geq 0$.

In the general case of both anisotropies $\Delta =1$ and $\Delta > 1$, it is useful to use the following 
simplified notations for the spin-current expectation values,
\begin{eqnarray}
\langle \hat{J}^z (l_{\rm r}^{\eta},S_q,S^z) \rangle & = &
\langle  l_{\rm r}^{\eta},S_q,S^z\vert\hat{J}^z\vert  l_{\rm r}^{\eta},S_q,S^z\rangle
\nonumber \\
\langle \hat{J}^z_{HWS} (l_{\rm r}^{\eta},S_q) \rangle & = &
\langle l_{\rm r}^{\eta},S_q,S_q\vert\hat{J}^z\vert  l_{\rm r}^{\eta},S_q,S_q\rangle \, ,
\label{not-currents}
\end{eqnarray}
where $l_{\rm r}^{\eta}=l_{\rm r}^{0}$ and $S_q=S$ for $\eta=0$ and thus anisotropy $\Delta = 1$. 

For the isotropic point, $\Delta = 1$,
the general matrix-elements relation, Eq. (\ref{currents-gen}), gives for $l_{\rm r}^0=l_{\rm r}^{0'}$
the following spin-current expectation values relation,
\begin{equation}
\langle \hat{J}^z (l_{\rm r}^0,S,S^z) \rangle =
{S^z\over S}\langle \hat{J}^z_{HWS} (l_{\rm r}^0,S) \rangle \, .
\label{rel-currents-XXX}
\end{equation}

As discussed in Sec. \ref{SECIIA}, the irreducible representations of the $\eta >0$ 
continuous $SU_q(2)$ symmetry \cite{Pasquier_90}
are isomorphic to those of the $\eta=0$ $SU(2)$ symmetry. The spin projection $S^z$ remains a good 
quantum number under the unitary transformation, Eq. (\ref{Ophi}). On the other hand,
in spite of the $\eta$-dependence of the $SU_q(2)$ symmetry generators, 
Eqs. (\ref{OneSOphipm})-(\ref{CasimirOknown}),
spin $S$ is under that unitary transformation mapped into $q$-spin $S_q$ such 
that $S_q=S$. 

It follows that under it the factor ${S^z\over S}$ in the exact relation, Eq. (\ref{rel-currents-XXX}), 
is mapped into a corresponding factor ${S^z\over S_ q}$ such that $S_q=S$. 
The $\Delta =1$ relation, Eq. (\ref{rel-currents-XXX}), is thus mapped into
a corresponding exact relation valid for $\Delta >1$ \cite{Carmelo_24},
\begin{equation}
\langle \hat{J}^z (l_{\rm r}^{\eta},S_q,S^z) \rangle =
{S^z\over S_ q}\langle \hat{J}^z_{HWS} (l_{\rm r}^{\eta},S_q) \rangle \, .
\label{rel-currents-XXZ}
\end{equation}

The spin-current expectation value $\langle \hat{J}^z_{HWS} (l_{\rm r}^{\eta},S_q) \rangle$
in this expression can be obtained for HWSs from the energy eigenvalues $E (l_{\rm r}^{\eta},S_q,\Phi/L)$ of
the Hamiltonian, Eq. (\ref{HD1}), in the presence of a vector potential,
$\hat{H}=\hat{H} (\Phi/N)$. It is given by,
\begin{equation}
\langle \hat{J}^z_{HWS} (l_{\rm r}^{\eta},S_q) \rangle  = \lim_{\Phi/N\rightarrow 0}
{d E (l_{\rm r}^{\eta},S_q,\Phi/L)\over d (\Phi/N)} \, .
\label{currentHWS}
\end{equation}
These spin-current expectation values are thus derived by taking the $\Phi/N\rightarrow 0$ limit after 
performing the $\Phi/N$-derivation in Eq. (\ref{currentHWS}). This involves the energy eigenvalues, Eq. (\ref{Energy}) 
for $M_{+1/2} - M_{-1/2} = M_{+1/2} = 2S_q$, and
Bethe-ansatz equations, Eq. (\ref{BAqn}), with $\varphi_{n} (q_{j})$ replaced
by $\varphi_{n} \left(q_{j}-{2n\Phi\over N}\right)$ and $q_n (\varphi)$ replaced 
by $q_n (\varphi) - {2n\Phi\over N}$, respectively. 

The use of the exact relation, Eq. (\ref{rel-currents-XXZ}), allows the evaluation of the elementary spin current $j^z_{\pm 1/2}$ 
carried by each spin carrier of projection $\pm 1/2$. 
Combination of that relation with both Eq. (\ref{2Sz2Sq}) and the generation of 
unpaired physical spin flips described by Eq. (\ref{state}), gives the following exact relation,
\begin{eqnarray}
\langle \hat{J}^z (l_{\rm r}^{\eta},S_q,S^z) \rangle & = & \sum_{\sigma =\pm 1/2}M_{\sigma}\,j^z_{\sigma} (l_{\rm r}^{\eta},S_q)
\nonumber \\
& = & j^z_{+1/2} (l_{\rm r}^{\eta},S_q) (M_{+1/2} - M_{-1/2}) 
\nonumber \\
& = &  j^z_{+1/2} (l_{\rm r}^{\eta},S_q)\sum_{\sigma =\pm 1/2} (2\sigma)M_{\sigma} \, ,
\label{rel-currents}
\end{eqnarray}
where in $\sum_{\sigma =\pm 1/2}M_{\sigma}\,j^z_{\sigma} (l_{\rm r}^{\eta},S_q)$ and as obtained in Ref. \onlinecite{Carmelo_24},
\begin{equation}
j^z_{\pm 1/2} = j^z_{\pm 1/2} (l_{\rm r}^{\eta},S_q) \equiv \pm {\langle \hat{J}^z_{HWS} (l_{\rm r}^{\eta},S_q) \rangle 
\over 2S_q} \, .
\label{elem-currents}
\end{equation}

To confirm that $j^z_{\pm 1/2} = j^z_{\pm 1/2} (l_{\rm r}^{\eta},S_q)$, Eq. (\ref{elem-currents}), is indeed the elementary spin current carried by
one spin carrier of projection $\pm 1/2$, note that under each unpaired physical-spin flip generated by 
the operator ${\hat{S}}^{-}_{\eta}$ in Eq. (\ref{state}) the spin-current 
expectation value $\langle \hat{J}^z\rangle = \langle \hat{J}^z (l_{\rm r}^{\eta},S_q,S^z) \rangle$, Eqs. (\ref{rel-currents-XXZ}) and
(\ref{rel-currents}), changes exactly by $-2j^z_{+1/2} = 2j^z_{-1/2}$ \cite{Carmelo_24}.

This reveals the deep physical meaning of the relation, Eqs. (\ref{rel-currents-XXZ}) and
(\ref{rel-currents}): It expresses the spin-current expectation value 
$\langle \hat{J}^z\rangle = \langle \hat{J}^z (l_{\rm r}^{\eta},S_q,S^z)\rangle$
of an energy eigenstate $\vert l_{\rm r}^{\eta},S_q,S^z\rangle$ in terms of the elementary spin currents
$j^z_{\pm 1/2}$ carried by each of the $M_{\pm 1/2} = S_q \pm S^z$ spin carriers of projection $\pm 1/2$ that
populate that state. 

Consistently with all $S=0$ current expectation values 
$\langle  l_{\rm r}^0,0,0\vert\hat{J}^z\vert  l_{\rm r}^0,0,0\rangle$ vanishing 
in the case of the isotropic point, $\Delta=1$, as shown above having
as starting point the spin-current matrix elements expression, Eq. (\ref{matrixele00})
for $l_{\rm r}^0=l_{\rm r}^{0'}$, the relation, Eqs. (\ref{rel-currents-XXZ}) and
(\ref{rel-currents}), confirms that when $M = M_{+1/2} + M_{-1/2} = 0$
the spin-current expectation value $\langle \hat{J}^z \rangle = \sum_{\sigma =\pm 1/2}M_{\sigma}\,j^z_{\sigma}$, 
Eq. (\ref{rel-currents}), vanishes, {\it i.e.} $S_q=0$ states have vanishing spin-current 
expectation value also for $\Delta >1$. This follows from their lack of spin
carriers: Only $S_q>0$ energy eigenstates can have finite spin-current expectation values 
as a result of being populated by a finite number $M=2S_q$ of spin carriers.

The spin stiffness at fixed value of $S^z$ associated with ballistic spin transport
can for finite temperatures $T>0$ be exactly expressed as follows \cite{Mukerjee_08},
\begin{equation}
D^z (T) = {1\over 2 N T}\sum_{S_q=\vert S^z\vert}^{N/2}\sum_{l_{\rm r}^{\eta}} 
p_{l_{\rm r}^{\eta},S_q,S^z}\vert \langle \hat{J}^z (l_{\rm r}^{\eta},S_q,S^z) \rangle  \vert^2 \, .
\label{D-all-T-simpA}
\end{equation}
Here the summations run over {\it all} quantum-problem energy eigenstates $\vert l_{\rm r}^{\eta},S_q,S^z\rangle$
with fixed $S^z$ value and $p_{l_{\rm r}^{\eta},S_q,S^z}$ are the corresponding Boltzmann weights.
Such states $q$-spin values belong to the range $S_q \in [\vert S^z\vert,N/2]$.
\begin{figure}
\begin{center}
\centerline{\includegraphics[width=8.50cm]{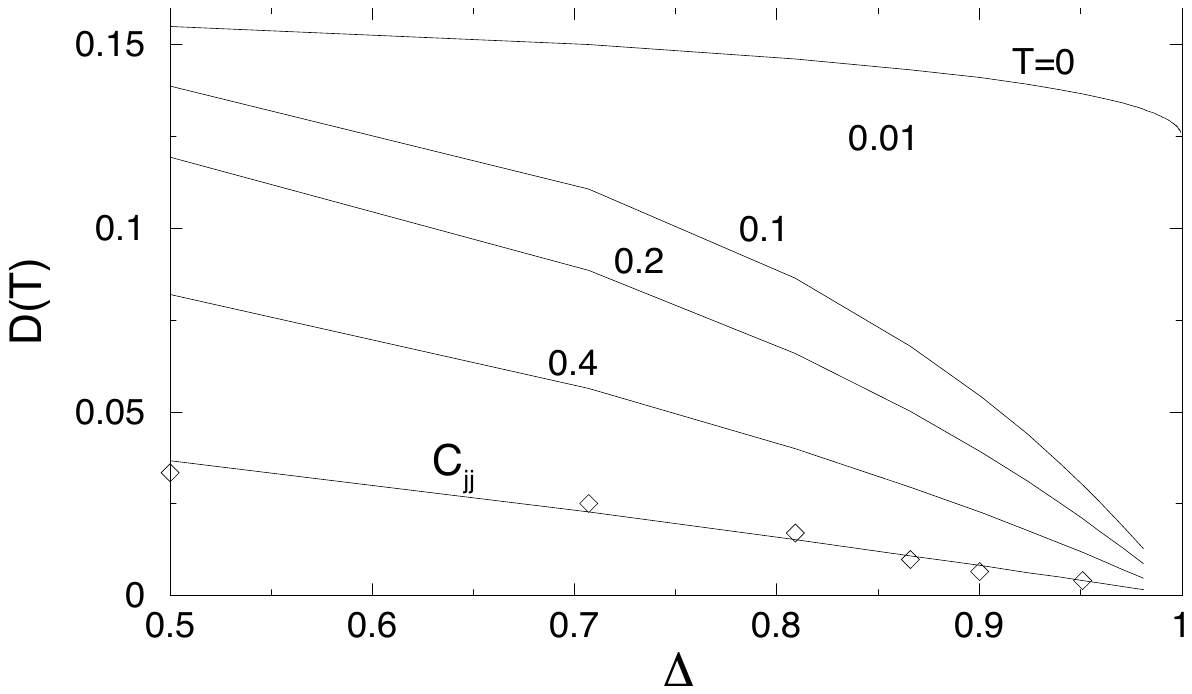}}
\caption{The spin stiffness of the spin-$1/2$ $XXZ$ chain at zero magnetic field,
denoted by $D$ in Refs. \onlinecite{Carmelo_18} and \onlinecite{Zotos_99},
as a function of anisotropy for $0.5<\Delta<1$ at various temperatures.
($C_{jj}$ is a high-temperature proportionality constant \cite{Zotos_99}.)
From Refs. \onlinecite{Carmelo_18} and \onlinecite{Zotos_99}.}
\label{FigChaos11}
\end{center}
\end{figure}

For anisotropy $0 < \Delta <1$ and zero magnetic field, the finite-temperature spin stiffness, Eq. (\ref{D-all-T-simpA}), 
is finite, so that the dominant spin transport is ballistic \cite{Zotos_99}.
It is plotted in Fig. \ref{FigChaos11} as a function of anisotropy for $0.5<\Delta<1$ at zero magnetic field and various temperatures.
As justified below in Sec. \ref{SECVB}, for anisotropy $\Delta >1$ and zero magnetic field it vanishes at any finite temperature \cite{Carmelo_94}.

Studies on the finite-temperature spin stiffness at zero magnetic field for anisotropy $\Delta >1$ use to rely on its expression,
Eq. (\ref{D-all-T-simpA}). It involves the absolute values $\vert \langle \hat{J}^z (l_{\rm r}^{\eta},S_q,S^z) \rangle\vert$
of the spin-current expectation values of the states that contribute to that transport. Whether
at zero magnetic field the finite-temperature spin stiffness vanishes
or is finite becomes then a very difficult problem. This is because for many such
states the absolute values $\vert \langle \hat{J}^z (l_{\rm r}^{\eta},S_q,S^z) \rangle\vert$
are proportional to $N$ and thus diverge in the thermodynamic limit.

On the other hand, as confirmed in the ensuing section, the problem simplifies if instead we express the spin-stiffness expression
in terms of the absolute values $\vert j^z_{+1/2} (l_{\rm r}^{\eta},S_q) \vert$ of the elementary currents carried
by the spin carriers. Note though that only the physical-spins representation used
in our studies naturally identifies such spin carriers, which are the $M=2S_q$ unpaired physical
spins of projections $\pm 1/2$ that populate finite-$S_q$ energy eigenstates.

\subsection{The contributions to ballistic spin transport for $\Delta >1$, $h=0$, and $T> 0$}
\label{SECVB}

Here we briefly review results of Ref. \onlinecite{Carmelo_24} concerning the vanishing
of the spin stiffness for anisotropy $\Delta >1$, vanishing magnetic field, $h=0$, and arbitrary finite temperature, $T> 0$.
The use of Eqs. (\ref{rel-currents}) and (\ref{elem-currents}) in Eq. (\ref{D-all-T-simpA})
allows to express the spin stiffness $D^z (T)$ directly in terms of
the elementary spin currents $j^z_{\pm 1/2}$ carried by the spin carriers as follows \cite{Carmelo_24},
\begin{equation}
D^z (T) = m\,{2S^z\over 2T} \langle\vert j^z_{\pm 1/2} (l_{\rm r}^{\eta},S_q) \vert^2\rangle_T \, ,
\label{D-all-T-simp-m}
\end{equation}
where $m = 2S^z/N$ and,
\begin{equation}
\langle\vert j^z_{\pm 1/2} (l_{\rm r}^{\eta},S_q) \vert^2\rangle_T = \sum_{S_q=\vert S^z\vert}^{N/2}\sum_{l_{\rm r}} 
p_{l_{\rm r}^{\eta},S_q,S^z}\vert j^z_{+1/2} (l_{\rm r}^{\eta},S_q) \vert^2 \, .
\label{jz2T}
\end{equation}
This is the thermal expectation value in the subspace spanned by states with fixed $S^z$
of the square of the absolute value $\vert j^z_{\pm 1/2}\vert = \vert j^z_{\pm 1/2} (l_{\rm r}^{\eta},S_q)\vert$ 
of the elementary spin current carried by one spin carrier of projection $\pm 1/2$. 

According to Eqs. (\ref{D-all-T-simp-m}) and (\ref{jz2T}), one has for finite temperatures $T>0$ that 
$\lim_{m\rightarrow 0}D^z (T) =0$ provided that $\langle\vert j^z_{\pm 1/2} (l_{\rm r}^{\eta},S_q) \vert^2\rangle_T$ is finite.
Accounting for the fixed-$S^z$ subspace normalization, $\sum_{S_q=\vert S^z\vert}^{N/2}\sum_{l_{\rm r}} 
p_{l_{\rm r}^{\eta},S_q,S^z} = 1$,
it thus follows from Eqs. (\ref{D-all-T-simp-m}) and (\ref{jz2T}) that the only property needed for showing that the 
spin stiffness $D^z (T)$ associated with the ballistic contributions to spin transport 
indeed vanishes in the $m\rightarrow 0$ limit for $\Delta >1$ and all finite temperatures $T>0$
is that the absolute value $\vert j^z_{\pm 1/2}\vert$
of the elementary spin current carried by one spin carrier of projection $\pm 1/2$ is finite
for all energy eigenstates for which $S^z/N\rightarrow 0$ for $N\rightarrow\infty$.

The studies of Ref. \onlinecite{Carmelo_24} use the fact that {\it all} finite-$S_q$
energy eigenstates belong for $\Delta >1$ to two classes (i) and (ii) of 
states whose concentration $M/N=2S_q/N$ of spin carriers
vanishes and is finite, respectively, in the thermodynamic limit, $N\rightarrow\infty$. For these
two classes of states, the absolute value of the ratio $\langle \hat{J}_{HWS}^z\rangle/N$
involving the spin-current expectation value, Eq. (\ref{currentHWS}), also vanishes and is finite, 
respectively, in that limit. 

That for class (i) and (ii) states both the ratio
$\langle \hat{J}^z_{HWS}\rangle/N$ and the concentration of spin carriers $M/N=2S_q/N$ 
(i) vanish and (ii) are finite, respectively, then implies that in both cases the 
ratio $\langle\hat{J}^z_{HWS}\rangle/(2S_q)$ that gives  $\vert j^z_{\pm 1/2}\vert$ is finite and does 
not diverge in the $N\rightarrow \infty$ limit. 

In Ref. \onlinecite{Carmelo_24} it is confirmed that this general argument is indeed an exact 
result for $\Delta >1$, yet fails in the $\Delta\rightarrow 1$ limit: A few states are found have
absolute values $\vert j^z_{\pm 1/2}\vert$ that diverge in that limit.
This may follow from anomalous spin transport associated with the expected isotropic point 
$\Delta =1$ super-diffusion.

However, our above analysis and statements refer to $\Delta >1$. 
That the absolute value $\vert j^z_{\pm 1/2}\vert$ of all finite-$S_q$
states is finite for anisotropy values $\Delta >1$ implies that 
$\langle\vert j^z_{\pm 1/2} (l_{\rm r}^{\eta},S_q) \vert^2\rangle_T$ is also finite in Eqs. 
(\ref{D-all-T-simp-m}) and (\ref{jz2T}), so that $\lim_{m\rightarrow 0}D^z (T) = 0$. In addition, the form of the 
exact spin stiffness expression, Eq. (\ref{D-all-T-simp-m}), shows that $D^z (T) = 0$ {\it at} $m=0$.

Accounting for the absence of phase transitions and critical points at $h=0$ in the
case of anisotropy $\Delta >1$, this reveals that in the thermodynamic limit one has within the 
grand-canonical ensemble both that $\lim_{h\rightarrow 0}D^z (T) = 0$ and $D^z (T) = 0$ at $h=0$.

Hence, for the spin-$1/2$ $XXZ$ chain at zero magnetic field the ballistic contributions to spin 
transport vanish in the thermodynamic limit for $\Delta >1$ and {\it all} finite temperatures $T>0$,
which implies non-ballistic spin transport.

The vanishing of the spin stiffness for anisotropy $\Delta >1$, vanishing magnetic field, $h=0$, and arbitrary finite temperature, $T> 0$
reveals that spin transport is normal diffusive {\rm provided} that the spin diffusion constant is finite.
The studies of Ref. \onlinecite{Carmelo_24} show that it is indeed finite.

The general $T>0$ spin stiffness expression, Eq. (\ref{D-all-T-simpA}), does not apply at $T=0$.
Finally, we briefly comment recent results on it for anisotropy $\Delta >1$ at zero temperature \cite{Carmelo_24}.
Spin transport at finite temperatures and zero magnetic field is qualitatively different at $\Delta =1$ and for $\Delta >1$, as it
is super diffusive and normal diffusive, respectively \cite{Carmelo_24}. Consistently, also at zero temperature
it is found to be qualitatively different (a) for $h=0$ and $\Delta =1$ and (b) for $h\in [0,h_{c1}]$ and $\Delta >1$, respectively.

In the case of the spin-$1/2$ $XXZ$ chain, the $T=0$ spin stiffness has been derived both for anisotropy $-1 < \Delta < 1$ 
and $\Delta =1$ by use of the Hamiltonian in the presence of a vector potential, $\hat{H}=\hat{H} (\Phi/L)$ (twisted boundary conditions) 
\cite{Shastry_90}. At the isotropic point, $\Delta = 1$, it was studied both at $h=0$ and $m=0$ \cite{Shastry_90,Zotos_99} 
and for $m=2S^z/N\in [0,1]$ \cite{Carmelo_15}.

The use of twisted boundary conditions leads for anisotropy $\Delta >1$ actually to the same general expression, 
$D^z = {\xi^2\over 2\pi}\,v_1(k_{F\downarrow})$, as for $\Delta =1$ \cite{Carmelo_24}. 
However, both the $1$-band group velocity $v_1 (k_{F\downarrow})$, Eqs. (\ref{v1q}) and (\ref{v1qm0kF}),
and the parameter $\xi$, Eqs. (\ref{x-aaPM}) and (\ref{xilimits}), have different values and behaviors at $\Delta =1$ 
and for $\Delta >1$, respectively.

For anisotropy $\Delta >1$ the $T=0$ spin stiffness vanishes both for fields $h\in [0,h_{c1}]$
and $h>h_{c2}$, whereas at $\Delta =1$ it only vanishes in the fully polarized ferromagnetic quantum phase.
This vanishing of the $T=0$ ballistic contribution to spin transport for fields $h\in [0,h_{c1}]$ and anisotropy
$\Delta >1$ is consistent with the occurrence of the corresponding spin-insulating quantum phase. In contrast,
at $\Delta =1$ spin transport is ballistic at zero magnetic field, $h=0$. 

The different behavior of the group velocity $v_{1} (k_{F\downarrow})$ in the $m=2S^x/N\rightarrow 0$ limit
at $\Delta =1$ and for $\Delta >1$, respectively, controls that of the $T=0$ spin stiffness. It reads 
$v_1(k_{F\downarrow}) = v_1(\pi/2) = 0$ for $\Delta >1$ and $v_1(k_{F\downarrow}) = 
v_1(\pi/2) = J{\pi\over 2}$ at $\Delta =1$, as given in Eq. (\ref{v1qm0kF}). This gives indeed
$D^z =0$ for $\Delta >1$ and $D^z = J/8$ at $\Delta =1$, respectively.

\section{Effects of selective interchain couplings}
\label{SECVI}

The previous sections of this paper were devoted to the physical-spins representation
of the purely 1D spin-$1/2$ $XXZ$ chain and the description of some physical properties of quasi-1D materials
by that 1D model. In contrast, here we discuss deviations from 1D physics of these materials.

Following the results of Sec. \ref{SECIVC}, in case that for fields $h\in [h_{c1},h_{c2}]$ and very low temperatures just above the 
small critical transition temperature $T_c (h)$, Eq. (\ref{TcTc}), the 1D physics fully applied to BaCo$_2$V$_2$O$_8$ 
and SrCo$_2$V$_2$O$_8$, the dependence on the magnetic field $h$ of the exponents $\zeta_{\parallel}$ and
$\zeta_{\perp}$ plotted Fig. \ref{FigChaos9} that appear in the expressions, Eq. (\ref{T1-1limit-1D}), would imply that the NMR 
spin-lattice relaxation rate $1/T_1 = 1/T_1^{\parallel} + 1/T_1^{\perp}$, Eq. (\ref{T1-1}), 
was dominated by its divergent longitudinal term $1/T_1^{\parallel}$ for fields $h\in [h_{c1},h_*]$ when $\zeta_{\parallel}<0$ and 
$\zeta_{\perp} >0$ and by its divergent transverse term $1/T_1^{\perp}$ for $h\in [h_*h_{c2}]$ when $\zeta_{\parallel}>0$ 
and $\zeta_{\perp} <0$. Here $h_* = 1.441$ for $\Delta =2$ and $h_* = 1.664$ for $\Delta =2.17$ in units of 
$J/g\mu_B$, which for $J = 3.55$\,meV gives $h_* = 14.25$\,T and for $J = 2.60$\,meV 
leads to $h_* = 12.06$\,T, respectively. 

In contrast to 1D physics, NMR experimental 
results of Ref. \onlinecite{Klanjsek_15} for BaCo$_2$V$_2$O$_8$ 
and of Ref. \onlinecite{Cui_22} for SrCo$_2$V$_2$O$_8$ though reveal that the longitudinal term
$1/T_1 = 1/T_1^{\parallel}\propto T^{\zeta_{\parallel}} = T^{2\xi^2 - 1}$ dominates for the {\it whole} 
magnetic field interval $h\in [h_{c1},h_{c2}]$ of the spin-conducting phases, including for
$h\in [h_{*},h_{c2}]$ when $1/T_1^{\perp}$ should dominate.

Note that the value of the intermediate magnetic field $h_*$ at which the exponents, Eq. (\ref{zetaLimits}), are
such that $\zeta_{\parallel}=\zeta_{\perp} =0$ for the purely 1D spin-$1/2$ $XXZ$ chain
in a longitudinal magnetic field is typically larger that than 
that of the field also denoted by $h_*$ at which the temperatures for longitudinal and transverse
orders in Eq. (\ref{TcTc}) obey the equality $T_c^{z} (h_*)=T_c^{x} (h_*)$ for a system of weakly coupled chains
\cite{Okunishi_07}. The experimental values of $h_*$ for BaCo$_2$V$_2$O$_8$ and 
SrCo$_2$V$_2$O$_8$ suggested by neutron scattering are indeed lower than
the predicted 1D values and read $h_* \approx 8.5$\,T and $h_* \approx 7.0$\,T, respectively \cite{Grenier_15,Shen_19}.

Here we review recent results of Ref. \onlinecite{Carmelo_23} that explain such a deviations
from the 1D physics by effects of selective interchain couplings.
Both such zigzag materials have similar chain structures along the $c$-axis,
being almost iso-structural: BaCo$_2$V$_2$O$_8$ has a centro-symmetric crystal structure ($I 4_{1}/acd$, nonpolar),
while SrCo$_2$V$_2$O$_8$ has a non-centro-symmetric crystal structure ($I 4_{1}/cd$, polar) \cite{Okutani_15}.

To introduce what is meant here by selective interchain couplings, note that
hopping-matrix elements associated with interchain couplings are obtained by the overlap between the 
wave functions of the excited states and the one-particle potential that transforms according to the underlying lattice symmetries. 
The corresponding quantum overlap is largest and spin states are coupled more strongly whenever they are connected by a symmetry operation of the 
underlying lattice. 

In the case of the above mentioned chain structures of the materials BaCo$_2$V$_2$O$_8$ and
SrCo$_2$V$_2$O$_8$, a four-fold rotation with additional translation of 1/4th of the unit cell 
allows for a coupling between different chains. On the other hand, antiferromagnetic intrachain 
coupling naturally leads to antiferromagnetic NN and NNN interchain couplings. 
The additional translation of 1/4th of the unit cell takes care of the change of chirality between adjacent 
chains and for an anti-ferromagnetic spin order. As a result, only states with the same spin-projection yield a finite overlap. 
In contrast, the interchain coupling should tend to zero for excitations that involve an electronic spin-flip. 

We thus conclude that in the case of excited states, the above symmetry operation involving the four-fold rotation with additional 
translation of 1/4th of the unit cell is only a symmetry in spin-space if {\it no} electronic spin flip is 
performed within the generation of such states. 

Recent results of Ref. \onlinecite{Carmelo_23} reviewed in the following provide strong evidence that this 
explains why interchain couplings can be neglected concerning the spin 
dynamical structure factor transverse components $S^{+-} (k,\omega)$ and $S^{-+} (k,\omega)$,
so that at very low temperatures just above the small critical temperature $T_c (h)$, Eq. (\ref{TcTc}), 
they can be described by 1D physics. The reason is that the
transverse excitations contributing to them involve an electronic spin flip.
This though does not apply to the longitudinal component $S^{zz} (k,\omega)$ whose longitudinal 
excitations do not involve such a spin flip. 

This selection rule is expected to be behind selective interchain couplings that both protect the 1D physics
of BaCo$_2$V$_2$O$_8$ and SrCo$_2$V$_2$O$_8$ and lead to deviations from it, mainly associated with the 
enhancement of the spectral-weight intensity of $S^{zz} (k,\omega)$. 

As reported in Sec. \ref{SECIV} and mentioned above, the 1D physics of quasi-1D spin-chain compounds occurs for the 
spin-conducting phases at very low temperatures just above a small critical temperature $T_c (h)$, Eq. (\ref{TcTc}), below 
which interchain couplings lead to 3D ordered phases \cite{Carmelo_23,Okunishi_07}. 
The use of interchain mean-field theory \cite{Okunishi_07} gives the expressions for 
the critical temperatures $T_c^{z} (h)$ and $T_c^{x} (h)$ provided in Eq. (\ref{TcTc}).

However and as justified below, the effects of selective interchain couplings are
consistent with the following alternative expression for $T_c^{z} (h)$,
\begin{eqnarray}
T_c^{z} (h) & = & {v_1(k_{F\downarrow})\over 2\pi}
\nonumber \\
& \times & \left(\Delta\,4J' \tilde{A}_1^z{\sin (\pi \xi^2)\over v_1 (k_{F\downarrow})}
B^2 \left({\xi^2\over 2}, 1 - \xi^2\right)\right)^{1\over 2 (1-\xi^2)} \, .
\nonumber \\
\label{Tc}
\end{eqnarray}
where we have replaced the TLL pre-factor $A_1^z$ plotted in 
Fig. \ref{FigChaos8} by a pre-factor $\tilde{A}_1^z$, which is sensitive to effects of selective interchain couplings.
The dependence on $J'$ of that pre-factor $\tilde{A}_1^z$ is beyond interchain mean-field theory. 

Such a replacement of $A_1^z$ by $\tilde{A}_1^z$ has a deep physically meaning.  
In the case of pure 1D physics, the relaxation rate does not depend on the effective interchain coupling $J'$. On the other hand,
that  $A_1^z$ is replaced by $\tilde{A}_1^z$ in the longitudinal critical transition temperature
$T_c^{z} (h)$, Eq. (\ref{Tc}), implies that in the expression, Eq. (\ref{T1-1limit-1D}), for the longitudinal relaxation 
rate term $1/T_1^{\parallel}$ the pre-factor $A_1^z$ 
must also also replaced by $\tilde{A}_1^z$, so that \cite{Carmelo_23},
\begin{eqnarray}
{1\over T_1^{\parallel}} & = & {\gamma^2\,\vert A_{\parallel} (2k_{F\downarrow})\vert^2\over 2}\,{\tilde{A}_1^z\cos (\pi \xi^2)\over v_1 (k_{F\downarrow})}
B (\xi^2, 1 - 2\xi^2)
\nonumber \\
& \times & \left({2\pi\,T\over v_1 (k_{F\downarrow})}\right)^{\zeta_{\parallel}} \, .
\label{T1-1limit}
\end{eqnarray}
In contrast to the 1D physics and beyond interchain mean-field theory, now the component $1/T_1^{\parallel}$ as given in Eq. (\ref{T1-1limit}) 
implicitly depends on $J'$ through the pre-factor $\tilde{A}_1^z = \tilde{A}_1^z (J')$, which obeys the 
boundary condition, $\tilde{A}_1^z (0) = A_1^z$.

Fits of the magnetization measurements \cite{Okunishi_07,Kimura_08,Canevet_13} 
lead to an effective interchain coupling given by $J'= 0.00138 J$ for BaCo$_2$V$_2$O$_8$. 
This is actually consistent with results found in Ref. \onlinecite{Klanjsek_15} by the use of the
expression for $T_c^{z} (h)$ provided in Eq. (\ref{TcTc}): For fields from $h=h_{c1}\approx 3.8$\,T
up to $h=h_* \approx 8.5$\,T the effective interchain coupling
in that expression was found to be given by $J'/K_B = 0.042$\,K, so that $J' = 0.0036$\,meV.
For the value $J = 2.60$\,meV suitable to BaCo$_2$V$_2$O$_8$, this gives $J' = 0.00139 J$,
very near the value $J' = 0.00138 J$ obtained by magnetization measurements. 

On the other hand, a giant variation of the effective interchain coupling $J'(h)$ by a factor up to 24 was found
upon increasing the magnetic field $h$ further from $h=h_* \approx 8.5$\,T towards $h=h_{c2}\approx 22.9$\,T \cite{Klanjsek_15}.
Note though that such an increase of the effective interchain coupling $J'(h)$ was obtained within
the use of the expression for the longitudinal critical transition temperature $T_c^{z} (h)$ given in Eq. (\ref{TcTc}),
which was obtained within interchain mean-field theory \cite{Okunishi_07}. 

We though argue that such an apparent huge increase of $J'(h)$ 
for fields in the interval $h\in [h_*,h_{c2}]$ renders that expression invalid.
As justified below, we argue that in the case of the longitudinal critical transition temperature $T_c^{z} (h)$ 
the effects of selective interchain couplings rather change that expression by increasing the 
values of {\it both} the pre-factor $\tilde{A}_1^z$ and the effective interchain coupling $J'(h)$, respectively, Eq. (\ref{Tc}).

In contrast to only the effective interchain coupling $J'(h)$ varying by a factor up to 24, which
would affect only the critical transition temperature $T_c^{z} (h)$, this 
implies that also the relaxation rate longitudinal component $1/T_1^{\parallel}$ is enhanced
due to the increase of the pre-factor $\tilde{A}_1^z$, as given in Eq. (\ref{T1-1limit}). And
this is consistent with the experimental data of Ref. \onlinecite{Klanjsek_15} for BaCo$_2$V$_2$O$_8$ 
and Ref. \onlinecite{Cui_22} for SrCo$_2$V$_2$O$_8$. 

The purely 1D pre-factors $A_1^z$ and $A_0^x$ in the expressions of $1/T_1^{\parallel}$ and $1/T_1^{\perp}$
given Eq. (\ref{T1-1limit-1D}) are controlled by matrix-element's overlaps within the
dynamic structure factor's components $S^{zz} (k,\omega_0)$ and $S^{+-} (k,\omega_0)+S^{-+} (k,\omega_0)$,
respectively, in the NMR relaxation expression, Eq. (\ref{T1-1}). 

According to the selection rule that determines the effects of selective interchain couplings, since $A_0^x$ is associated 
with transverse excitations that {\it involve} an electronic spin flip, it remains insensitive to such effects.
They though affect the spin-states quantum overlaps that control 
the pre-factor $A_1^z$ associated with $S^{zz} (k,\omega)$, which are associated 
with longitudinal excitations that {\it do not involve} an electronic spin flip and thus are 
sensitive to the selective interchain couplings. This is why in Ref. \onlinecite{Carmelo_23} 
it was proposed that, beyond interchain mean-field theory \cite{Okunishi_07}, 
in the expression for $T_c^{z} (h)$ in Eq. (\ref{TcTc}), the giant enhancement of $J'(h)$ for 
for fields in the interval $h\in [h_*,h_{c2}]$ detected in Ref. \onlinecite{Klanjsek_15} is actually distributed between 
$J'$ and $\tilde{A}_1^z$, as given in Eq. (\ref{Tc}). 

This implies that such a giant variation refers in in Eq. (\ref{Tc}) to the {\it product} $J' \times \tilde{A}_1^z$
rather than to $J'$ alone. It then follows that the effective interchain coupling of Ref. \onlinecite{Klanjsek_15},
which we denote by $J'_{\rm Ref. 39} (h)$, is replaced by the quantity \cite{Carmelo_23},
\begin{eqnarray}
J'_{\rm Ref. 39} (h) & \rightarrow & C_1^z\,C'\,J'_{\rm min}\hspace{0.20cm}{\rm where}
\nonumber \\
C_1^z & = & {\tilde{A}_1^z\over A_1^z}
\hspace{0.20cm}{\rm and}\hspace{0.20cm}C' = {J'\over J'_{\rm min}} \, .
\label{subs}
\end{eqnarray}
Here the value of $C_1^z\,C'\,J'_{\rm min}$ is exactly the same as that of $J'_{\rm Ref. 39} (h)$,
so that the equality $C_1^z\,C'\,J'_{\rm min}=J'_{\rm Ref. 39} (h)$ holds. In addition,
$J'_{\rm min} = 0.00139 J$, the enhanced effective coupling $J'= J'(h)$ is such that $J'_{\rm min}<J'<J'_{\rm Ref. 39} (h)$ 
for magnetic fields in the interval $h\in [h_*,h_{c2}]$, and $A_1^z$ is the non-universal TLL longitudinal pre-factor 
of the static spin correlation function plotted in Fig. \ref{FigChaos8}. 

While both $\tilde{A}_1^z$ and $J'$ are enhanced, we cannot access the precise values of their separate 
enhancement factors $C_1^z=\tilde{A}_1^z/A_1^z$ and $C'=J'/J'_{\rm min}$, respectively.
Nonetheless, we know that their product gives $C_1^z (h)\times C'(h)= J'_{\rm Ref. 39} (h)/J'_{\rm min}\in [1,24]$ for $h\in [h_*,h_{c2}]$.

That magnetic field interval $h\in [h_*,h_{c2}]$ for which the enhancement of $C_1^z\,C'\,J'_{\rm min}=J'_{\rm Ref. 39} (h)$ 
was found in Ref. \onlinecite{Klanjsek_15} is precisely that for which {\it in contrast to the 1D physics} there is 
unexpected experimental dominance of the relaxation rate longitudinal component $1/T_1^{\parallel}\propto T^{\zeta_{\parallel}}$ 
relative to $1/T_1^{\perp}\propto T^{\zeta_{\perp}}$. This is in spite of for the
purely 1D model one having that $\zeta_{\parallel} >0$ and $\zeta_{\perp} < 0$
in Eqs. (\ref{T1-1limit-1D}) and (\ref{exppnNMR}), as shown in Fig. \ref{FigChaos9}.

This is thus consistent with the enhancement by $C_1^z = \tilde{A}_1^z/A_1^z$ of the pre-factor 
$\tilde{A}_1^z$ in the $1/T_1^{\parallel}$'s expression, Eq. (\ref{T1-1limit}).
Indeed, due to selective interchain couplings that act on $S^{zz} (k,\omega)$,
also the ratio $\tilde{A}_1^z/A_0^x$ of the pre-factors $\tilde{A}_1^z$ and $A_0^x$ of
the expressions of $T_1^{\parallel}$ and $1/T_1^{\perp}$ in Eq. (\ref{T1-1limit}), respectively,
is enhanced relative to the corresponding ratio of the 1D physics, $A_1^z/A_0^x$.

The unexpected experimental low-temperature dominance of the longitudinal NMR relaxation rate term 
$1/T_1 = T_1^{\parallel}\propto T^{\zeta_{\parallel}}$ for magnetic fields $h\in [h_*,h_{c2}]$ found both
in BaCo$_2$V$_2$O$_8$ \cite{Klanjsek_15} and in SrCo$_2$V$_2$O$_8$ \cite{Cui_22} is thus 
associated in Ref. \onlinecite{Carmelo_23} with the enhancement of 
$\tilde{A}_1^z$ by $C_1^z = \tilde{A}_1^z/A_1^z$ in both such zigzag materials.
That dominance is not mainly due to the relative values of the hyperfine form 
factors $A_{\parallel} (k)$ and $A_{\perp} (k)$ in Eqs. (\ref{T1-1}) and (\ref{T1-1limit}): It rather 
mainly follows from the effects of selective interchain couplings
on the quantum overlaps within the matrix elements of $S^{zz} (k,\omega)$.

Note though that the weaker effects of transverse staggered fluctuations 
are behind the experimental studies of SrCo$_2$V$_2$O$_8$ showing a NMR line splitting that indicates 
the onset of transverse fluctuations \cite{Cui_22} at $h=h_*\approx 7.0$\,T. This confirms that the transverse NMR form factor $A_{\perp} (k)$ 
does not vanish. Consistently, transverse fluctuations and corresponding peaks have been
observed by neutron scattering for magnetic fields $h\in [h_*,h_{c2}]$ both in BaCo$_2$V$_2$O$_8$ 
\cite{Grenier_15} and in SrCo$_2$V$_2$O$_8$ \cite{Shen_19}. This suggests some degree of coexistence of 
both longitudinal and transverse orders \cite{Grenier_15}, in spite of the experimental dominance of the longitudinal
NMR relaxation rate term $1/T_1 = T_1^{\parallel}\propto T^{2\xi^2 - 1}$. 
 
The above results refer to the limit of low energy, $\omega/(k_B T)\ll 1$, for which the expressions
of the NMR relaxation rate given in Eqs. (\ref{T1-1limit-1D}) and (\ref{T1-1limit}) are valid.
However, the additional $S^{zz} (k,\omega)$'s spectral-weight intensity brought about by selective interchain couplings
also applies to higher energy scales. 

This is indeed also clearly visible by SrCo$_2$V$_2$O$_8$'s neutron scattering in 
$S^{zz} (k,\omega)$ for larger $\omega$ values, as shown in Fig. 5-b of Ref. \onlinecite{Bera_20} 
for the magnetic field interval $h\in [3.8\,{\rm T},15\,{\rm T}]$, in what the longitudinal 
sharp peak called $R^{\rm PAP(zz)}_{\pi}$ in that reference is concerned. 

The intensity of such a sharp peak's spectral weight and that of the longitudinal sharp peak 
called $R^{\rm PAP(zz)}_{\pi/2}$ shown in Fig. 5-a of that reference for fields larger than $h_{c1}$, called $B_c$ in these figures, is 
larger than that of the transverse sharp peaks.  The data shown in Fig. 5 of Ref. \onlinecite{Bera_20} 
also reveal that for higher energies the enhancement occurs for a larger magnetic-field interval than 
the interval $h\in [h_*,h_{c2}]$ found for low energy.

\section{Concluding remarks}
\label{SECVII}

The usual spinon and alike representations are unsuitable to describe some quantum problems
associated with the spin-$1/2$ $XXZ$ chain with anisotropy $\Delta >1$. Two examples are the Bethe strings that 
contribute to the spin dynamical structure factor and have no spinon representation and quantities associated with
finite-temperature spin transport, such as the spin stiffness at zero magnetic field. In this paper an alternative
representation in terms of physical spins \cite{Carmelo_15A, Carmelo_23,Carmelo_22,Carmelo_24} was reviewed and 
used to handle these two quantum problems beyond the spinon paradigm.

The contribution to the low-temperature dynamical properties of both the spin-$1/2$ $XXZ$ chain in
a longitudinal magnetic field and the zigzag material SrCo$_2$V$_2$O$_8$ \cite{Wang_18} of the Bethe strings
of lengths $n=1,2,3$ \cite{Carmelo_23,Carmelo_22} was one of the issues revisited in
this paper. Such $n$-Bethe strings describe a number $n$ of singlet pairs of physical spins.

The identification of the spin carriers that populate all finite-$S_q$ energy eigenstates becomes 
within the physical-spins representation a simple problem. We have briefly reviewed recent results of Ref. \onlinecite{Carmelo_24}
that show that for the spin-$1/2$ $XXZ$ chain with anisotropy $\Delta >1$ at zero magnetic field 
only one condition should be fulfilled for the finite-temperature spin stiffness vanishing: The absolute
value of the spin elementary current carried by the spin carriers that populate all finite-$S_q$ energy eigenstates 
that contribute to it must be finite. The finiteness of such absolute values is confirmed
for $\Delta >1$, which shows that the contributions to ballistic spin transport indeed vanish at zero magnetic field 
for anisotropy $\Delta >1$ and {\it all} finite temperatures. 
In Ref. \onlinecite{Carmelo_24} its is found that the
spin diffusion constant is finite at $h=0$, $\Delta >1$, and $T>0$, so that spin transport is
normal diffusive. 

All above reviewed results involve the use of the physical-spins representation
for the purely 1D spin-$1/2$ $XXZ$ chain and the description by that model of physical properties of quasi-1D materials.
Finally, we have also briefly discussed the effects of selective interchain couplings \cite{Carmelo_23}
that are behind deviations form the 1D physics of the quasi-1D materials BaCo$_2$V$_2$O$_8$ and SrCo$_2$V$_2$O$_8$.

The goal of this paper is not an exhaustive review on spin-chain models, spin-based technologies, and experimental techniques 
to study spin-chain systems and materials. Reference to recent reviews on some such issues is provided in the following, 
for the reader interested in them.

Magnetism at low dimensions is actually a thriving field of research with exciting opportunities in technology. 
From the emulation of 1D quantum phases to the potential realization 
of Majorana edge states, spin chains are unique systems to study. For a review on how
the advent of scanning tunneling microscope (STM) based techniques has permitted to engineer spin 
chains in an atom-by-atom fashion via atom manipulation and 
to access their spin states on the ultimate atomic scale, see Ref. \onlinecite{Choi_19}. 
An account on recent research on spin correlations and 
dynamics of atomic spin chains as studied by the STM is presented in that review paper.

Laser-cooled and trapped atomic ions form an ideal standard for the simulation of interacting 
quantum spin models. Effective spins are represented by appropriate internal energy levels within 
each ion, and the spins can be measured with near-perfect efficiency using state-dependent fluorescence 
techniques. For a review on programmable quantum simulations of spin systems with trapped ions,
see Ref. \onlinecite{Monroe_21}.

Concerning spintronic effects described based on theoretical and experimental analysis of 
antiferromagnetic materials, see a review in Ref. \onlinecite{Baltz_18}. 
For applications to quantum technologies involving magnetic potentials and corresponding magnetic traps, 
atom chips, adiabatic magnetic potentials, and time-averaged adiabatic potentials, a recent
review is given in Ref. \onlinecite{Amico_22}.
For a review on the quantum limits to the energy resolution of magnetic field sensors, including the
standard quantum limit, Heisenberg limit, and amplification quantum noise, see Ref. \onlinecite{Mitchell_20}.

For a description of each major spin-qubit type, the present limits of fidelity, 
and an overview of alternative spin-qubit platforms, see Ref. \onlinecite{Burkard_23}.
Finally, a review on the basic principles, methods, and concepts of quantum sensing from the viewpoint of the 
interested experimentalist, including its platform being spin qubits is given in Ref. \onlinecite{Degen_17}.

%%%%%%%%%%%%%%%%%%%%%%%%%%%%%%%%%%%%%%%%%%%%%%%%%%%%%%%%%%%%%%%%%%%%%
\acknowledgements
We thank David K. Campbell for the many collaborations and illuminating discussions over the years
and Toma\v{z} Prosen and Tobias Stauber for stimulating discussions. J. M. P. C. acknowledges support from 
the Portuguese FCT through the Grants UIDB/04650/2020 and UID/CTM/04540/2019.
P. D. S. acknowledges the support from FCT through the Grant UID/CTM/04540/2019.\\ \\ 
%%%%%%%%%%%%%%%%%%%%%%%%%%%%%%%%%%%%%%%%%%%%%%%%%%%%%%%%%%%%%%%%%%%%%
\appendix


\begin{references}
\bibitem{Carmelo_93}
	J. M. P. Carmelo, P. Horsch, D. K. Campbell, and A. H. Castro Neto, 
	``Magnetic effects, dynamical form factors, and electronic instabilities in the Hubbard chain,"
	Phys. Rev. B {\bf 48}, 4200-4203(R) (1993).
\bibitem{Carmelo_94}
	J. M. P. Carmelo, A. H. Castro Neto, and D. K. Campbell,	
         ``Perturbation theory of low-dimensional quantum liquids I. The pseudoparticle-operator basis,"
	Phys. Rev. B {\bf 50}, 3667-3682 (1994).
\bibitem{Carmelo_94A}
	J. M. P. Carmelo, A. H. Castro Neto, and D. K. Campbell,	
         ``Perturbation theory of low-dimensional quantum liquids. II. Operator description of Virasoro algebras in integrable systems,"
	Phys. Rev. B {\bf 50}, 3683-3695 (1994).	
\bibitem{Carmelo_94B}
	J. M. P. Carmelo, A. H. Castro Neto, and D. K. Campbell,	
         ``Conservation laws and bosonization in integrable Luttinger liquids,"
	Phys. Rev. Lett. {\bf 73}, 926-929 (1994); {\bf 74}, 3089 (1995), Erratum.
\bibitem{Carmelo_95}
	J. M. P. Carmelo, A. H. Castro Neto, and D. K. Campbell,	
         ``Exotic low-energy separation in 1D quantum liquids,"
	J. Low Temp. Phys. {\bf 99}, 577-582 (1995).
\bibitem{Baeriswyl_95}
	D. Baeriswyl, D. K. Campbell, J. M. P. Carmelo, F. Guinea, and E. Louis,
	{\it The Hubbard Model -- Its Physics and Mathematical Physics,"}
	(NATO ASI Series, Series B: Physics, Vol. 343, Plenum Press, Nova Iorque,1995).
\bibitem{Carmelo_95A}	
	J. M. P. Carmelo, A. H. Castro Neto, and D. K. Campbell,	
	``New operator algebra for the Hubbard chain" in 
	{\it The Hubbard model its physics and mathematical physics},
	edited by D. Baeriswyl, D. Campbell, J. M. P. Carmelo, F. Guinea, and E. Louis
	(NATO ASI Series, Series B: Physics -- Vol. 343, Plenum Press, New York,1995), pp 117-124.	
\bibitem{Peres_97}	
	N. M. R. Peres, J. M. P. Carmelo, D. K. Campbell, and A. W. Sandvik, 
	``Pseudoparticle description of the 1D Hubbard model electronic transport properties,"
	Zeits. f\"ur Phys, B {\bf 103}, 217-220 (1997).	
\bibitem{Carmelo_98}	
	J. M. P. Carmelo, P. Horsch, A. A. Ovchinnikov, D. K. Campbell, A. H. Castro Neto, and 
	N. M. R. Peres,
	``Comment on "Generalization of a Fermi liquid to a liquid with fractional exclusion 
	statistics in arbitrary dimensions: theory of a Haldane liquid,"
	Phys. Rev. Lett. {\bf 81}, 489 (1998).	
\bibitem{Peres_99}
	N. M. R. Peres, P. D. Sacramento, D. K. Campbell, and J. M. P. Carmelo, 
	``Curvature of levels and charge stiffness of one-dimensional spinless fermions,"
	Phys. Rev. B  {\bf 59}, 7382-7392 (1999).
\bibitem{Carmelo_01}
	J. M. P. Carmelo, T. Prosen, and D. K. Campbell, 
	``Conservation laws in the one-dimensional Hubbard model,"
	Phys. Rev. B {\bf 63}, 205114 (2001).
\bibitem{Carmelo_15}	
	J. M. P. Carmelo, P. D. Sacramento, J. D. P. Machado, and D. K. Campbell, 
	``Singularities of the dynamical structure factors of the spin-$1/2$ $XXX$ chain at finite magnetic field,"
	J. Phys: Cond. Matt. {\bf 27}, 406001 (2015); {\bf 33}, 069501 (2021), Corrigendum.
\bibitem{Carmelo_15A}
	J. M. P. Carmelo, T. Prosen, and D. K. Campbell,
	``Vanishing spin stiffness in the spin-${1\over 2}$ Heisenberg chain for any nonzero temperature,"
	Phys. Rev. B {\bf 92}, 165133 (2015).	
\bibitem{Carmelo_19}
	J. M. P. Carmelo, T. \v{C}ade\v{z}, Y. Ohtsubo, S.-i. Kimura, and D. K. Campbell,
	``Effects of finite-range interactions on the one-electron spectral properties of 
	one-dimensional metals: Application to Bi/InSb(001),"
	Phys. Rev. B {\bf 100}, 035105 (2019).
\bibitem{Carmelo_19A}
	J. M. P. Carmelo, T. \v{C}ade\v{z}, D. K. Campbell, M. Sing, and R. Claessen,
	``Effects of finite-range interactions on the one-electron spectral properties of TTF-TCNQ,''
	Phys. Rev. B {\bf 100}, 245202 (2019).
\bibitem{Carmelo_23}
	J. M. P. Carmelo, P. D. Sacramento, T. Stauber, and D. K. Campbell,
	``Zigzag materials: Selective interchain couplings control the coexistence of one-dimensional physics and deviations from it,"
	Phys. Rev. Res. {\bf 5}, 043058 (2023).
\bibitem{Baeriswyl_87}
	D. Baeriswyl, J. Carmelo, and K. Maki,
	``Coulomb correlations in one-dimensional conductors with incommensurate band fillings and
	semiconductor-metal transition in polyaceteylene,''
	Synth. Metals {\bf 21}, 271-278 (1987).	
\bibitem{Heisenberg}
	W. Heisenberg, 
	``Zur Theorie des Ferromagnetismus,"
	Z. Phys. {\bf 49}, 619-636 (1928).
\bibitem{Bethe}
	H. Bethe, 
	``Zur Theorie der Metalle. I. Eigenwerte und Eigenfunktionen der linearen Atomkette,"
	Z. Phys. {\bf 71}, 205-226 (1931).
\bibitem{Gaudin_71}	
	M. Gaudin, 
	``Thermodynamics of the Heisenberg-Ising ring for $\Delta\geq 1$,"
	Phys. Rev. Lett. {\bf 26}, 1301-1304 (1971).		
\bibitem{Carmelo_22} 
	J. M. P. Carmelo and P. D. Sacramento, 
        ``The role of $q$-spin singlet pairs of physical spins in the dynamical properties of the spin-$1/2$ Heisenberg-Ising $XXZ$ chain,"
        Nucl. Phys. B {\bf 974}, 115610 (2022). Nucl. Phys. B {\bf 997}, 116385 (2023) (Corrigendum).	
\bibitem{Carmelo_24} 
	J. M. P. Carmelo and P. D. Sacramento, 
%CORR CHANGE OF TITLE
	``Diffusive spin transport of the spin-$1/2$ $XXZ$ chain in the Ising regime at zero magnetic field and finite temperature,''
%``Non-ballistic spin transport of the spin-1/2 $XXZ$ chain in the Ising regime at zero magnetic field and temperatures $T\geq 0$,"
        submitted for publication (2024).	
\bibitem{Gaudin_14}	
	M. Gaudin, 
	``The Bethe wavefunction"
	(Cambridge University Press, 2014).
\bibitem{Takahashi_71}
        M. Takahashi,
        ``One-dimensional Heisenberg model at finite temperature,"
	Progr. Theor. Phys. {\bf 46}, 401-415 (1971).
\bibitem{Takahashi_99}	
	M. Takahashi,
	``Thermodynamics of one-dimensional solvable models,"
	(Cambridge University Press, 1999).   	
\bibitem{Imambekov_12}  
	A. Imambekov, T. L. Schmidt, and L. I. Glazman, 
	``One-dimensional quantum liquids: Beyond the Luttinger liquid paradigm,"
	Rev. Mod. Phys. {\bf 84}, 1253-1306 (2012).
\bibitem{Carmelo_18}
	J. M. P. Carmelo and P. D. Sacramento, 
	``Pseudoparticle approach to 1D integrable quantum models,"
	Phys. Reports {\bf 749}, 1-90 (2018).				
\bibitem{Carmelo_17}
	J. M. P. Carmelo and T. Prosen, 
	``Absence of high-temperature ballistic transport in the spin-$1/2$ $XXX$ chain within the grand-canonical ensemble,"
	Nucl. Phys. B {\bf 914}, 62-98 (2017).		
\bibitem{Carmelo_20} 
	J. M. P. Carmelo, T. \v{C}ade\v{z}, and P. D. Sacramento, 
	``Bethe strings in the dynamical structure factor of the spin-$1/2$ Heisenberg $XXX$ chain,"
	Nucl. Phys. B {\bf 960}, 115175 (2020).
\bibitem{Hikihara_04}
	T. Hikihara and A. Furusaki,
	{\it Correlation amplitudes for the spin-${1\over 2}$ $XXZ$ chain in a magnetic field},
	Phys. Rev. B {\bf 69}, 064427 (2004).		
\bibitem{Kimura_07}
	S. Kimura, H. Yashiro, K. Okunishi, M. Hagiwara, Z. He, K. Kindo, T. Taniyama, and M. Itoh,
	{\it Field-induced order-disorder transition in antiferromagnetic BaCo$_2$V$_2$O$_8$ driven by a softening of spinon excitation},
	Phys. Rev. Lett. {\bf 99}, 087602 (2007).	
\bibitem{Okunishi_07}
	Okunishi, K. and Suzuki, T.
	{\it Field-induced incommensurate order for the quasi-one-dimensional $XXZ$ model in a magnetic field},
	Phys. Rev. B {\bf 76}, 224411 (2007).
\bibitem{Kimura_08}
	S. Kimura, T. Takeuchi, K. Okunishi, M. Hagiwara, Z. He, K. Kindo, T. Taniyama, and M. Itoh,
	{\it Novel ordering of an $S=1/2$ quasi-1d Ising-like antiferromagnet in magnetic field},
	Phys. Rev. Lett. {\bf 100}, 057202 (2008).	
\bibitem{Canevet_13}
	E. Can\'evet, B. Grenier, M. Klanj\v{s}ek, C. Berthier, M. Horvati\'c, V. Simonet, and P. Lejay,
	{\it Field-induced magnetic behavior in quasi-one-dimensional Ising-like antiferromagnet BaCo$_2$V$_2$O$_8$:
	A single-crystal neutron diffraction study},
	Phys. Rev. B {\bf 87}, 054408 (2013).
\bibitem{Okutani_15}
	A. Okutani, T. Kida, T. Usui, T. Kimura, K. Okunishi, and M. Hagiwara, 
	{\it High field magnetization of single crystals of the $S=1/2$ quasi-1D Ising-like Antiferromagnet SrCo$_2$V$_2$O$_8$},
	Phys. Procedia {\bf 75}, 779 (2015).	 	
\bibitem{Klanjsek_15}
	M. Klanj\v{s}ek, M. Horvati\'c, S. Kr\"amer, S. Mukhopadhyay, H. Mayaffre, C. Berthier, E. Can\'evet,
	B. Grenier, P. Lejay, and E. Orignac,
	{\it Giant magnetic field dependence of the coupling between spin chains in BaCo$_2$V$_2$O$_8$},
	Phys. Rev. B {\bf 92}, 060408(R) (2015).
\bibitem{Grenier_15}
	B. Grenier, V. Simonet, B. Canals, P. Lejay, M. Klanj\v{s}ek, M. Horvati\'c, and C. Berthier,
	{\it Neutron diffraction investigation of the $H-T$ phase diagram above the longitudinal incommensurate phase of BaCo$_2$V$_2$O$_8$},
	Phys. Rev. B {\bf 92}, 134416 (2015).	
\bibitem{Dupont_16} 
	M. Dupont, S. Capponi, and N. Laflorencie,
	{\it Temperature dependence of the NMR relaxation rate $1/T_1$ for quantum spin chains},
	Phys. Rev. B {\bf 94}, 144409 (2016).			
\bibitem{Shen_19}
	L. Shen, O. Zaharko, J. O. Birk, E. Jellyman, Z. He, and E. Blackburn, 
	{\it Magnetic phase diagram of the quantumspin chains compound SrCo$_2$V$_2$O$_8$: 
	a single-crystal neutron diffraction study},
	New J. Phys. {\bf 21}, 073014 (2019).
\bibitem{Han_21}
	Y. Han, S. Kimura, K. Okunishi, and M. Hagiwara,	
	{\it Unconventional magnetic excitations and spin dynamics of exotic quantum spin systems 
	BaCo$_2$V$_2$O$_8$ and Ba$_3$CuSb$_2$O$_9$},
	App. Magn. Reson. {\bf 52}, 349-362 (2021).
\bibitem{Scheie_21}
	A. Scheie, N. E. Sherman, M. Dupont, S. E. Nagler, M. B. Stone, G. E. Granroth, J. E. Moore, and D. A. Tennant,
	``Detection of Kardar-Parisi-Zhang hydrodynamics in a quantum Heisenberg spin-$1/2$ chain,"
	Nat. Phys. {\bf 17}, 726-730 (2021).
\bibitem{Cui_22}
	Y. Cui, Y. Fan, Z. Hu, Z. He, W. Yu, and R. Yu,  
	{\it Field-induced antiferromagnetism and Tomonaga-Luttinger liquid behavior in the quasi-one-dimensional Ising 
	antiferromagnet SrCo$_2$V$_2$O$_8$},
	Phys. Rev. B {\bf 105}, 174428 (2022).
\bibitem{Shen_22}
	L. Shen, E. Campillo, O. Zaharko, P. Steffens, M. Boehm, K. Beauvois, B. Ouladdiaf, Z. He, D. Prabhakaran, A. T. Boothroyd, and E. Blackburn,
	``Inhomogeneous spin excitations in weakly coupled spin-$1/2$ chains,''
	Phys. Rev. Res. {\bf 4}, 013111 (2022).
\bibitem{Wang_18}
	Z. Wang, J. Wu, W. Yang, A. K. Bera, D. Kamenskyi, A. T. M. N. Islam, S. Xu, J. M. Law, B. Lake, C. Wu, and A. Loidl,
	``Experimental observation of Bethe strings,"
	Nature {\bf 554}, 219-223 (2018).	
\bibitem{Wang_19}
	Z. Wang, M. Schmidt, A. Loidl, J. Wu,3, H. Zou, W. Yang, C. Dong, Y. Kohama, 
	K. Kindo, D. I. Gorbunov, S. Niesen, O. Breunig, J. Engelmayer, and T. Lorenz,
	{\it Quantum critical dynamics of a Heisenberg-Ising chain in a longitudinal field: 
	many-body strings versus fractional excitations},
	Phys. Rev. Lett. {\bf 123}, 067202 (2019).
\bibitem{Bera_20}
	A. K. Bera, J. Wu, W. Yang, R. Bewley, M. Boehm, J. Xu, M. Bartkowiak, O. Prokhnenko, B. Klemke, A. T. M. N. Islam, 
	J. M. Law, Z. Wang, and B. Lake, 
	``Dispersions of many-body Bethe strings,"
	Nature Phys. {\bf 16}, 625-630 (2020).	
\bibitem{Karbach_02}	
	M. Karbach, D. Biegel, and G. M\"uller,
	``Quasiparticles governing the zero-temperature dynamics of the one-dimensional spin-$1/2$ 
	Heisenberg antiferromagnet in a magnetic field,"
	Phys. Rev. B {\bf 66}, 054405 (2002).        			
\bibitem{Pasquier_90}
	V. Pasquier and H. Saleur,
	``Common structures between finite systems and conformal field theories through quantum groups,"
	Nucl. Phys. B {\bf 330}, 523-556 (1990). 
\bibitem{Prosen_13}
	T. Prosen and E. Ilievski,
	``Families of quasilocal conservation laws and quantum spin transport,"
	Phys. Rev. Lett. {\bf 111}, 057203 (2013).	
\bibitem{Bertini_21}
	B. Bertini, F. Heidrich-Meisner, C. Karrasch, T. Prosen, R. Steinigeweg, and M. \v{Z}nidari\v{c},
	``Finite-temperature transport in one-dimensional quantum lattice models,''
	Rev. Mod. Phys. {\bf 93}, 025003-71 (2021).
\bibitem{Znidaric_11}
	M. \v{Z}nidari\v{c}
	``Spin transport in a one-dimensional anisotropic Heisenberg model,"
	Phys. Rev. Lett. {\bf 106}, 220601 (2011).
\bibitem{Ljubotina_17}
	M. Ljubotina, M. \v{Z}nidari\v{c}, and T. Prosen,
	``Spin diffusion from an inhomogeneous quench in an integrable system,"
	Nat. Comm. {\bf 8}, 16117 (2017).
\bibitem{Medenjak_17}
	M. Medenjak, C. Karrasch, and T. Prosen,
	``Lower bounding diffusion constant by the curvature of Drude weight,"
	Phys. Rev. Lett. {\bf 119}, 080602 (2017).
\bibitem{Ilievski_18}
	E. Ilievski, J. De Nardis, M. Medenjak, and T. Prosen,
	``Superdiffusion in one-dimensional quantum lattice models,"
	Phys. Rev. Lett. {\bf 121}, 230602 (2018).
\bibitem{Gopalakrishnan_19}	
	S. Gopalakrishnan and R. Vasseur,
	``Kinetic theory of spin diffusion and superdiffusion in XXZ spin chains,"	
	Phys. Rev. Lett. {\bf 122}, 127202 (2019).
\bibitem{Weiner_20}	
	F. Weiner, P. Schmitteckert. S. Bera, and F. Evers,
	``High-temperature spin dynamics in the Heisenberg chain: Magnon propagation and emerging Kardar-Parisi-Zhang 
	scaling in the zero-magnetization limit,"
	Phys. Rev. B {\bf 101}, 045115 (2020).	
\bibitem{Gopalakrishnan_23}	
	S. Gopalakrishnan and R. Vasseur,
	``Anomalous transport from hot quasiparticles in interacting spin chains,"
	Rep. Prog. Phys. {\bf 86}, 036502 (2023).	
\bibitem{Jepsen_20}
	P. N. Jepsen, J. Amato-Grill, I. Dimitrova, W. W. Ho, E. Demler, and W. Ketterle,
	``Spin transport in a tunable Heisenberg model realized with ultracold atoms,"
	Nature {\bf 588}, 403-407 (2020).			
\bibitem{Shastry_90}
	B. S. Shastry and B. Sutherland, 
	``Twisted boundary conditions and effective mass in Heisenberg-Ising and Hubbard rings,"
	Phys. Rev. Lett. {\bf 65}, 243-246 (1990).	
\bibitem{Zotos_99}	
	X. Zotos,  
	``Finite temperature Drude weight of the one-dimensional spin-$1/2$ Heisenberg model,"
	Phys. Rev. Lett. {\bf 82}, 1764-1767 (1999).		
\bibitem{Kruis_04}	
	H. V. Kruis, I. P. McCulloch, Z. Nussinov, and J. Zaanen, 
	``Geometry and the hidden order of Luttinger liquids:?The universality of squeezed space,"
	Phys. Rev. B {\bf 70}, 075109 (2004).
\bibitem{Horvatic_20}
	M. Horvati\'c, M. Klanj\v{s}ek, and E. Orignac, 
	``Direct determination of the Tomonaga-Luttinger parameter $K$ in quasi-one-dimensional spin systems,"
	Phys. Rev. B {\bf 101}, 220406(R) (2020).			
\bibitem{Mukerjee_08}
	S. Mukerjee and B. S. Shastry, 
	``Signatures of diffusion and ballistic transport in the stiffness, dynamical correlation functions, and statistics 
	of one-dimensional systems,"
	Phys. Rev. B {\bf 77}, 245131 (2008).	
\bibitem{Choi_19}
	D.-J. Choi, N. Lorente, J. Wiebe, K. von Bergmann, A. F. Otte, and A. J. Heinrich,
	``Atomic spin chains on surfaces,''
	Rev. Mod. Phys. {\bf 91}, 041001-24 (2019).
\bibitem{Monroe_21}
	C. Monroe, W. C. Campbell, L.-M. Duan, Z.-X. Gong, A. V. Gorshkov, P. W. Hess, R. Islam, K. Kim, N. M. Linke,
	G. Pagano, P. Richerme, C. Senko, and N. Y. Yao,
	``Programmable quantum simulations of spin systems with trapped ions,''
	Rev. Mod. Phys. {\bf 93}, 025001-57 (2021).
\bibitem{Baltz_18}
	V. Baltz, A. Manchon, M. Tsoi, T. Moriyama, T. Ono, and Y. Tserkovnyak,
	``Antiferromagnetic spintronics,''
	Rev. Mod. Phys. {\bf 90}, 015005-57 (2018).
\bibitem{Amico_22}	
	L. Amico, D. Anderson, M. Boshier, J.-P. Brantut, L.-C. Kwek, A. Minguzzi, and W. von Klitzing,
	``Atomtronic circuits: From many-body physics to quantum technologies,''
	Rev. Mod. Phys. {\bf 94}, 041001-30 (2022).
\bibitem{Mitchell_20}	
	M. W. Mitchell and S. P. Alvarez,
	``Quantum limits to the energy resolution of magnetic field sensors,''
	Rev. Mod. Phys. {\bf 92}, 021001-17 (2020).
\bibitem{Burkard_23}	 
	G. Burkard, T. D. Ladd, A. Pan, J. M. Nichol, and J. R. Petta,
	``Semiconductor spin qubits,''
	Rev. Mod. Phys. {\bf 95}, 025003-58 (2023)
\bibitem{Degen_17}	
	C. L. Degen, F. Reinhard, and P. Cappellaro,
	``Quantum sensing,''
	Rev. Mod. Phys. {\bf 89}, 035002-39 (2017).	
\end{references}
\end{document}